\documentclass[12pt]{iopart}
\usepackage{iopams}

\usepackage{graphicx}
\usepackage{dcolumn}
\usepackage{bm}

\graphicspath{{figures/}}

\newcommand{\beq}[0]{\begin{equation}}
\newcommand{\eeq}[0]{\end{equation}}
\newcommand{\beqfn}[0]{\begin{footnotesize}\[}
\newcommand{\eeqfn}[0]{\]\end{footnotesize}}
\newcommand{\beqa}[0]{\begin{small}\begin{eqnarray*}}
\newcommand{\eeqa}[0]{\end{eqnarray*}\end{small}}
\newcommand{\bsm}[0]{\begin{small}}
\newcommand{\esm}[0]{\end{small}}
\newcommand{\bfn}[0]{\begin{footnotesize}}
\newcommand{\efn}[0]{\end{footnotesize}}
\newcommand{\bst}[0]{\begin{small}}
\newcommand{\est}[0]{\end{small}}

\newcommand{\be}{\begin{enumerate}}
\newcommand{\ee}{\end{enumerate}}
\newcommand{\I}{\item}

\newcommand{\vlowk}{V_{{\rm low\,}k}}

\newcommand{\ad}{a^{\dagger}}

\newcommand{\flow}{s}

\newcommand{\bea}{\begin{eqnarray}}
\newcommand{\eea}{\end{eqnarray}}
\newcommand{\fmi}{\mbox{\,fm}^{-1}}

\newcommand{\la}{\langle}
\newcommand{\ra}{\rangle}

\newcommand{\amps}[1]{&#1&}
\newcommand{\ampseq}{\amps{=}}
\newcommand{\eqref}[1]{(\ref{#1})}

\newcommand{\Trel}{T_{\rm rel}}

\newcommand{\Vtwo}[2]{V_{#1#2}}
\newcommand{\Vthree}{V_{123}}
\newcommand{\Ttwo}[2]{T_{#1#2}}

\newcommand{\kf}[0]{k_{\rm F}}

\begin{document}

\title{New applications of renormalization group methods in nuclear physics}
\author{R.J. Furnstahl}
\address{Department of Physics, Ohio State University, Columbus OH 43210, USA}
\eads{furnstahl.1@osu.edu}

\author{K. Hebeler}
\address{Department of Physics, Ohio State University, Columbus OH 43210, USA}
\address{Institut f\"ur Kernphysik, Technische Universit\"at Darmstadt, 64289 Darmstadt, Germany}
\address{ExtreMe Matter Institute EMMI, GSI Helmholtzzentrum f\"ur Schwerionenforschung GmbH, 64291 Darmstadt, Germany}
\eads{kai.hebeler@physik.tu-darmstadt.de}

\date{\today}

\begin{abstract}
We review recent developments in the use of renormalization
group (RG) methods in low-energy nuclear physics.
These advances include enhanced RG technology, particularly for
three-nucleon forces, which greatly extends the reach and accuracy
of microscopic calculations.
We discuss new results for the nucleonic equation
of state with applications to astrophysical systems such as neutron
stars, new calculations of the structure and reactions of finite nuclei,
and new explorations of correlations in nuclear systems.    
\end{abstract}


\tableofcontents



\section{Introduction}\label{sec:Intro}
 
\begin{figure}[!tbp]
  \begin{center}
   \includegraphics[width=3in]{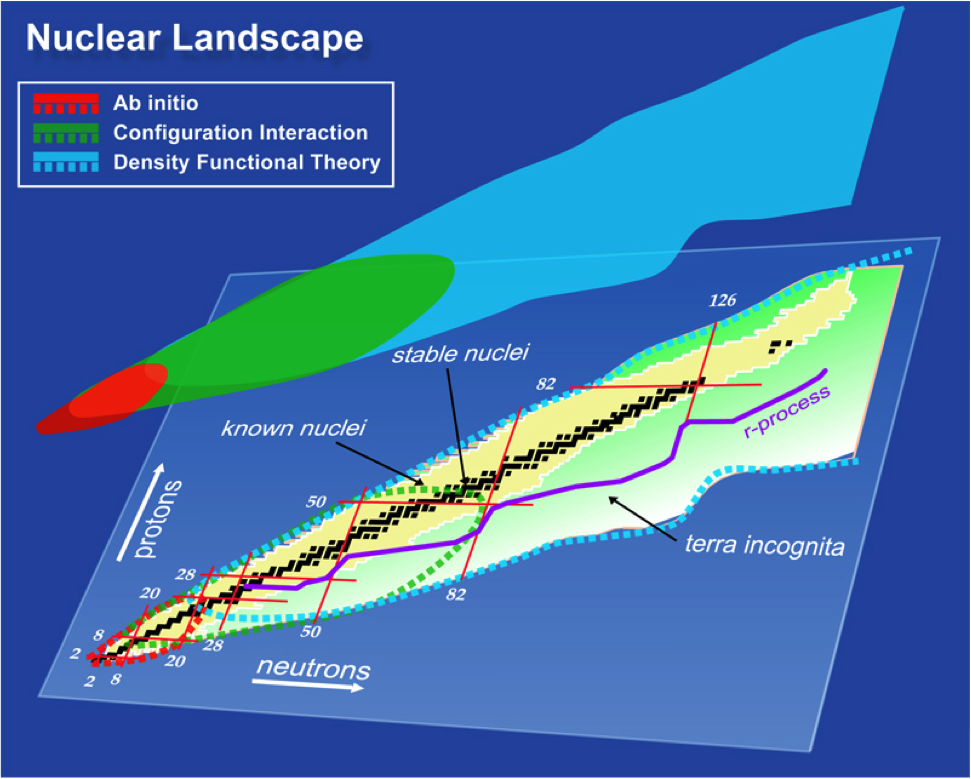}
   \caption{The nuclear landscape. A nuclide is specified by the number
   of protons and neutrons (figure from Ref.\cite{LRP:2007}). Most are unstable, 
   e.g., to radioactive decay via the weak interaction. Fewer than half of the
   estimated total have been measured by experiment. The overlapping domains of 
   theoretical methods are also indicated.}
   \label{fig:landscape}
  \end{center}
\end{figure}

The principal domain of low-energy nuclear physics is the table of the nuclides,
shown in figure~\ref{fig:landscape}.
There are several hundred stable nuclei (black squares) 
but also several thousand \emph{unstable}
nuclei are known through experimental measurements.  
However, the total 
number of nuclides is unknown (see the region marked ``terra incognita''), with theoretical
estimates suggesting approximately seven thousand total~\cite{citeulike:10848356}.
Many of these unstable nuclei (``rare isotopes'') 
will be created and studied in new and planned
experimental facilities around the world. 
An on-going
challenge for low-energy nuclear theory is to describe the structure and
reactions of all nuclei, whether measured or not.

A wide range of questions drive low-energy nuclear physics 
research~\cite{LRP:2007}.  
At the fundamental level: How do protons and neutrons make stable
nuclei and rare isotopes and where are the limits of nuclear existence?
What is the equation of state of nucleonic matter?  What is the origin
of simple patterns observed in complex nuclei?
How do we describe fission, fusion, and other nuclear reactions?
These topics inform and are in turn illuminated by applications
to other fields, such as astrophysics, where one can ask:
How did the elements from iron to uranium originate? 
How do stars explode? What is the nature of neutron star matter?
There are also connections to fundamental symmetries:    
Why is there now more matter than antimatter in the universe?
What is the nature of the neutrinos, what are their masses, 
and how have they shaped the evolution of the universe?
Finally, there are applications, for which we are led to ask: How can our knowledge of 
nuclei and our ability to produce them benefit humankind? The impact is 
very broad, encompassing the Life Sciences, Material Sciences, Nuclear 
Energy, and National Security.

\begin{figure}[!tbp]
  \begin{center}
   \includegraphics[width=2.5in]{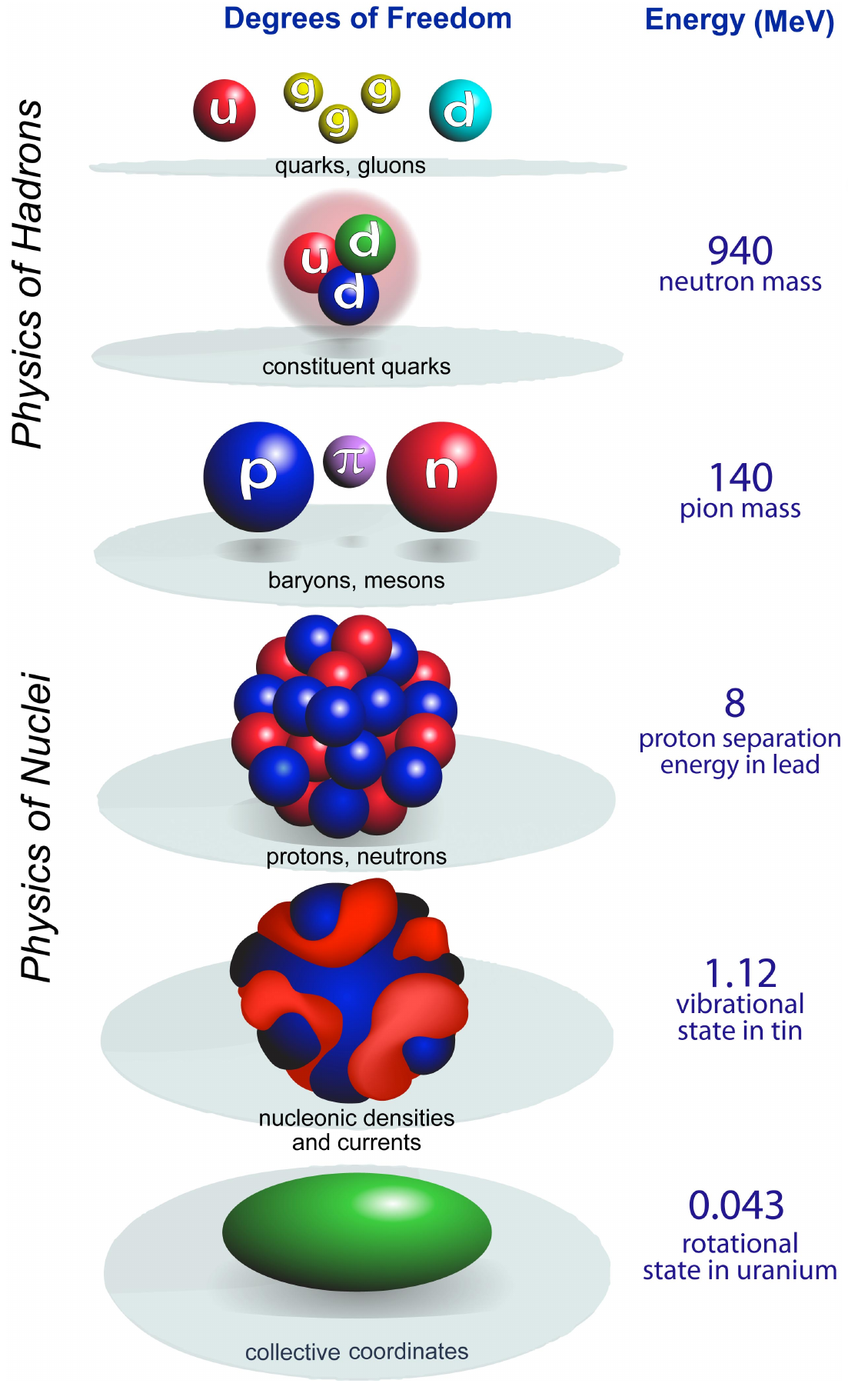}
   \caption{Hierarchy of nuclear degrees of freedom and associated
   energy scales~\cite{LRP:2007}.}
   \label{fig:scales}
  \end{center}
\end{figure}

In figure~\ref{fig:scales}, the energy scales of nuclear physics are shown 
schematically. The extended hierarchy provides both challenges and opportunities. 
We can exploit the hierarchy by treating the ratio of scales as an expansion parameter, 
leading to a systematic, model-independent treatment at lower energies using 
\emph{effective field theory} (EFT). The progression from top to bottom can be viewed 
as a reduction in resolution, which can be carried out theoretically using renormalization group (RG)
methods. Our focus in this review is on the intermediate region only, where 
protons and neutrons are the relevant degrees of freedom. But even within this limited 
scope, the concept of reducing resolution by RG methods is extremely powerful and fruitful.

\begin{figure}[t!]
  \begin{center}
   \includegraphics[height=2.5in]{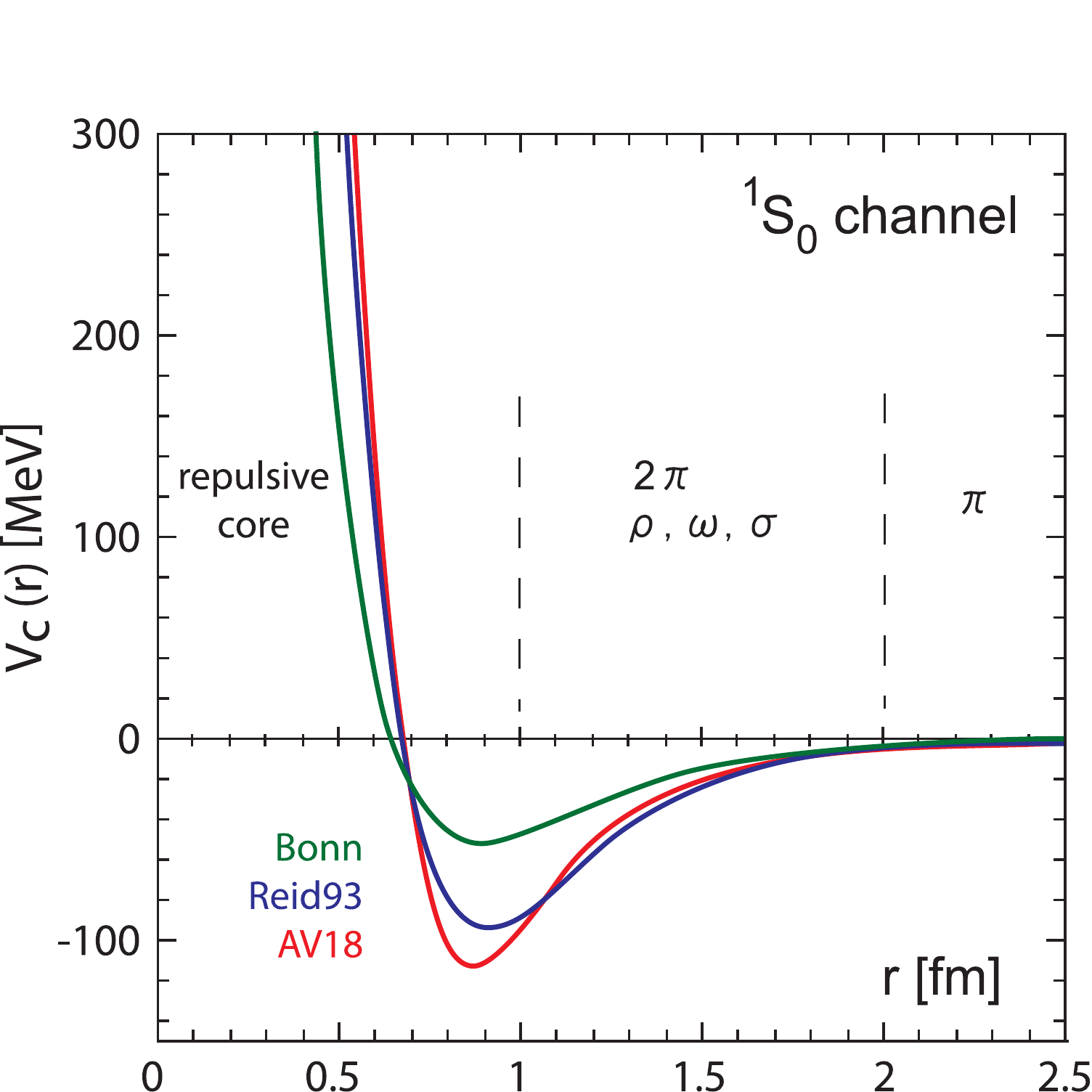}
   \hspace{1cm}
   \includegraphics[height=2.5in]{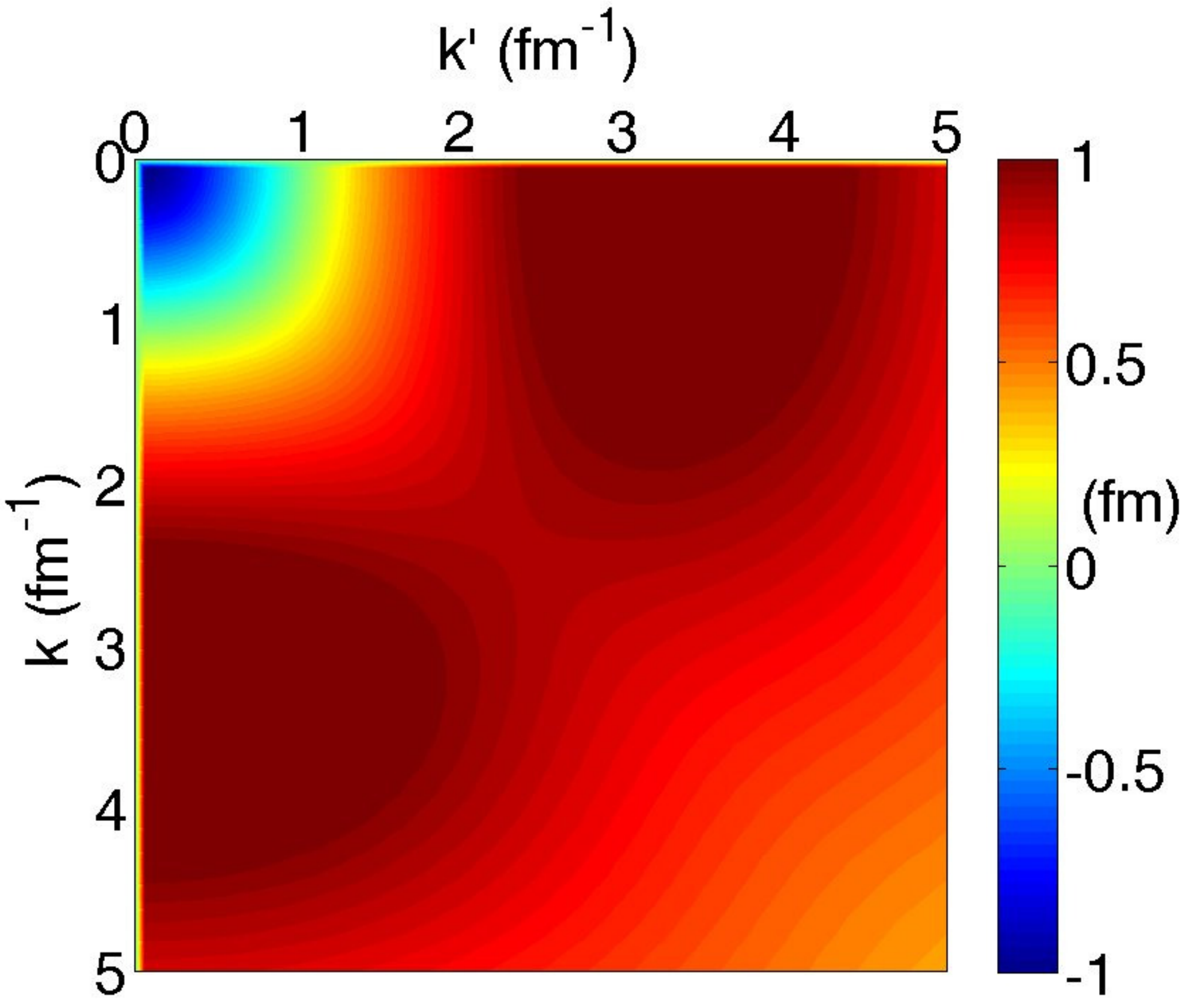}
   \caption{Left panel: three phenomenological potentials as functions of interparticle distance 
   that accurately describe proton-neutron scattering up to laboratory energies of 300\,MeV~\cite{Aoki:2008hh}. 
   Right panel: alternative momentum space representation of the AV18 potential
   $\langle k | v_{18} | k' \rangle$
   in the $^1$S$_0$ channel~\cite{Bogner:2009bt}.}
   \label{fig:phenpots}
  \end{center}
\end{figure}

The role of resolution scales can be illustrated using the phenomenon of diffraction: if the 
wavelength of light is comparable to or larger than an aperture, then diffraction is significant. 
Since two objects can only be resolved if the diffracted images do not overlap too much, the 
level of resolvable details depends crucially on the wavelength used. Being unable to resolve 
details at long wavelength is generally considered to be a disadvantage, but, as we will 
show in this review, it can be turned into an advantage.

A fundamental principle of \emph{any} effective low-energy description
is that if a system is probed at low energies, fine details are not 
resolved, and one can instead use low-energy variables. Short-distance structures 
can then be replaced by something simpler without affecting low-energy observables. 
This is analogous to using a truncated multipole expansion for a complicated charge 
or current distribution in classical electrodynamics. In the quantum case, EFT provides 
a systematic framework for such an expansion. We emphasize that while observable quantities
do not change, the physics interpretation can (and generally does) change with resolution.

In many-body systems the natural resolution scale is set by the
de Broglie wavelengths, i.e. by the typical momenta of the particles.
By using units for which $\hbar = c = 1$, the typical relative momentum in the 
Fermi sea of any large nucleus is of order $1\fmi$ or 200\,MeV. For our discussion, 
we will adopt $2\fmi$ as the (arbitrary but reasonable)
dividing line between low and high momentum for nuclei.

The left panel of figure~\ref{fig:phenpots} shows several phenomenological potentials as functions
of the interparticle distance.  Each was fit to reproduce nucleon-nucleon scattering phase 
shifts up to about 300\,MeV in lab energy. They are each characterized by a long-range
attractive tail from one-pion exchange, intermediate attraction,
and a strongly repulsive short-range ``core''. Alternatively, nuclear interactions can also be visualized in
momentum space. In the right panel the potential ${\rm AV} 18$ is 
shown as a function of the initial and final relative momenta $k$ and $k'$ of the two nucleons.
Here the color coding indicates the strength and sign of the coupling of different momentum states. 

It is evident from figure~\ref{fig:phenpots} that there are large matrix elements 
connecting low momenta with momenta much larger than $2\fmi$. This is directly associated with 
the strong repulsive core of the potential. The consequences of these couplings can also 
be seen in wave functions and probability densities. Generally, these repulsive couplings lead to a 
significant suppression of the probability density at small separations. This 
suppression, called ``short-range correlations'' (SRC) in this context, significantly complicates basis expansions.
For example, in a harmonic oscillator basis, which is frequently the choice for self-bound 
nuclei, convergence is substantially slowed by the need to accommodate these correlations. The factorial 
growth of the basis size with the number of nucleons then greatly limits the reach of calculations.

\begin{figure}[tb!]
  \begin{center}
   \includegraphics[width=2.8in]{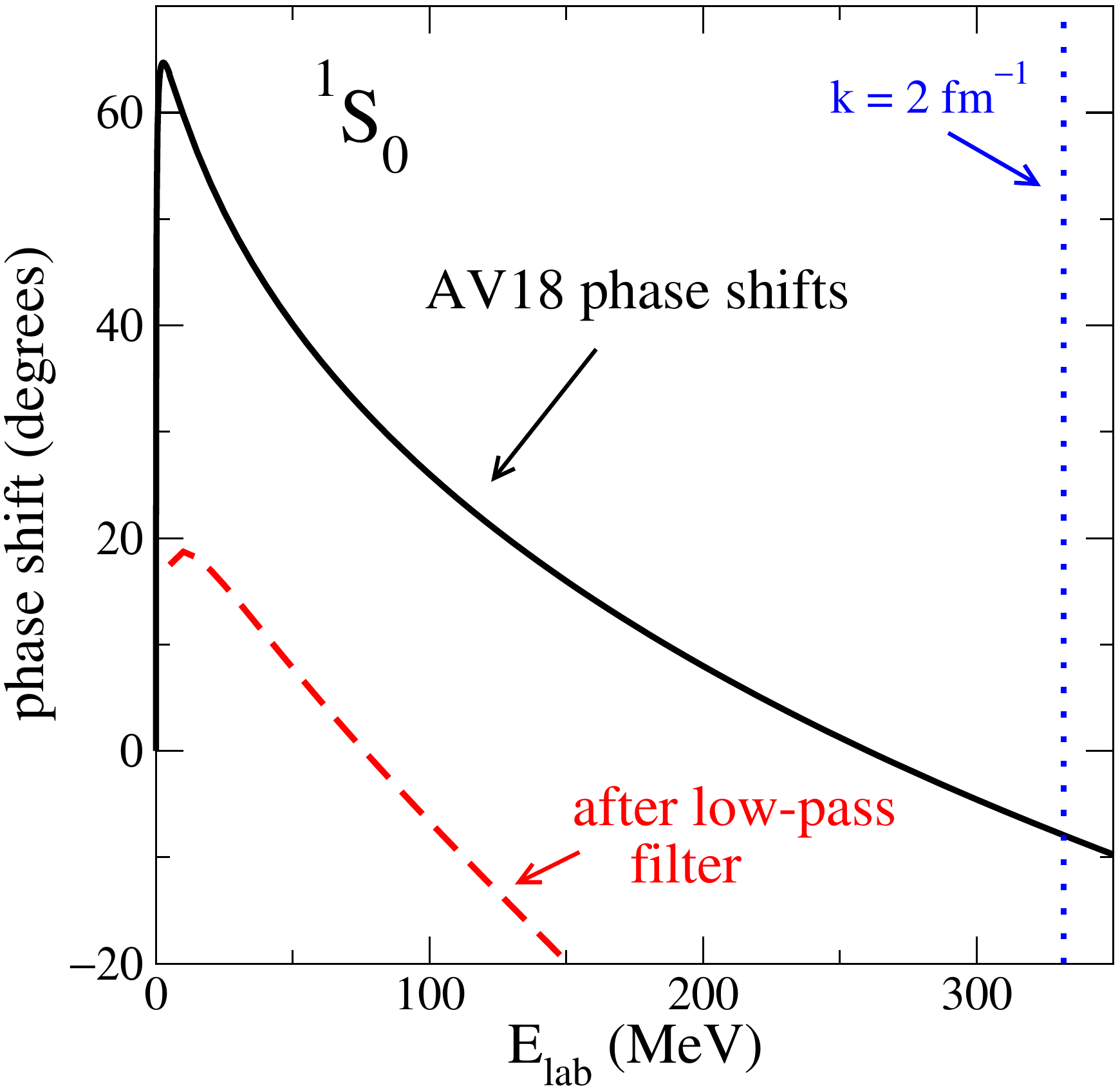}
   \caption{Effect of a low-pass filter on observables: the $^1$S$_0$
   phase shifts.  The unaltered
   AV18 phase shifts reproduce experimentally extracted phase shifts
   in this energy range.}
   \label{fig:lowpass}
  \end{center}
\end{figure}

The underlying problem is that the resolution scale
induced by the potential is mismatched with the scale of the
low-energy nuclear states. This might seem to be analogous to compressing a digital
photograph, which is readily accomplished by Fourier transforming and
then applying a low-pass filter, i.e. simply setting the short wavelength
parts to zero, and then transforming back. However, this strategy fails
for nuclear potentials: figure~\ref{fig:lowpass} shows
the results for the scattering phase shifts based on the original $\rm{AV} 18$ potential
compared to the potential where all matrix elements for $k > 2\fmi$ have been set to zero.
It is evident that the truncated potential fails completely to reproduce 
the phase shifts at all energies. The basic problem is that low
and high momenta are coupled by the potential when solving quantum mechanically
for observables.  

Our solution to this problem is to \emph{decouple} low and high energies rather than just
setting the high-energy parts to zero. This can be achieved by a short-distance unitary transformation 
$U$. For example, for the evaluation of energy expectation values we can insert the relation $U^\dagger U = 1$ twice:
\bea
  E_n \ampseq \langle \Psi_n | H | \Psi_n \rangle = \langle \Psi_n | U^\dagger U H U^\dagger U | \Psi_n \rangle  \equiv \langle \widetilde\Psi_n | \widetilde H | \widetilde \Psi_n \rangle \;.
    \label{eq:Utrans}  
\eea
with $\widetilde{H} = U H U^\dagger$ and $| \widetilde \Psi_n \rangle = U | \Psi_n \rangle$. 
In doing so operators \emph{and} wavefunctions get modified but 
observables remain unchanged. An appropriate choice of 
the unitary transformation can, in principle, achieve the
desired decoupling. This general approach has long been used in
nuclear structure physics and for other many-body applications~\cite{Ring:2005,preston1975structure,Navratil:2009ut}.
The new feature here is the use of RG flow
equations to create the unitary transformation successively via a series
of infinitesimal transformations.

The RG is well suited to this purpose and is more powerful and
versatile than many other approaches. The common features of RG for critical 
phenomena and high-energy scattering are discussed by Steven Weinberg in an essay in
Ref.~\cite{Guth:1984rq}. He summarizes:
\begin{quote}
``The method in its most general form can I think be understood
as a way to arrange in various theories that the degrees of freedom
that you're talking about are the relevant degrees of freedom for the
problem at hand.''
\end{quote}
This is the essence of what is done with the low-momentum interaction
approaches considered here: arrange for the degrees of freedom for nuclear structure
to be the relevant ones. This does not mean that other degrees of
freedom cannot be used, but to again quote
Weinberg~\cite{Guth:1984rq}: 
\begin{quote}
``You can use any degrees of freedom you want, but if you use the wrong ones, you'll be sorry.''
\end{quote}
The consequences of using RG for high-energy (particle) physics include improving perturbation
theory, e.g., in QCD. A mismatch of energy scales can generate large logarithms
that ruins perturbative convergence even when couplings by
themselves are small.
The RG shifts strength between loop integrals and coupling constants
to reduce these logs.  For critical phenomena in condensed matter
systems, the RG reveals the
nature of observed
universal behavior by filtering out short-distance degrees
of freedom.  

We see both these aspects in our applications
of RG to nuclear structure and reactions.  
As the resolution is lowered, nuclear calculations become more perturbative
(e.g., see figures~\ref{fig:EOS_SNM_lambda} and \ref{fig:roth1}) and the potentials flow toward universal form (e.g., see figure~\ref{fig:universality}).
The end result can be said
to make nuclear physics look more like quantum chemistry calculationally, opening 
the door to a wider variety of techniques (such as many-body perturbation
theory) and simplifying calculations (e.g., by improving convergence
of basis expansions).
On the other hand, microscopic
three-nucleon forces (3NF) have been found to be essential for accurate results,
and developing RG technology to handle them is an on-going challenge.
 
Over the last decade there have been increasing applications of RG
technology to low-energy physics.
This brief review focuses on the most
recent developments and therefore details of prior work are
necessarily limited.  More extensive reviews of the earlier progress
can be found in Refs.~\cite{Bogner:2009bt} and \cite{Furnstahl:2012fn} and references therein.
In Section~\ref{sec:RG_tech} we present the basics of RG
technology for evolving two- and three-body nuclear potentials, concentrating on new developments.  
Our starting potentials are generally taken from chiral EFT, which provides
a systematic hierarchy of initial two- and higher-body interactions.
Recent progress in calculating the
equation of state of nucleonic matter with applications to neutron
stars is discussed in Section~\ref{sec:NM}.   There are many new
results for both the structure and reactions of finite nuclei,
which are reviewed in Section~\ref{sec:Finite}.  In Section~\ref{Sect:Correlations} we discuss the use of RG with external
probes of nuclear correlations.  
We conclude in Section~\ref{Sect:Summary} with a summary of the main points, on-going developments, and important open questions.



\section{Renormalization group technology}
\label{sec:RG_tech}

In this section, we give a brief overview of the equations and techniques used to derive low-momentum 
nuclear interactions. We illustrate how RG methods decouple low- and high-energy degrees of freedom and how this
leads to simplified many-body calculations. We emphasize recent developments, such as the evolution of 
many-body forces, the use of local projections for visualization of interactions, and the phenomenon of 
universality in nuclear many-body forces at low resolution.

\subsection{Flow equations}

\begin{figure}
  \begin{center}
    \includegraphics[width=0.3 \textwidth]{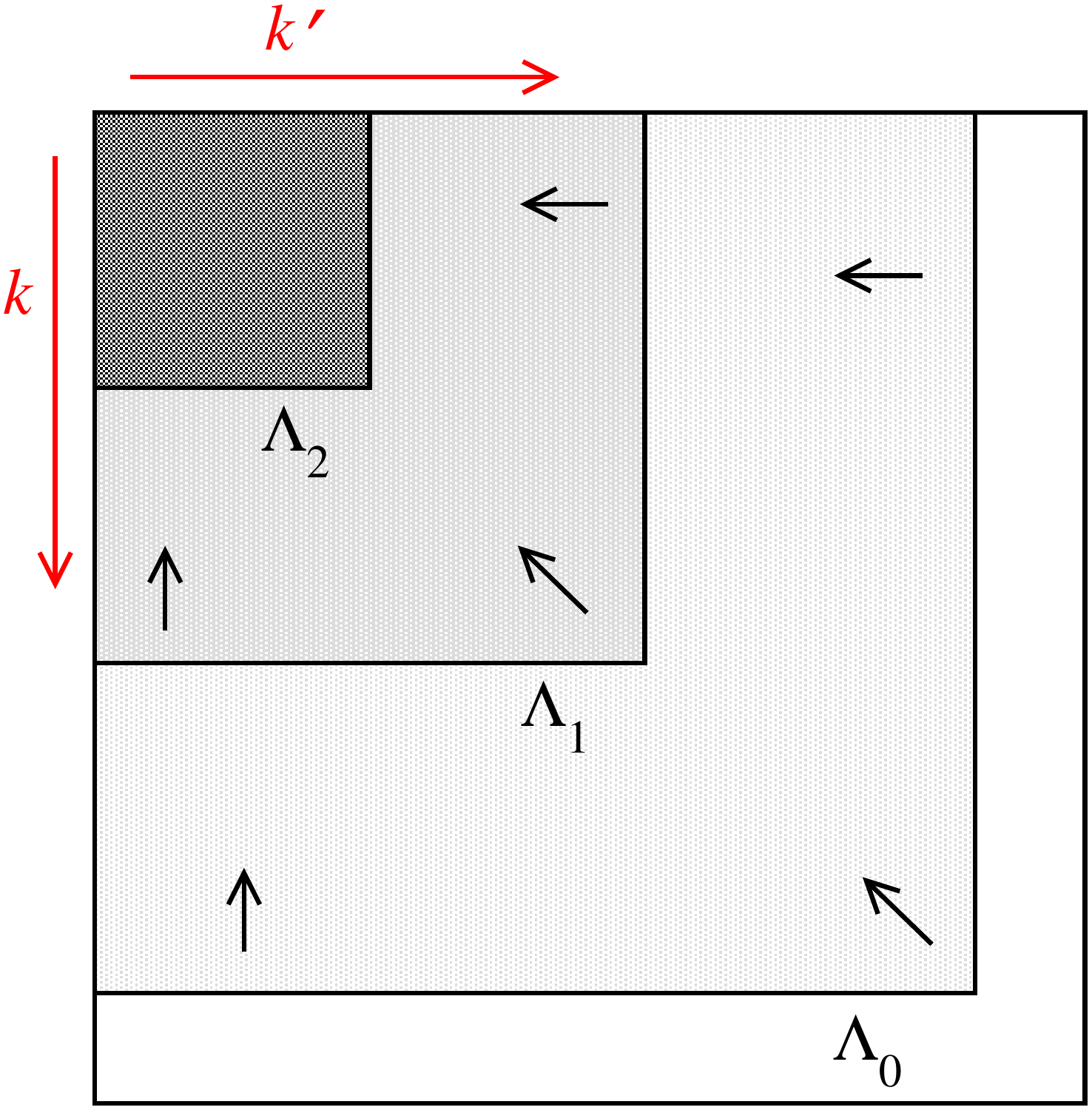}
    \hspace{1.0cm}
    \includegraphics[width=0.3 \textwidth]{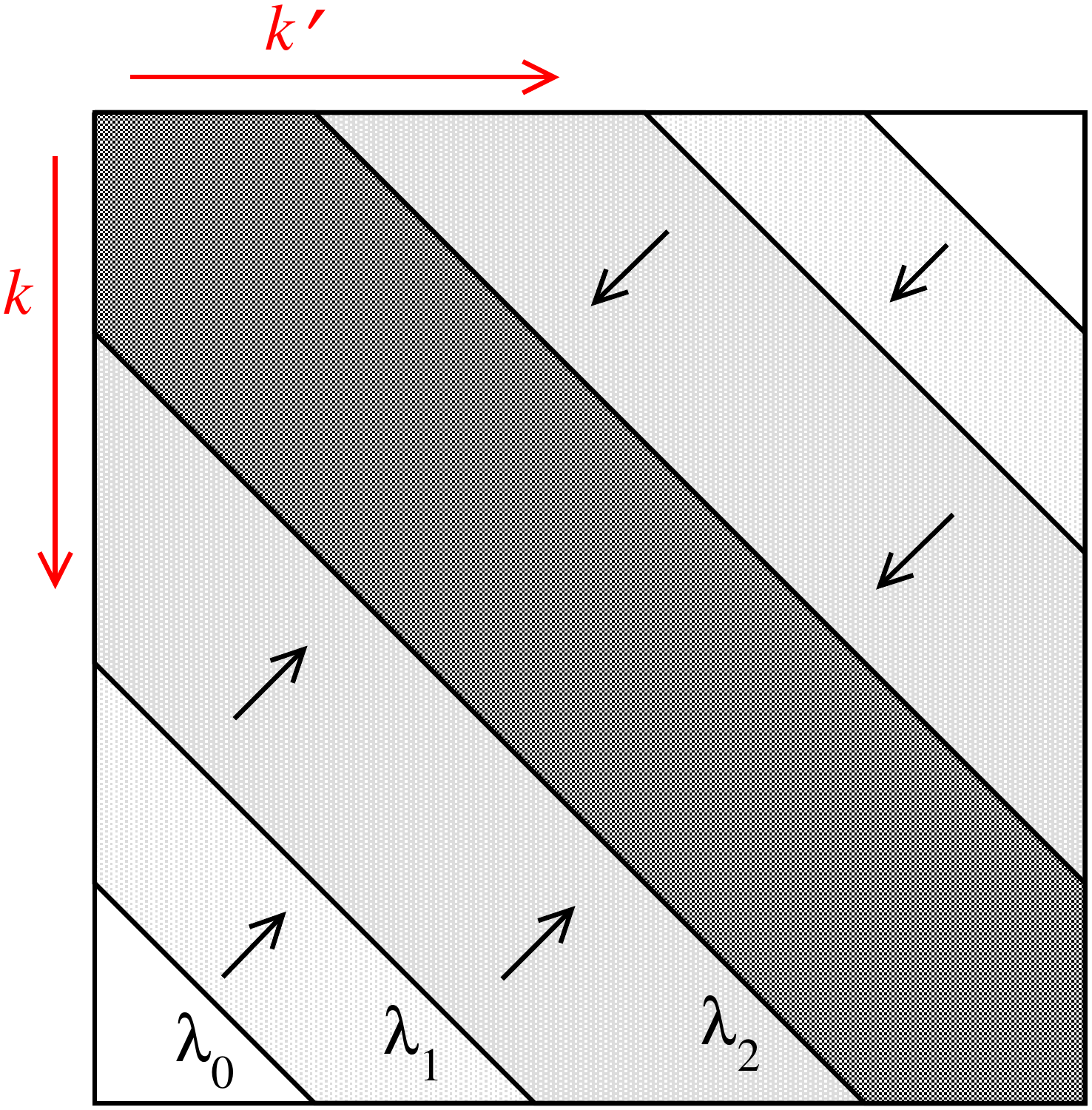}
  \end{center}
  \caption{Schematic illustration of two types of RG evolution for NN potentials in momentum space: (a)~$\vlowk$ running in $\Lambda$, and (b)~SRG running in $\lambda$ for $G_s = T$ (see main text). Here $k$ $(k')$ denote the relative momenta of the initial (final) state. At each $\Lambda_i$ or $\lambda_i$, the matrix elements outside of the corresponding lines are negligible, so that high- and low-momentum states are decoupled.}
 \label{fig:vlowkschematic}
\end{figure}

At the heart of every RG framework is a flow equation. In general, the flow equation
is a set of coupled differential equations for operators and couplings. In 
low-energy nuclear physics, these equations specify how matrix elements of the nuclear interactions change
when the RG resolution scale is varied by an infinitesimal amount. In figure~\ref{fig:vlowkschematic}, two common 
options are shown for how the RG can be used to decouple a two-body Hamiltonian. As in figure \ref{fig:phenpots}, $k$ 
$(k')$ denotes the initial (final) relative momentum of the interacting nucleons. The more traditional approach in 
the left panel lowers a momentum cutoff $\Lambda$ in 
small steps, with the matrix elements adjusted by requiring some observables such as nucleon-nucleon scattering phase shifts remain 
invariant up to the momentum scale $\Lambda$. Matrix elements well above $\Lambda$ are zero and are therefore trivially decoupled. 
In low-energy nuclear physics this approach is typically referred to as ``$\vlowk$'' \cite{Bogner:2009bt,Bogner:2003wn}, and shares many features 
with the traditional Wilsonian RG approach in field theory, corresponding to successively integrating out ``momentum shells''~\cite{peskin1995introduction}. In recent years this $\vlowk$ approach has been applied very successfully to various two-nucleon forces~\cite{Bogner:2009bt}.
However, as discussed in Section~\ref{sec:Intro}, 
three- and higher-body forces have been shown to play an important role in nuclear systems and there are unsolved technical complications
in systematically treating such many-body forces within the $\vlowk$ framework.

A more recent RG approach to the nuclear Hamiltonian is illustrated in the right panel of
figure~\ref{fig:vlowkschematic}, in which the matrix is driven toward band-diagonal form to achieve
decoupling of low- and high momenta. This RG framework was originally developed in the early 1990's by 
Wegner~\cite{Wegner:1994,Wegner:2000gi,Wegner:2006zz} for condensed matter applications under 
the name ``Hamiltonian flow equations'' and independently by Glazek and Wilson~\cite{Glazek:1993rc} for 
solving quantum chromodynamics in light-front formalism under the name ``similarity renormalization group''
(SRG). Only in the last five years was it realized that this approach is particularly well suited for low-energy 
nuclear physics, where it is technically simpler and more versatile than other methods such as the $\vlowk$ 
approach~\cite{Bogner:2009bt,Bogner:2006pc}.

The basic idea of the SRG is to apply a unitary transformation 
as in Eq.~\eqref{eq:Utrans} to an initial Hamiltonian, $H = T_{\rm{rel}} + V$,
in a series of infinitesimal steps, labeled by the flow parameter $s$ increasing
from $s=0$:
\begin{equation}
   H_\flow = U_\flow H U^\dagger_\flow \equiv T_{\rm{rel}} + V_\flow 
\end{equation}
with $U^\dagger_\flow U_\flow = U_\flow U^\dagger_\flow = 1$. Here, $T_{\rm{rel}}$ is the relative kinetic energy operator, which is
chosen to be invariant under the unitary transformation and $V_s$ indicates all two-nucleon and higher-body interactions. 
Any unitary transformation can be recast in form of a flow equation~\cite{Bogner:2006pc}:
\begin{equation}
  \frac{dH_\flow}{d\flow} = [\eta_\flow,H_\flow] \quad \textrm{with}  \quad \eta_\flow \equiv \frac{dU_\flow}{d\flow} U^\dagger_\flow.
  \label{eq:SRG}
\end{equation}
The anti-Hermitian generator $\eta_\flow$ can be specified by a commutator of $H_\flow$ with a 
Hermitian operator $G_s$, i.e. $\eta_\flow = [G_s, H_\flow]$, to obtain a transformation that tends to diagonalize $H_\flow$ in the eigenbasis of $G_s$.

\begin{figure}
  \begin{center}
  \includegraphics[width=1.0\textwidth]{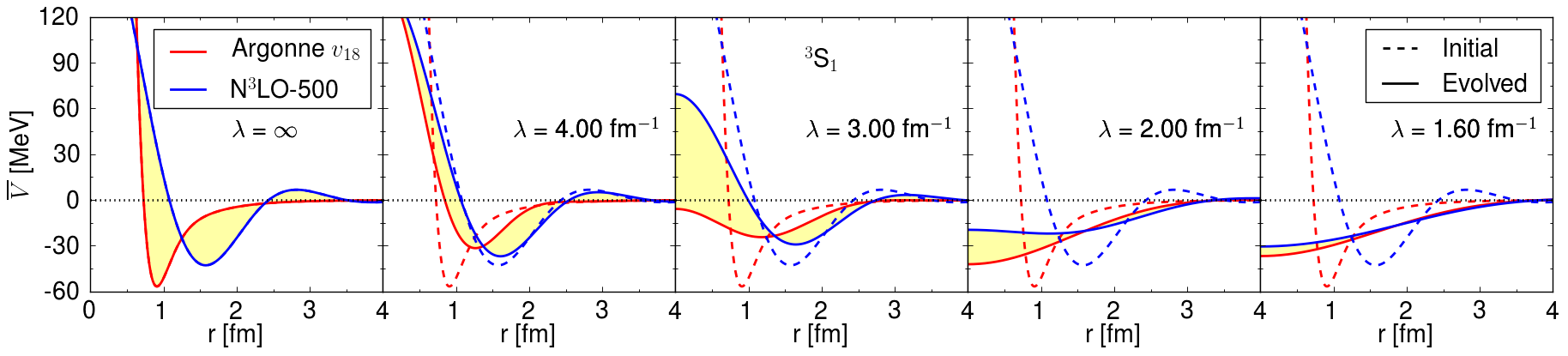}    
  \end{center}
  \caption{Local projection of AV18 and N3LO(500\,MeV) potentials in $^3$S$_1$ channel at different resolutions~\cite{Wendt:2012fs}. The dashed lines show the matrix elements of the initial unevolved potentials.}
  \label{fig:LPsideSwave}
\end{figure}

\subsection{RG evolution of nucleon-nucleon interactions}
\label{subsec:NN_evolv}

The operator $G_s$ completely defines the flow of a given initial Hamiltonian and there are many possible choices
one can consider. The most common choice in recent applications for its
simplicity and effectiveness is $G_s = T_{\rm{rel}}$. The evolution of a two-body force as 
$s$ increases or as $\lambda \equiv s^{-1/4}$ decreases is illustrated for this case schematically in the right panel of figure~\ref{fig:vlowkschematic}. 
The off-diagonal strength gets successively suppressed and the matrix is driven towards a band-diagonal form 
with $\lambda$ a measure of the degree of decoupling (see \cite{Glazek:2008pg,Wendt:2011qj} for exceptions).
During the RG evolution the low-energy physics is shifted to the low-momentum part of the Hamiltonian 
and other observables. The preservation of the low-energy physics \emph{at all scales} 
under the RG flow is guaranteed by the unitarity of the transformation $U_s$. 

The effects of the RG evolution can also be visualized in configuration space. Since SRG-evolved interactions 
are in general non-local, i.e. non-diagonal in coordinate representation, it is convenient to 
consider for visualization purposes a local projection of the nuclear potentials
(given here for S-waves)~\cite{Wendt:2012fs}:
\beq
\overline V_\lambda(r) = \int_0^\infty \! r'{}^2\,dr'\, V_\lambda(r,r')
  \;.
\eeq
The local projections for two realistic NN potentials at different SRG resolution scales
are shown in figure~\ref{fig:LPsideSwave}. As already shown in figure \ref{fig:phenpots}, the 
potential ``AV18'' shows a very strong repulsive short-range part at high resolution scales (left panel). 
The potential ``N$^3$LO'' denotes a very commonly used high-precision potential which has been derived 
within chiral EFT~\cite{Entem:2003ft}. For this potential the short-range repulsion is much less pronounced, but 
still significant. Clearly, during the flow to lower scales the short-range repulsive parts get gradually 
dissolved~\cite{Wendt:2012fs}. Also evident is the flow of the two potentials, initially quite different, toward 
a universal form at the lower values of $\lambda$. This phenomenon is referred to
as ``universality'' of nucleon-nucleon forces at low resolution and will be discussed in more detail below.

Besides the canonical choice $G_s = T_{\rm{rel}}$, it is also possible to choose a 
generator that reproduces the block diagonal (as opposed to band diagonal) form of the $\vlowk$ RG
shown schematically in figure~\ref{fig:vlowkschematic}, except that
the transformation will be unitary~\cite{Anderson:2008mu}.
Recently, alternative generators  that allow computationally much faster evolution have been explored,
see Ref.~\cite{Li:2011sr} for details.  

\begin{figure}
  \begin{center}
    \includegraphics[width=0.9 \textwidth]{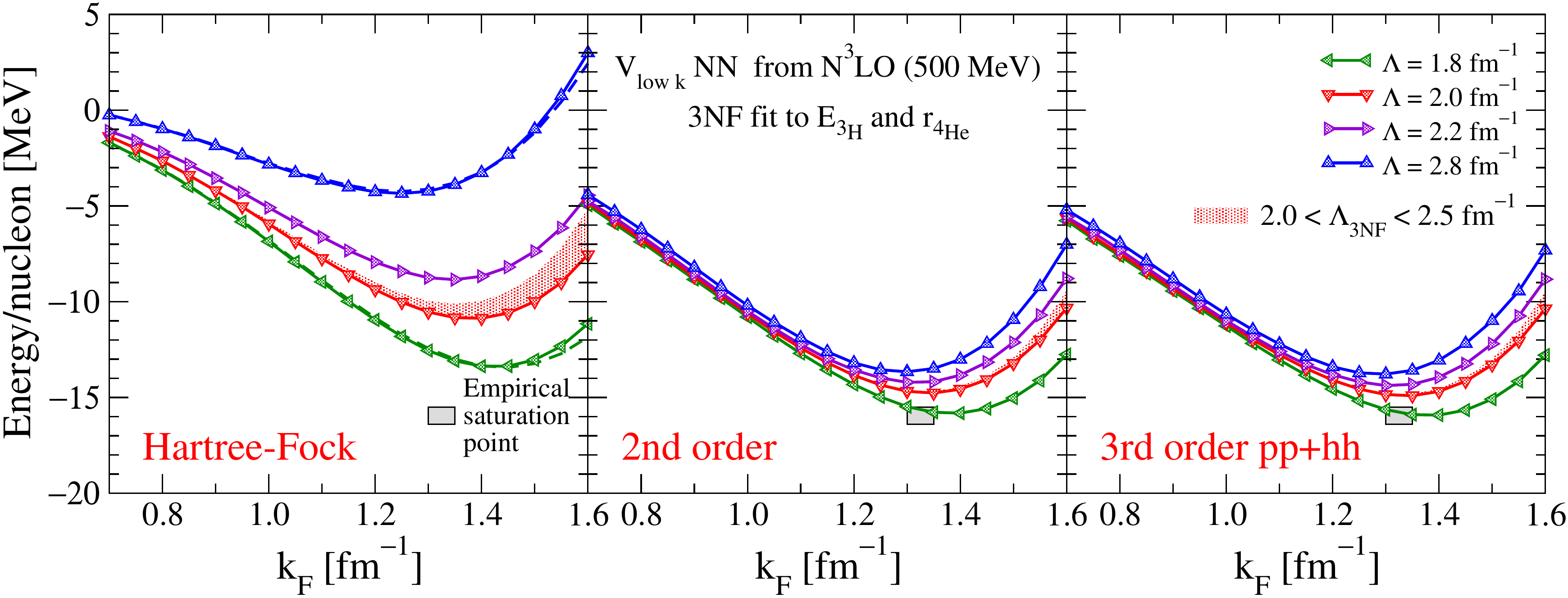}      
  \end{center}
  \caption{Nuclear matter energy per particle versus Fermi momentum $\kf$ at the Hartree-Fock level (left) and including second-order (middle) and third-order particle-particle/hole-hole contributions (right), based on evolved N$^3$LO NN potentials and 3NF fit to $E_{\rm^3H}$ and $r_{\rm^4He}$. Theoretical uncertainties are estimated by the NN (lines)/3N (band) cutoff variations. See Ref.~\cite{Hebeler:2010xb} for details.}
  \label{fig:EOS_SNM_lambda}
\end{figure}

The evolution to lower resolution scales is accompanied by a shift of physics. In particular, 
effects of short-range two-body interactions are replaced by new longer-range two- and many-body forces. 
We emphasize that the relative importance of separate contributions to observables from different sectors 
of the Hamiltonian or from different orders in perturbation theory are 
resolution dependent and consequently are not themselves observables.

We illustrate the improved perturbativeness at low resolution scales using results for infinite nuclear matter. We 
will discuss the physics of nuclear matter and its relevance for astrophysical applications in more detail in Section~\ref{sec:NM}. Here, we 
only illustrate the convergence pattern of the many-body expansion as a function of the RG resolution scale. In figure~\ref{fig:EOS_SNM_lambda} the 
energy per particle of symmetric nuclear matter is shown
as a function of Fermi momentum $\kf$, with density $\rho = 2 \kf^3/(3\pi^2)$. The 
grey square represents the empirical saturation point in each panel. Its boundaries reflect the 
ranges of nuclear matter saturation properties predicted by phenomenological Skyrme energy functionals that most accurately 
reproduce properties of finite nuclei~\cite{Bender:2003jk}. The figure shows results based on $\vlowk$-evolved NN interactions plus contributions from 
3N interactions in three many-body approximations: Hartree-Fock (left), Hartree-Fock plus second-order contributions (middle), and 
additionally summing selected third-order contributions (right). Evidently, the size of the higher-order contributions in the many-body expansion become 
smaller with decreasing cutoff $\Lambda$; for all values considered the third-order diagrams provide only very small contributions. 

A calculation without approximations should be independent of the RG resolution scale. In practice, there will be approximations 
both in the implementation of the RG and in the subsequent calculations of nuclear structure observables. That is, cutoff 
dependence arises because of the truncation of induced many-body forces (see below) or because of 
many-body approximations. Hence it is possible to use the cutoff-dependence as a diagnostic of approximations and 
to estimate theoretical errors. For example, in figure~\ref{fig:EOS_SNM_lambda} the second-order results show a significant narrowing of the spread over a large
density region compared to Hartree-Fock results. The observed remaining cutoff dependence in the right panel provides a scale for neglected four-body 
interactions. Indeed, this size is consistent with the expected size based on chiral EFT~\cite{Kaiser:2012ex} (see Section~\ref{sec:NM}).

\begin{figure}
  \begin{center}
    \includegraphics[width=0.8 \textwidth]{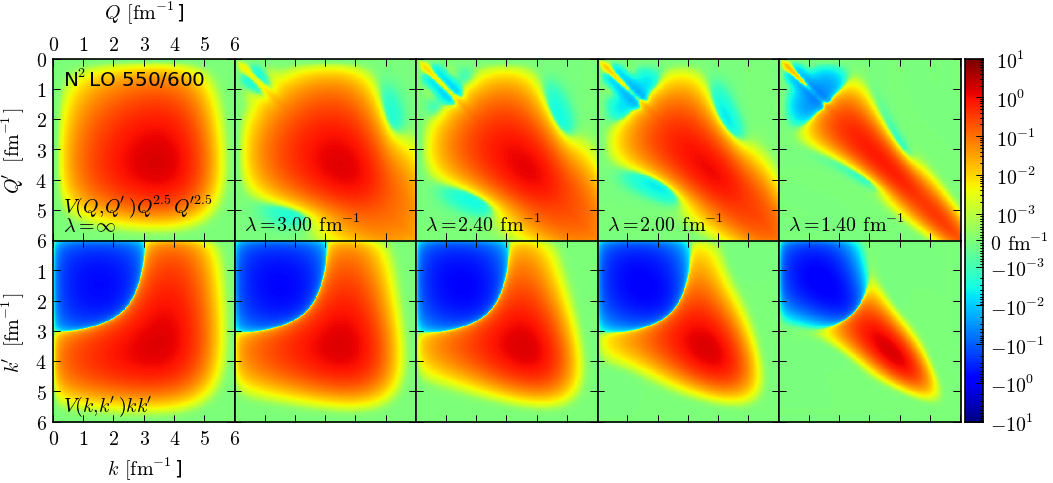}
  \end{center}
  \caption{Contour plot of the 3NF (upper row) and 2NF (lower row) as a function of $\lambda$ for the 550/600 MeV chiral EFT potential. The lowest antisymmetric hyperspherical partial wave is plotted as well as the two-body partial waves that are embedded in this three-body partial wave. See Ref.~\cite{Wendt:2013bla} for details.}
  \label{fig:kyleNN_NNN}
\end{figure}
 
\subsection{RG evolution of three-nucleon interactions}
When evolving nuclear interactions to lower resolution, it is inevitable that many-body interactions and operators 
are induced even if initially absent~\cite{Jurgenson:2009qs}. This might be considered problematic if nuclei could be accurately 
calculated based on only NN interactions, as was assumed for much of the history of nuclear structure calculations. However, chiral EFT has revealed 
a natural scale and hierarchy of many-body forces, which dictates that more than NN 
be included in modern calculations of nuclei and nucleonic matter~\cite{Epelbaum:2008ga,Hammer:2012id}. 
Thus, the real concern is whether this hierarchy is maintained as nuclear interactions are evolved~(see Section~\ref{sec:NM}).
The consistent treatment of three-body forces in the RG evolution and in many-body calculations is a complex task and currently one of 
the key frontiers in nuclear physics. 

Currently, there exist different ways to treat 3N forces in the RG framework:

\begin{itemize} 
\item[(a)] Starting from nuclear NN and 3N forces, derived and fitted in chiral EFT, it is 
possible to systematically evolve the full Hamiltonian. For calculations of light and medium mass nuclei this has 
been achieved by representing Eq.~(\ref{eq:SRG}) using a discrete harmonic oscillator basis~\cite{Jurgenson:2009qs}.
Results for light nuclei based on this approach are very promising~\cite{Jurgenson:2010wy, Roth:2011ar}. 
For heavier nuclei however, significant scale dependencies have been found~\cite{Roth:2011ar, Roth:2011vt} (see also Section~\ref{sec:Finite}),
which suggest that infinite matter will not be realistic. These could be indications of significant induced 
4N forces or possibly an insufficient evolution of 3N forces due to basis truncations.

\item[(b)] Only the NN interactions are evolved 
with RG methods and then the chiral EFT
N$^2$LO 3N force is added with its short-range parameters determined at the low-momentum scale from fits to 
few-body systems. This procedure assumes that the long-range part of the 3N forces remains invariant under the RG 
transformations and that the N$^2$LO operator structure is a sufficiently complete operator basis that induced contributions can
be absorbed to good approximation. The results shown in figure~\ref{fig:EOS_SNM_lambda} are based on this strategy (and 
also Refs.~\cite{Otsuka:2009cs, Holt:2010yb}), and are found to be in agreement with nuclear phenomenology within the 
theoretical uncertainties~\cite{Hebeler:2010xb,Hebeler:2009iv}.

\item[(c)] Recently a complementary framework to consistently evolve 3N forces in a continuous 
plane-wave basis has been developed~\cite{Hebeler:2012pr}. In this approach the evolution of NN and 3N forces is separated explicitly, which allows the
subtraction of NN interactions in a three-body basis to be avoided. Such momentum-space interactions can be directly used
for calculations of infinite systems and finite nuclei. First results for neutron matter 
based on such interactions are presented in Section~\ref{sec:NM}. Since SRG transformations are usually characterized 
by the coupling patterns of momentum eigenstates, the momentum basis is a natural basis in which to 
construct the SRG generator $\eta_s$. The construction of optimized generators for suppressing the growth of 
many-body forces is currently under active investigation.

\begin{figure}
  \begin{center}
   \includegraphics[width=0.4 \textwidth]{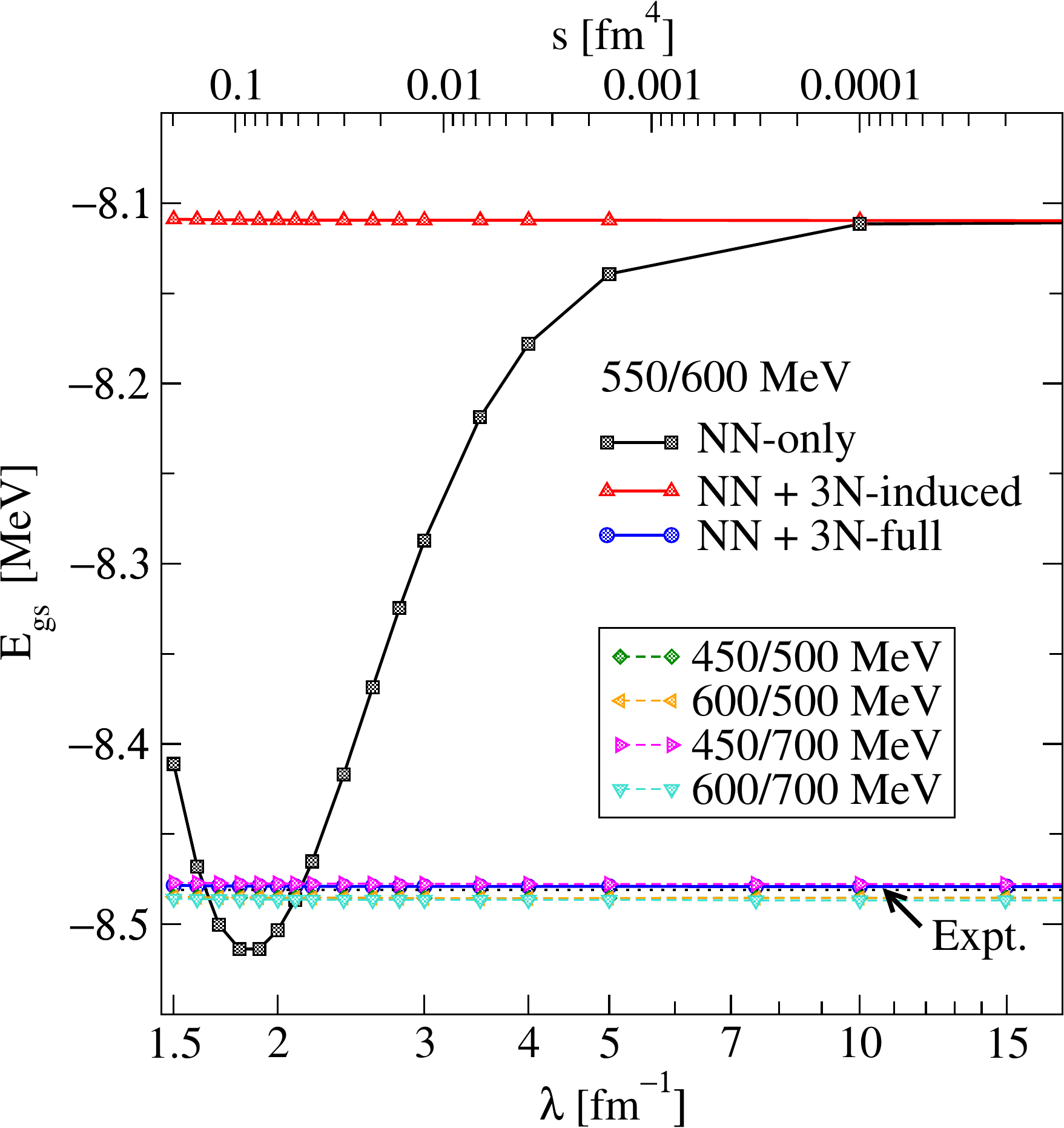}
  \end{center}
  \caption{Ground state energy of $^3$H as a function of the flow parameter $\lambda = s^{-1/4}$ for different initial chiral interactions. NN-only means initial and induced 3N interactions are discarded, NN+3N-induced takes only induced 3N interactions into account, and 3N-full contains initial and induces 3N interactions. The black dotted line shows the experimental binding energy.}
  \label{fig:Triton}
\end{figure}

\item[(d)] Finally, there is a new framework for evolving NN and 3N forces in a hyperspherical momentum representation~\cite{Wendt:2013bla}. This framework represents a hybrid approach in the sense that it is based on a continuous momentum basis like 
(c), but the RG evolution is performed for the entire Hamiltonian as in (a). The hyperspherical basis is particularly useful 
for visualizing matrix elements of interactions. In figure~\ref{fig:kyleNN_NNN} are shown some representative matrix elements 
of NN (bottom) and 3N (top) interactions at different resolution scales. It is evident that both display the characteristic SRG decoupling pattern
as they are evolved to lower scales.

\end{itemize}

\begin{figure}
  \begin{center}
    \includegraphics[width=0.8 \textwidth]{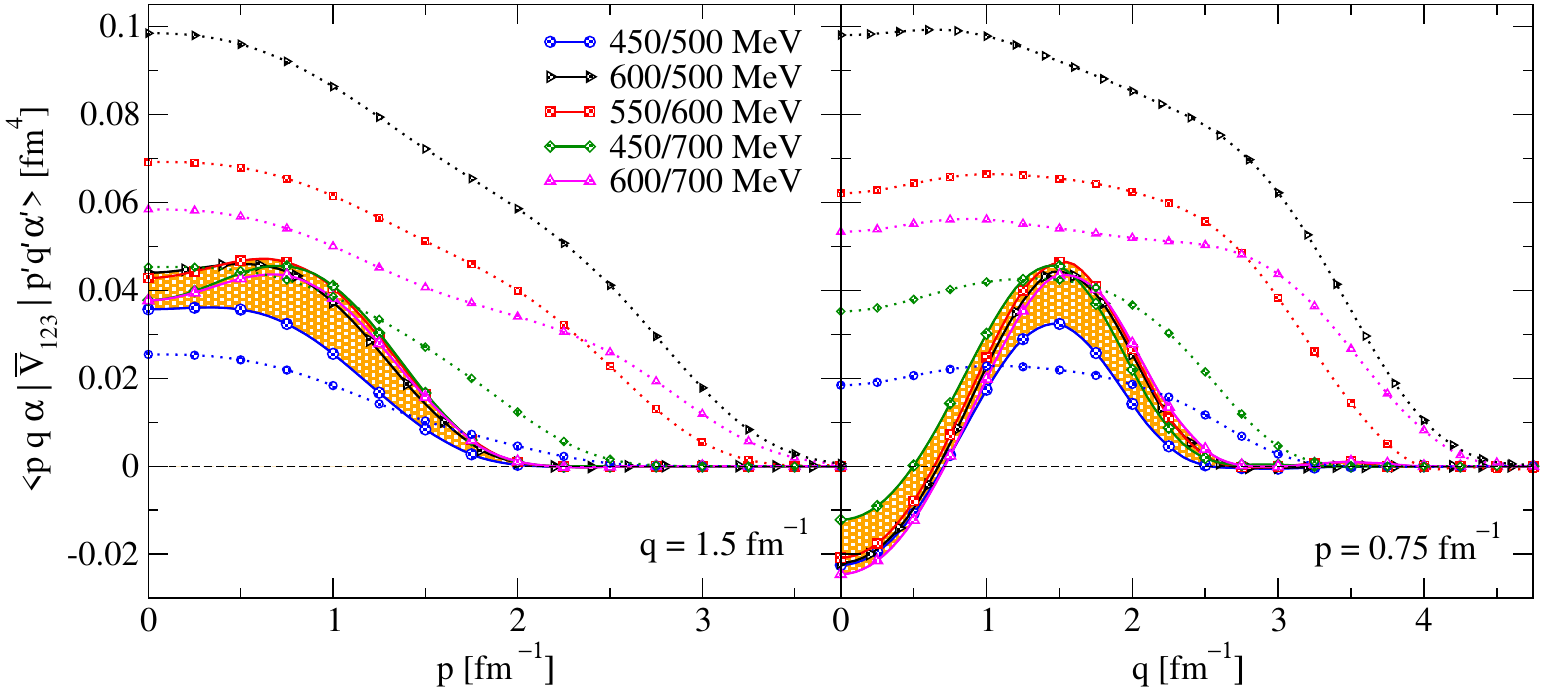}
  \end{center}
  \caption{Matrix elements of the initial 3N forces (dotted lines) compared to evolved 3N forces (solid lines) 
  at $\lambda=1.5\,\rm{fm}^{-1}$ for different interactions labeled by the values of the cutoffs 
  $\Lambda/\tilde{\Lambda}$~\cite{Epelbaum:2005pn}. For details see Ref.~\cite{Hebeler:2012pr}. The shaded 
  band marks the maximal variation between different evolved 3N force matrix elements.}
  \label{fig:universality}
\end{figure}

To illustrate the basic ideas behind evolving three-body forces within the SRG framework for $G_s=\Trel$,
we adopt a notation in which $\Vtwo12$ means the two-body interaction between particles 1 and 2
while $\Vthree$ is the irreducible three-body potential.  We start
with the Hamiltonian in the three-particle space:
\begin{equation}
  H_s = \Trel + \Vtwo12 + \Vtwo13 + \Vtwo23 + \Vthree \equiv \Trel + V_s
  \,.
  \label{eq:H}
\end{equation}
The SRG flow equation (\ref{eq:SRG}) in this space is
\begin{equation}
  \frac{dH_s}{ds} = \frac{dV_s}{ds} = \frac{d\Vtwo12}{ds} + \frac{d\Vtwo13}{ds} 
                      + \frac{d\Vtwo23}{ds}
                      + \frac{d\Vthree}{ds} = [[\Trel, V_s], H_s] \,,
  \label{eq:dVs}
\end{equation}
where $dH_s/ds = dV_s/ds$ because we define $\Trel$ to be 
independent of $s$.
The equations for each of the two-body potentials (which are completely determined
by their evolved matrix elements in the two-particle space) are
\begin{equation}
  \frac{d\Vtwo12}{ds} = [[\Ttwo12, \Vtwo12], \Ttwo12 + \Vtwo12] 
  \,, \label{eq:Vtwoa} 
\end{equation}
were $T_{12}$ denotes the
relative kinetic energy of particle $1$ and $2$,
and similarly for the $\Vtwo13$ and $\Vtwo23$ equations.  
When Eq.~\eqref{eq:Vtwoa} is used in Eq.~\eqref{eq:dVs}, particle~3 is 
a \emph{spectator}.
If we use a discrete basis this is not a problem, and 
it is straightforward to represent the Hamiltonian
as a matrix and evolve the entire matrix in a three-body basis via 
Eq.~(\ref{eq:dVs}). 
But in 
a continuous basis, delta functions associated with spectator particles in two-body interaction processes
make this representation problematic. 
However, 
it is straightforward to show that the
derivatives of two-body potentials on the left side cancel precisely
with terms on the right side, leaving an explicit
equation for evolving the three-body interaction separately
from the two-body interaction~\cite{Bogner:2006pc},
\begin{eqnarray}
  \frac{d\Vthree}{ds} &=
  [[\Ttwo12,\Vtwo12], \Vtwo13 + \Vtwo23 + \Vthree]
  + [[\Ttwo13,\Vtwo13], \Vtwo12 + \Vtwo23 + \Vthree]
  \nonumber \\
    &+ [[\Ttwo23,\Vtwo23], \Vtwo12 + \Vtwo13 + \Vthree]
    +  [[\Trel,\Vthree],H_s] \,.
  \label{eq:diffeqp}
\end{eqnarray}
The cancellations eliminate the
disconnected spectator contributions (all 3 indices appear in each term on the right
side) and the ``dangerous'' delta functions, so 
Eq.~\eqref{eq:diffeqp} can now be solved directly in a continuous basis.
From this equation it is manifest that the matrix elements of $V_{123}$ will change even if initially $V_{123}=0$ on the right side.

In figure~\ref{fig:Triton} we show for illustration the ground state energy of the triton for different initial chiral interactions at
different SRG resolution scales based on method (b) (very similar results have been found using strategies (a)~\cite{Jurgenson:2009qs} 
and (d)~\cite{Wendt:2013bla}). For one interaction we show three different cases: ``NN-only'' corresponds to taking only NN interactions into account
and discard all 3N contributions, in the case ``NN+3N-induced'' we start with only NN forces at $\lambda=\infty$ but we keep all 
3N contributions which are ``induced'' during the SRG evolution and finally for ``NN+3N-full'' we include in addition 
3N interactions at $\lambda=\infty$ and retain all induced 3N contributions during the RG flow. Evidently, neglecting induced 3N forces results in a significant variation of the binding energy. 
Only after retaining consistently all 3N contributions does the binding energy remain invariant under changes in the SRG resolution scale.
Of course, this strict invariance holds only for three-body systems since neglected higher-body forces cannot contribute. It is one of the main frontiers
to understand the nature and importance of many-body forces in four- and higher-body systems.

\subsection{Universality of low-momentum nuclear interactions}
As already illustrated in figure~\ref{fig:LPsideSwave}, low-resolution NN interactions are found to be quantitatively very 
similar~\cite{Bogner:2009bt,Bogner:2006pc}. This universality can be attributed to common long-range pion physics and 
phase-shift equivalence of all potentials, which is reflected in the matrix elements at 
low resolution. 
It has been an open question whether the same is true for 3N forces since there are important differences: 
First, chiral 3N forces are fixed by fitting only two low-energy constants ($c_D$ and $c_E$), in contrast to numerous couplings in
NN interactions~\cite{Epelbaum:2008ga}. Second, 3N forces give only subleading contributions to observables. Since universality is only 
approximate in NN interactions, it is not obvious to what extent 3N forces are constrained by long-range physics at low resolution. 
In figure~\ref{fig:universality} we illustrate explicitly the form of the matrix elements of 3N forces at two resolution scales for 
five different chiral interactions (see Ref.~\cite{Wendt:2013bla} for another example of 3NF universality). We also 
find a remarkably reduced model dependence for evolved 3N interactions in the dominant 
kinematical region. This suggests that the chiral low-energy coupling constants $c_D$ and $c_E$ are flowing to an approximately universal 
value at low resolution. In addition, new momentum-dependent universal structures are induced at low resolution, as can be seen in the right panel. 
Future plans involve the explicit extraction of the low-energy constants from the evolved matrix elements and the investigation
of 3NF universality, in particular the role of the contributions at N$^3$LO~\cite{Bernard:2007sp, Bernard:2011zr}.



\section{Nuclear equation of state and astrophysical applications}
\label{sec:NM}

\begin{figure}
  \begin{center}
    \parbox[c]{2.5cm}{\centering \includegraphics[scale=0.7]{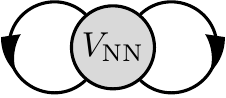}}
    \parbox[c]{2.5cm}{\centering \includegraphics[scale=0.7]{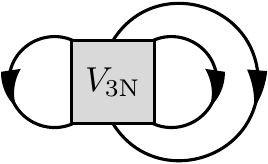}}
    \parbox[c]{2.5cm}{\centering \includegraphics[scale=0.7]{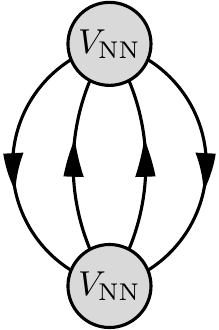}}
    \parbox[c]{2.5cm}{\centering \includegraphics[scale=0.7]{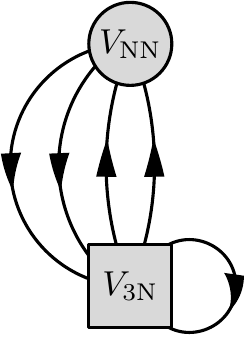}}
    \parbox[c]{2.5cm}{\centering \includegraphics[scale=0.7]{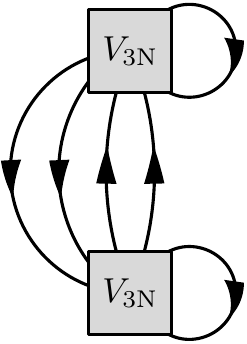}}
    \parbox[c]{2.5cm}{\centering \includegraphics[scale=0.7]{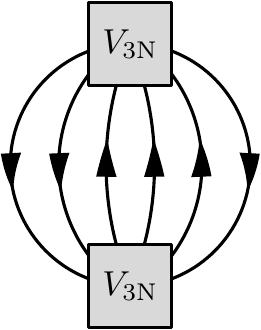}}
  \end{center}
  \caption{Diagrams contributing to the energy per particle up to second order in MBPT, taking two- and three-body interactions into account.}
  \label{fig:EOS_diagrams}
\end{figure}

\begin{figure}[t]
  \begin{center}
    \includegraphics[width=0.5 \textwidth]{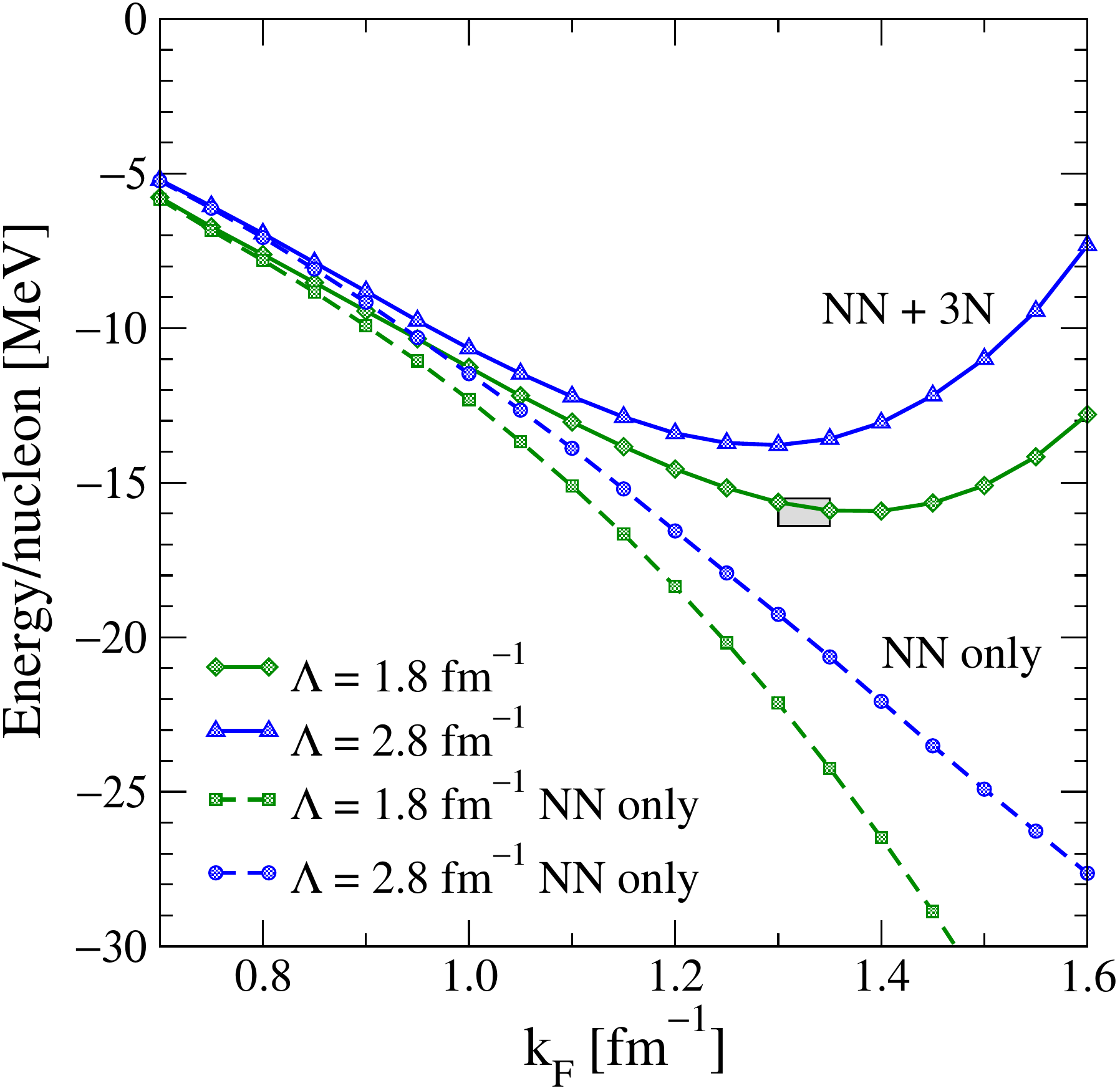}
  \end{center}
  \caption{Energy per particle in symmetric nuclear matter as a function of the Fermi momentum based on NN+3NF forces compared to NN-only results for two representative NN cutoffs and a fixed 3N cutoff. For details see Ref.~\cite{Hebeler:2010xb}.}
  \label{fig:SNM_3N}
\end{figure}

Nuclear matter describes an idealized infinite system consisting of neutrons and protons in the thermodynamic limit, interacting only via the 
strong nuclear force and neglecting the Coulomb forces between protons. Neutron matter in particular consists only of neutrons, whereas symmetric 
nuclear matter refers to the special case with equal neutron and proton densities. The interplay of chiral EFT and
RG methods offers new opportunities for efficient and simplified microscopic calculations of the nuclear equation of state. RG-evolved 
interactions enable the application of perturbative methods, which also provide improved estimates of theoretical uncertainties 
(see also the discussion of in-medium chiral perturbation theory in
Ref.~\cite{Holt:2013fwa}.)
Figure~\ref{fig:EOS_diagrams} shows all interaction diagrams contributing to the energy up to second order in many-body perturbation theory (MBPT), taking 
two- and three-body interactions into account. These diagrams have been found to already provide well-converged results for the energy at small resolution scales
(a nonperturbative validation of MBPT for two-body interactions in neutron
matter has recently been given in \cite{Gezerlis:2013ipa}).

The physics of nuclear matter covers a wide range of extremes. At very low densities, the interparticle distances are sufficiently large 
that details of the nuclear interaction are not resolved and all properties of the system are governed by the large s-wave 
scattering length. In this universal regime neutron matter shares many features with properties 
of atomic gases close to the unitary limit, which are currently the subject of active theoretical and experimental investigations~\cite{Giorgini:2008zz}. 
At intermediate densities, which are most relevant for finite nuclei, nuclear matter properties are used to guide the development of nuclear 
energy density functionals and in particular to constrain the physics of neutron-rich nuclei close to the limit of stability, which are key for 
understanding the synthesis of heavy nuclei in the universe. At very high densities, the composition and properties of nuclear matter are still 
unknown. Exotic states of matter containing strange particles or isolated quarks might be present under such conditions and possibly exist in the 
interior of neutron stars~\cite{Haensel_nstar_book}.

\subsection{Symmetric nuclear matter}

Over the last decades, an accurate prediction of symmetric nuclear matter at intermediate densities starting from microscopic nuclear forces has been a 
theoretical milestone on the way to finite nuclei close to the valley of stability, but has proved to be an elusive target. Progress for controlled calculations 
has long been hindered by the difficulty and the non-perturbative nature of the nuclear many-body problem when conventional nuclear interactions are used.

Most advances in microscopic nuclear structure theory over the last decade have been through expanding the reach of few-body calculations. This has clearly 
established the quantitative role of 3N forces for light nuclei (see Section \ref{sec:Finite}). However, until recently few-body fits have not 
sufficiently constrained 3N force contributions at higher density such that calculations of symmetric nuclear matter calculations are predictive. One key challenge
is the correct reproduction of nuclear saturation. It has long been known 
that the particle density in the center of atomic nuclei 
is approximately $n_s = 0.16$ fm$^{-3}$ over a wide range of masses~\cite{preston1975structure}, which means that the density in equilibrium approaches a constant 
value in the thermodynamic limit. 
Historically, when a quantitative reproduction of empirical saturation 
properties has been obtained, it was imposed by hand through adjusting phenomenological short-range three-body forces (see e.g., \cite{Akmal:1998cf, Lejeune:2001bg}).

\begin{figure}
  \begin{center}
    \includegraphics[width=0.5 \textwidth]{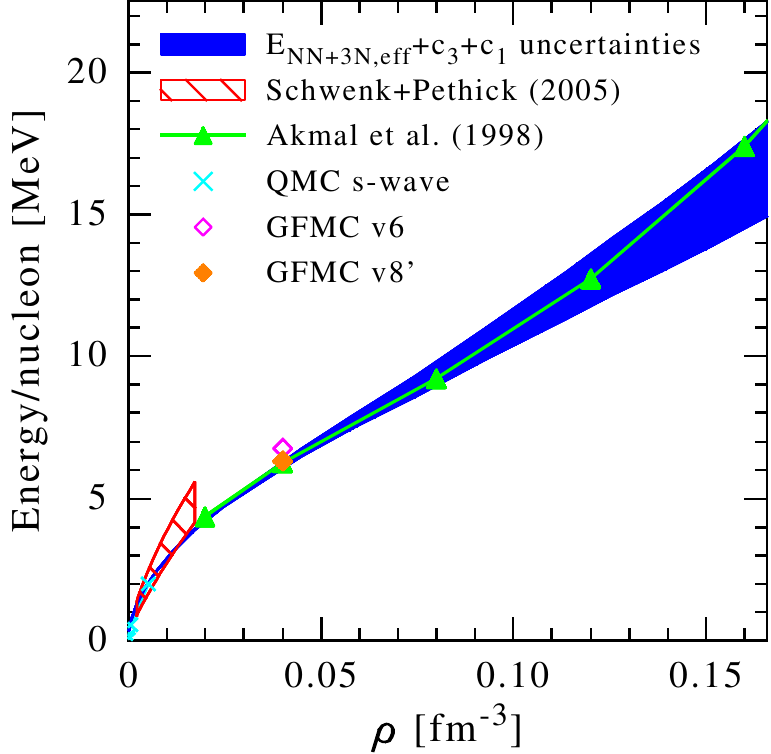}
    
  \end{center}
  \caption{ Neutron matter energy as a function of density including NN, 3N and 4N forces up to N$^3$LO in comparison to other studies
  (from \cite{Hebeler:2009iv}).}
  \label{fig:PNM_RG}
\end{figure}

In chiral EFT all short-range couplings of the Hamiltonian are fixed in two- and few-body systems and then used to predict properties of 
many-body systems. In Section 2 (see figure~\ref{fig:EOS_SNM_lambda}) the convergence pattern of such calculations for the equation 
of state of symmetric nuclear matter was already discussed. For these calculations the short-range 3N forces have been fitted to the experimental 
values of the $^3$H binding energy and the radius of $^4$He. The Hartree-Fock results show that nuclear matter is bound even at the simplest level 
in the many-body expansion. It is encouraging that the results agree with the empirical saturation point, indicated by the grey square, within 
the uncertainty in the many-body calculation and omitted higher-order many-body forces implied by the cutoff variation. We stress that the cutoff dependence of 
order 3 MeV around saturation density is small compared to the total size of the kinetic energy ($\approx$ 23 MeV) and potential energy ($\approx -44$ MeV) at this
density. Moreover, the cutoff dependence is smaller at $k_F=1.1\fmi$, which is the typical density in the interior of medium-mass to heavy nuclei. 
For all cases in the right panel of figure~\ref{fig:EOS_SNM_lambda} the nuclear compressibility $K = 175 - 210$ MeV is also in the empirical range.

\begin{figure}
  \begin{center}
    \includegraphics[width=0.95 \textwidth]{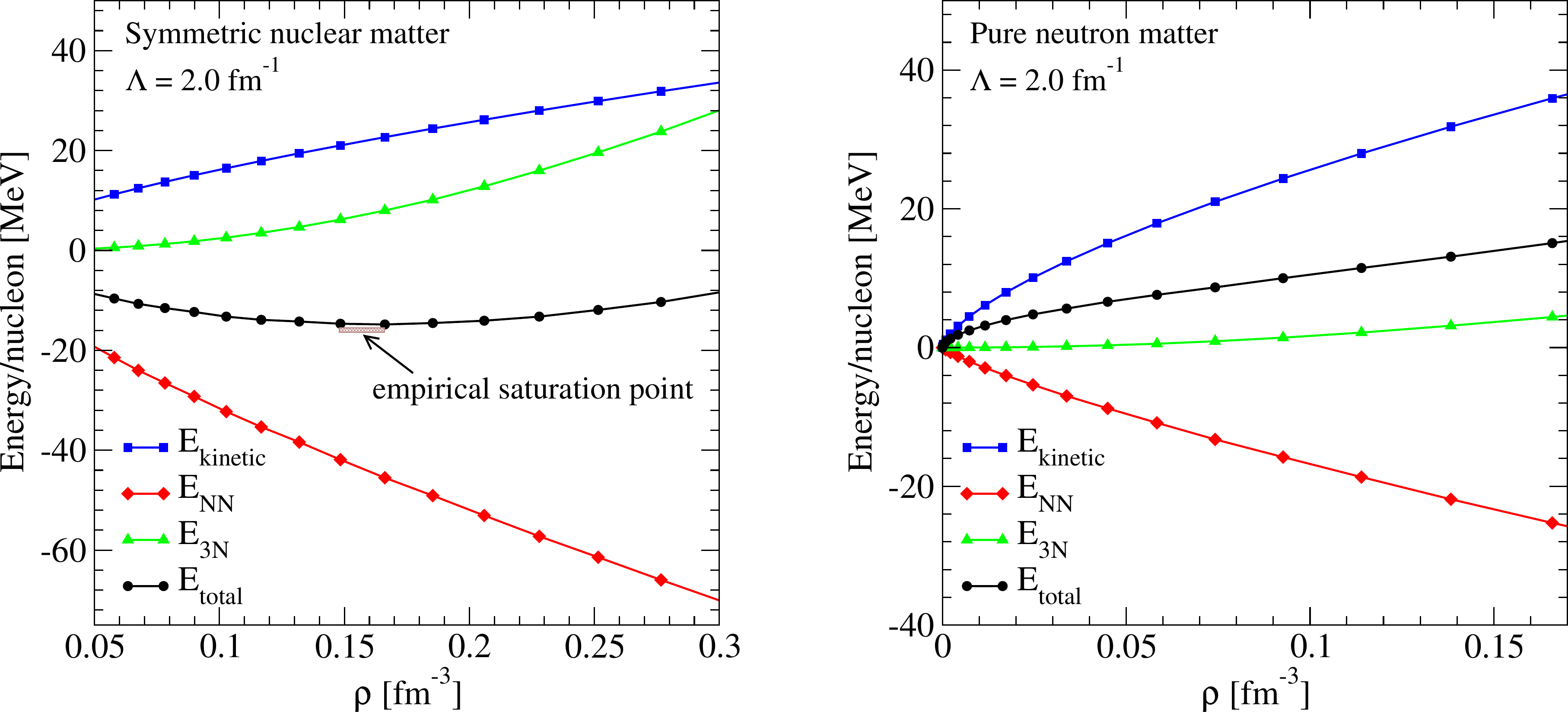}
  \end{center}
  \caption{Hierarchy of many-body contributions as function of density for symmetric nuclear 
  matter~\cite{Hebeler:2010xb} and pure neutron matter~\cite{Hebeler:2009iv}. $E_{\rm{NN}}$ denotes the energy contributions from NN interactions and 
  $E_{\rm{3N}}$ all contributions which include at least one 3N interaction.}
  \label{fig:EOS_PNM_SNM}
\end{figure}

The role of 3N forces for saturation is demonstrated in figure~\ref{fig:SNM_3N}. The two pairs of curves show the differences between the 
nuclear matter results for NN-only and NN plus 3N interactions. It is evident that saturation is driven by 3N forces. Even 
for $\Lambda = 2.8$ fm$^{-1}$, which is similar to the lower cutoffs in chiral EFT potentials, symmetric nuclear matter does not even saturate 
in the plotted density range. While 3N forces drive saturation for low-momentum interactions, the 3N contributions are not unnaturally 
large (see also figure~\ref{fig:EOS_PNM_SNM} and discussion below).

\subsection{Neutron matter}

Neutron matter is a particularly useful testing ground for
chiral forces because
only the long-range 2$\pi$-exchange 3N forces contribute~\cite{Hebeler:2009iv}, 
which implies that all three- and four-neutron (4N) forces are predicted up to N$^3$LO. In addition, as a result of weaker tensor forces between neutrons 
and the absence of short-range 3N forces, neutron matter behaves more perturbatively than symmetric nuclear matter. This results in very small theoretical
uncertainties of the neutron matter calculations, mainly due to uncertainties of the low-energy couplings in chiral EFT~\cite{Epelbaum:2008ga,Hebeler:2009iv,Hammer:2012id}. In 
figure~\ref{fig:PNM_RG} we present the neutron matter energy including the uncertainties of the low-energy constants, indicated by the blue band, in comparison to
other approaches. These include Green’s Function Monte Carlo (GFMC)~\cite{Carlson:2003zz}, Quantum Monte Carlo (QMC)~\cite{Gezerlis:2007fs}, results 
of Akmal et al.~\cite{Akmal:1998cf}, and difermion EFT results for lower densities~\cite{Schwenk:2005ka}.
The results derived from chiral EFT interactions correspond to a nuclear symmetry energy of $S_v=30.4$--$33.6$ MeV. Compared to the empirical 
range $S_v=25$--$35$ MeV~\cite{Steiner:2004fi}, the microscopic range of $\approx 3$ MeV is very useful and comparison to experiment could also provide guidance for improved constraints on the values of the low-energy constants.

\begin{figure}
  \begin{center}
    \includegraphics[width=0.5 \textwidth]{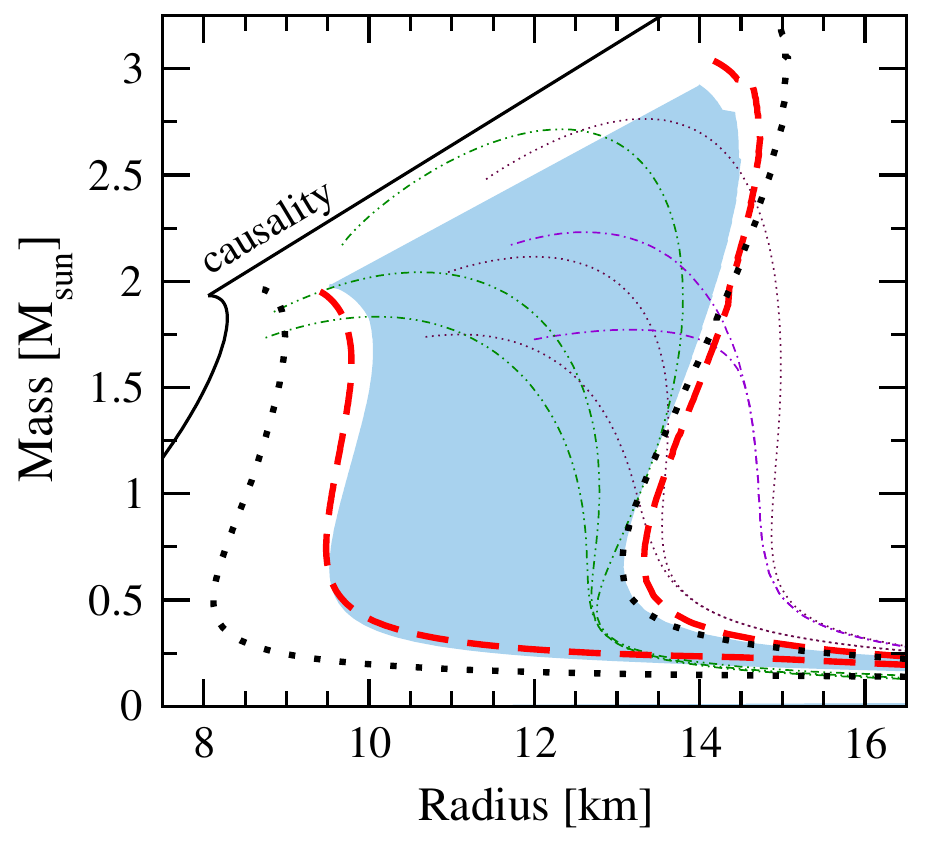}
  \end{center}
  \caption{Neutron star mass-radius results for the equations of state based for $M > 1.97 \:M_{\odot}$. The blue band shows the radius constraint based on RG-evolved interactions plus 3N interactions based on chiral EFT. The dashed and dotted lines represent a selection of other EOSs which are currently used in astrophysical simulations and neutron star calculations. See Ref.~\cite{Hebeler:2013nza} for details.}
 \label{fig:MvR}
 \end{figure}

The size of the separate contributions from the kinetic energy, NN forces, and 3N forces for symmetric nuclear matter and neutron matter as a function 
of density for a RG cutoff of $\Lambda = 2.0$ fm$^{-1}$ are shown in figure~\ref{fig:EOS_PNM_SNM}. The empirical saturation point for symmetric nuclear matter 
is again indicated by the rectangle in the left panel. Here it is obvious that the separate contributions are much larger than the sum of all terms 
(black line) and that saturation is the result of a delicate interplay of these terms. Furthermore, the contributions from 3N forces grow faster
with density than those from NN forces due to the additional nucleon involved. This implies that beyond some critical density region the chiral hierarchy 
of many-body forces will break down and Hamiltonians based on chiral EFT will not be useful anymore. However, for symmetric nuclear matter we still find that the
contributions involving 3N forces are about a factor 3 smaller than those from NN forces around twice nuclear saturation density. For neutron matter the
3N forces play a less significant role and give only moderate repulsive contributions.

\subsection{Applications to neutron stars}

The neutron matter results have direct implications for the properties of neutron stars. Since core densities inside a neutron star can reach 
several times the interior density of heavy nuclei, much higher than the maximal density up to which the equation of state can be calculated reliably, it is 
necessary to extend the microscopic results to higher densities. This can be achieved by employing a general strategy that does not 
rely on assumptions about the nature of the nuclear constituents and their interactions at high densities: by choosing a piecewise 
polytropic ansatz~\cite{Read:2009yp} and limiting the range of the free parameters by physics and constraints from neutron star 
observations~\cite{Hebeler:2010jx}. In particular, the following two constraints have been used: (a) the speed of sound remains smaller 
than the speed of light for all densities, and (b) the EOS is able to support a neutron star of mass $M \ge M_{\rm{min}}= 1.97\,M_{\odot}$, which 
is currently the heaviest confirmed observed neutron star mass~\cite{Demorest:2010bx}. This results in uncertainty bands for the equation 
of state and for neutron star radii. The constraints for neutron stars are shown in figure~\ref{fig:MvR}. For a typical neutron star of 
mass $M=1.4\,M_{\odot}$ a radius range $R=10.0-13.7\:\rm{km}$ is found. 
For comparison, a selected set of alternative EOSs are shown that are currently used
in astrophysical simulations~(for details see Ref.~\cite{Kruger:2013kua}). It is evident that many of these are inconsistent 
with constraints derived from interactions based on chiral EFT.

\begin{figure}
  \begin{center}
    \includegraphics[width=0.4 \textwidth]{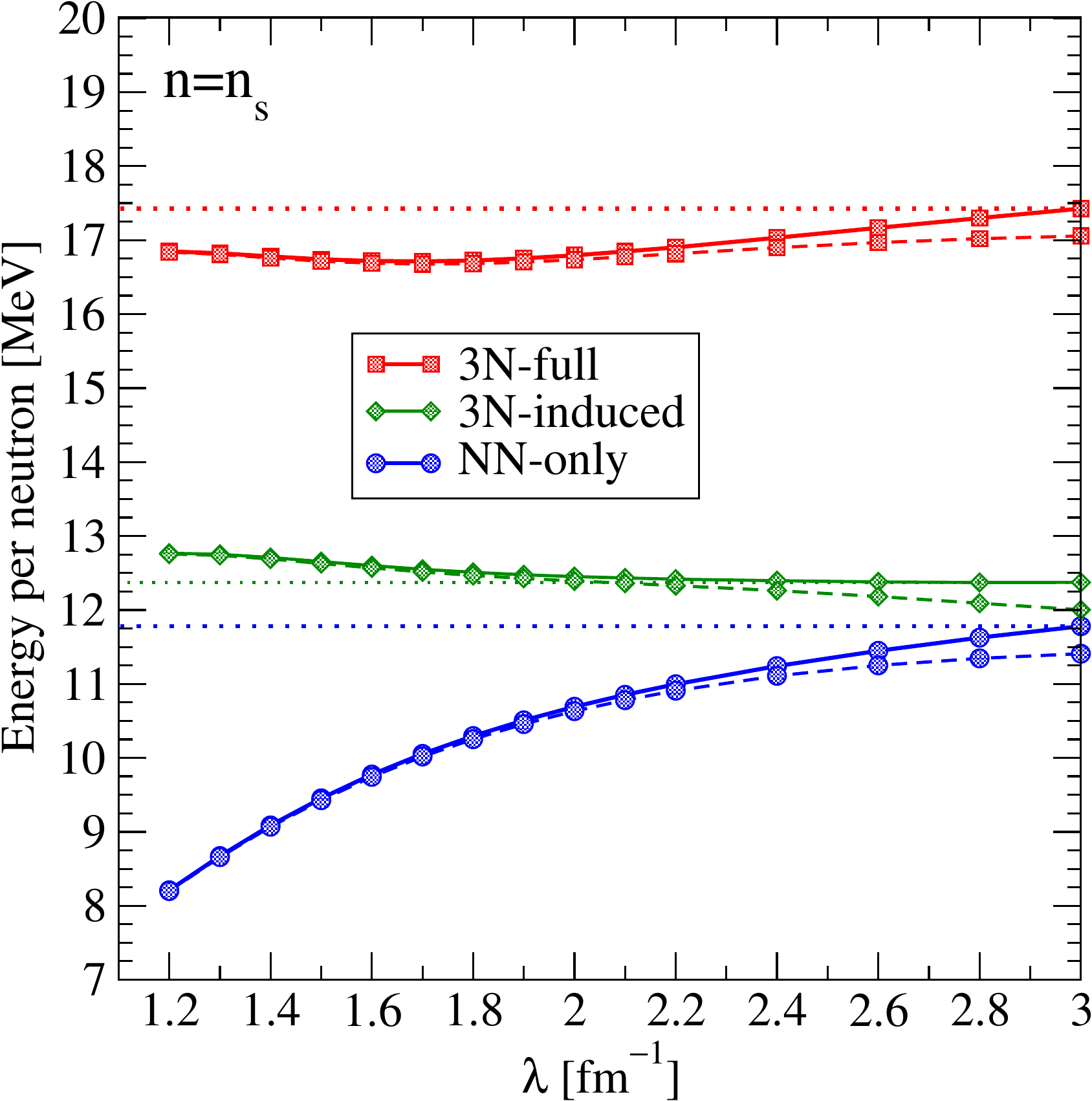}
    \hspace{1cm}
    \includegraphics[width=0.4 \textwidth]{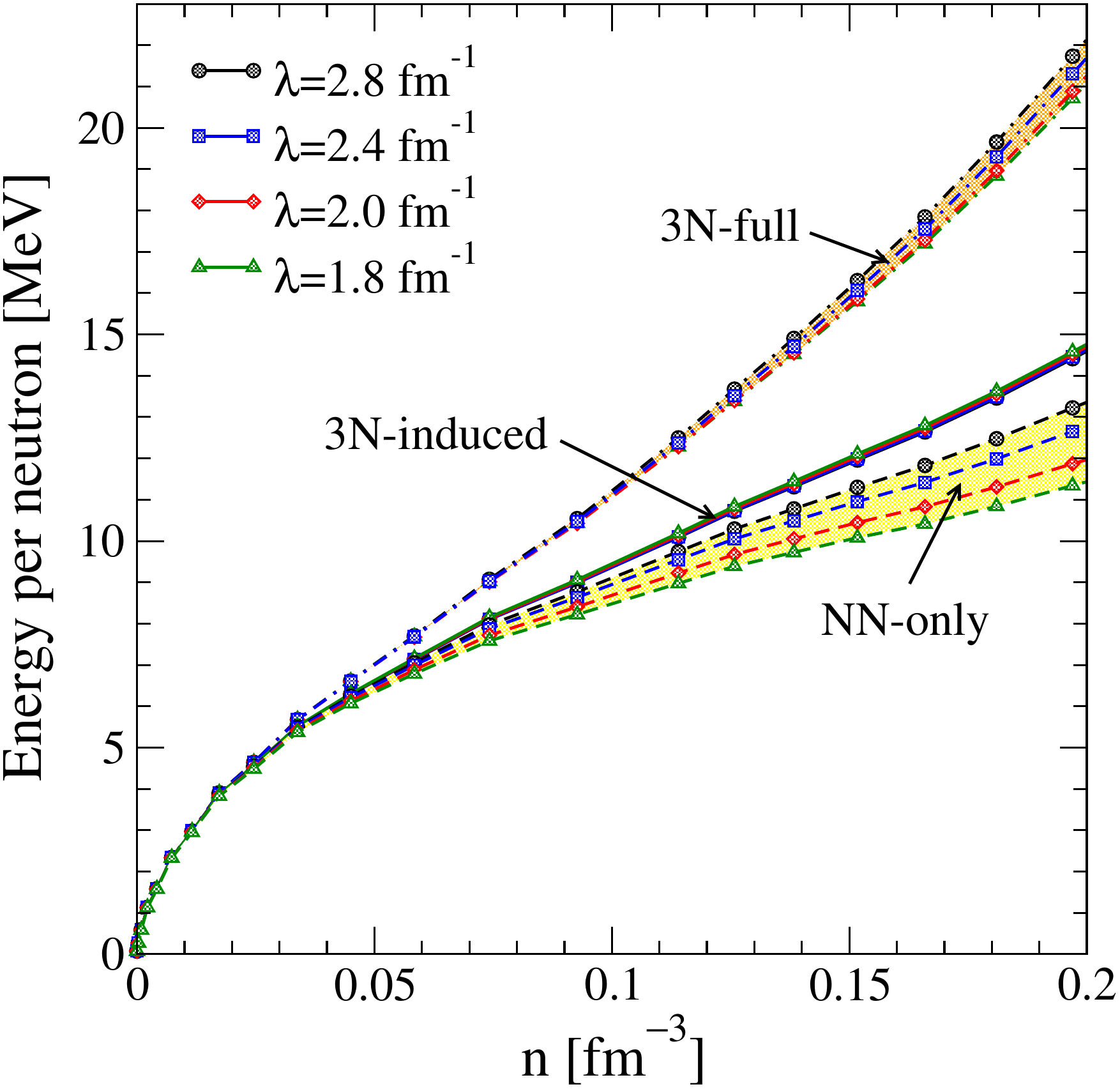}  
  \end{center}
  \caption{Neutron matter equation of state based on consistently-evolved NN+3N interactions. See Ref.~\cite{Hebeler:2013ri} for details.}
  \label{fig:PNM_evolved3NF}
\end{figure}

\subsection{First results based on consistently evolved three-nucleon forces}

All results for neutron and symmetric nuclear matter presented
so far were based on evolved NN interactions plus 3N interactions that were fixed 
at the low-momentum scale rather than the chiral EFT cutoff scale (strategy (b) in section~\ref{sec:RG_tech}). Thanks to recent developments 
it is now also possible to consistently evolve 3N interactions~\cite{Hebeler:2012pr}, which can be used directly in microscopic calculations of infinite nuclear
matter~(strategy (c) in section~\ref{sec:RG_tech}). In figure~\ref{fig:PNM_evolved3NF} we present very recent results for neutron matter based on 
interactions derived in this framework~\cite{Hebeler:2013ri}. For these calculations the 3N contributions to the equation of state have been 
calculated in Hartree-Fock approximation. This approximation is expected to be reliable at small $\lambda$, whereas at larger scale higher-order
contributions are known to be important~\cite{Hebeler:2009iv}. The left panel shows results for the energy per neutron at saturation density as a function of the 
SRG resolution scale $\lambda$ using the three approximations ``NN-only'', ``3N-induced'' and ``3N-full''~(see also figure~\ref{fig:Triton}). Evidently,
neglecting all 3N interactions in the RG evolution results in a significant resolution-scale dependence with a energy variation of about 3.5 MeV. 
By including the induced contributions the variation is significantly reduced to about 400 keV, with the major part of this variation
happening at small $\lambda$. By including also initial 3N interactions, a total variation of about 600 keV is found. The yellow band indicates the size of the
second-order contributions at $\lambda=\infty$, which can be calculated using the framework from~\cite{Hebeler:2009iv}. The fact that the observed variation 
of the energy is within this band suggests that the results can be systematically improved by including higher-order contributions in the many-body expansion. 
This work is currently in progress.

In the future it will also be possible to systematically study response functions of nuclear matter within this framework. For this it will 
be essential to also consistently evolve other operators (see Section~\ref{Sect:Correlations}).



\section{Applications to finite nuclei}\label{sec:Finite}

The softening of nuclear forces by RG evolution has made possible
many new calculations of finite nuclei with a wide variety of techniques.
In this section we present a selection of recent results, highlighting the present successes, future
potential, and open problems.

\subsection{Many-body perturbation theory in finite nuclei}
\label{subsec:MBPT_finnucl}

\begin{figure}[tb!]
\begin{center}
 \includegraphics[width=4.0in]{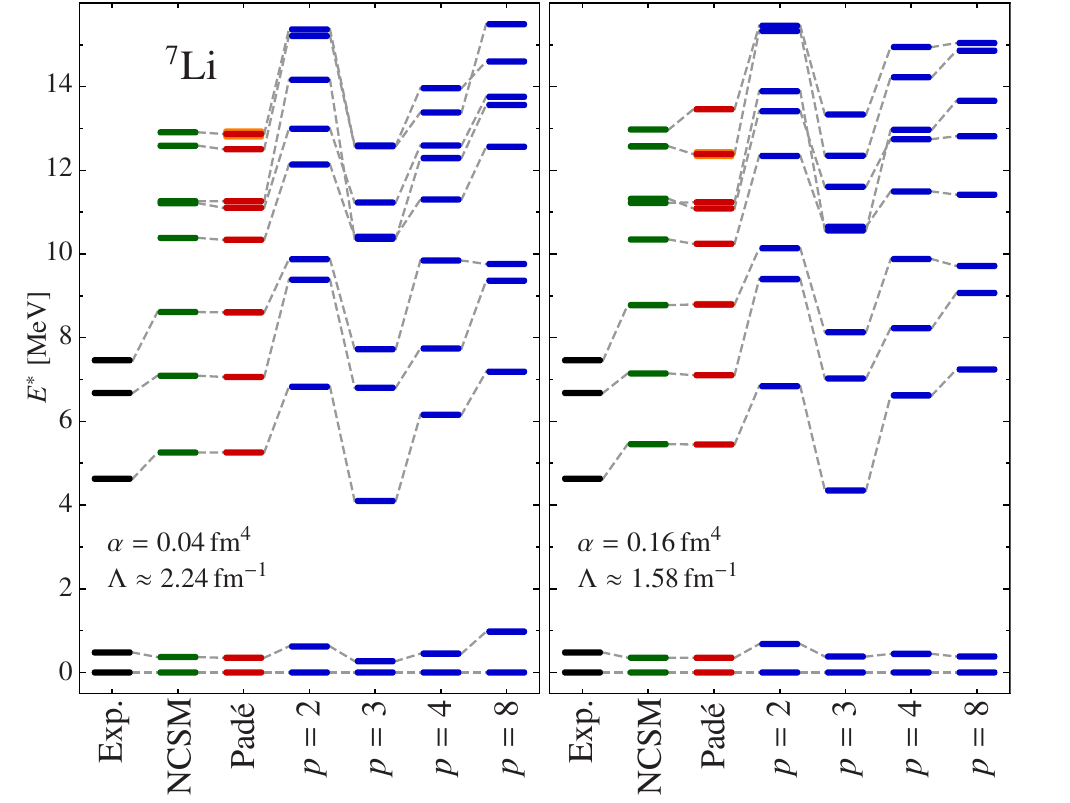}
  \caption{Excitation energies in $^7$Li calculated in degenerate Rayleigh-Schr\"odinger MBPT at order
 $p=2$, 3, 4, and 8 for a N$^3$LO two-body interaction evolved using the
 SRG to two different resolutions ($\Lambda$ here is the same as $\lambda$
 used elsewhere) compared to experiment, exact NCSM calculations, and the Pad\'e resummed
 result~\cite{Langhammer:2012jx}  A fixed harmonic oscillator
 model space is used for all calculations.}.
 \label{fig:rothmbpt}
\end{center}
\end{figure}

The apparent success of low-order many-body perturbation theory (MBPT) in infinite nuclear matter with low-momentum potentials has been
tested for finite nuclei by Roth and collaborators, who have 
performed calculations in high-order Rayleigh-Schr\"odinger MBPT using SRG-evolved two-body interactions
(based on an initial N$^3$LO interaction)
for both closed-shell~\cite{Roth:2009up} and
open-shell~\cite{Langhammer:2012jx} nuclei.  The calculations for $^7$Li in degenerate
MBPT in a fixed harmonic oscillator model space 
shown in figure~\ref{fig:rothmbpt} are typical.  Even for very soft potentials (e.g.,
the right panel) the perturbation series diverges. However, a simple resummation with
Pad\'e approximants (see Refs.~\cite{Roth:2009up,Langhammer:2012jx} for details) results
in stable energies in very good agreement with exact no-core shell-model (NCSM) calculations
using the same model space.
Future work will include three-body forces and study applications to heavier open-shell nuclei 
and alternative partitionings of the Hamiltonian (e.g., using a Hartree-Fock unperturbed basis).

\begin{figure*}[tb!]
\begin{center}
 \includegraphics[width=5.2in]{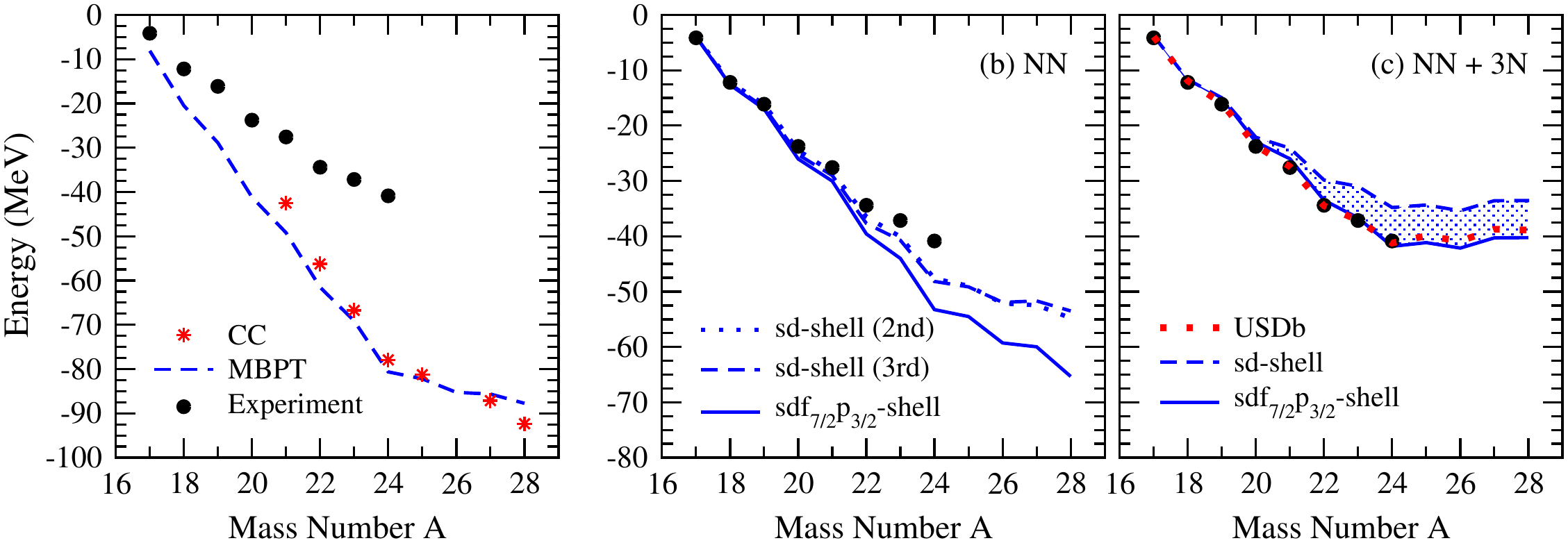}
   \caption{Calculations of oxygen isotope ground-state energies
   compared to experiment~\cite{Holt:2011fj}.
   Left: MBPT compared to coupled cluster theory with the same low-momentum
   NN-only interaction; middle: MBPT NN-only at 2nd and 3rd order, with
   empirical SPEs; right: MBPT at 3rd order from NN and 3N forces with calculated
   MBPT SPEs, plus the phenomenological USDb model.}
   \label{fig:oxygen_sm}
\end{center}
\end{figure*}

\begin{figure}[b]
\begin{center}
 \includegraphics[width=2.2in]{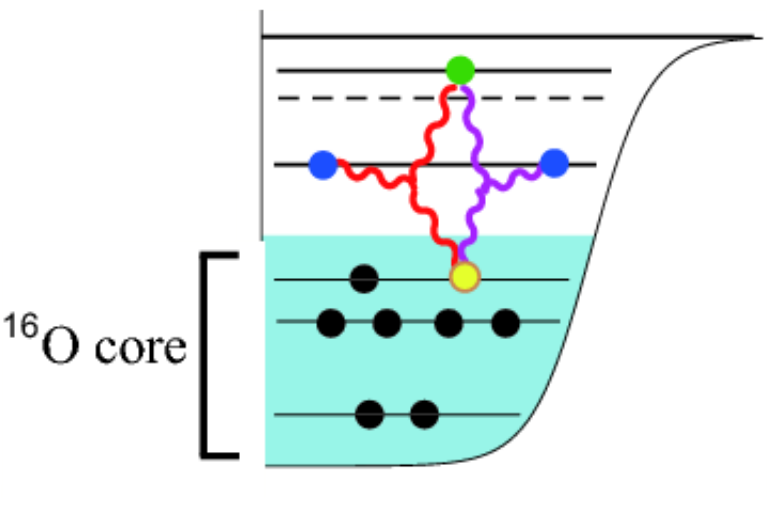}
   \caption{Interaction between valence neutrons and a core nucleon in
   an oxygen isotope through
   a three-body force~\cite{Otsuka:2009cs}.}
   \label{fig:oxygen_sm2}
\end{center}
\end{figure}

The softening of potentials also enables the direct use of perturbative methods
in microscopic valence-shell calculations, in which a small number
of nucleons outside a closed-shell
core interact via an effective interaction treated in MBPT (with a nonperturbative transformation to remove the energy dependence of the MBPT effective Hamiltonian~\cite{HjorthJensen:1995ap}).
One application is to identify the effects of 3NF
on the location of the neutron dripline: the limits of nuclear existence
where an added neutron is no longer bound---it ``drips'' away.
This limit is not always easily understood.
For example, experiment shows that as neutrons are added to
stable $^{16}$O, the neutrons stay bound until $^{24}$O.  But adding
one more proton to get fluorine extends the dripline all the
way to $^{31}$F.  
A valence-shell MBPT calculation building on $^{16}$O and using an RG-softened NN potential
is shown in the left panel of figure~\ref{fig:oxygen_sm}, where it is validated 
against nonperturbative coupled-cluster (CC) calculations with the same interaction
and consistent single-particle energies (SPEs)~\cite{Holt:2011fj}.  Using empirical SPEs improves the agreement with experimental
energies (middle panel), but
these microscopic NN-only calculations show too much attraction beyond $A=24$,
so that $^{28}$O is the predicted dripline.
The missing physics from 3NF is indicated schematically
in figure~\ref{fig:oxygen_sm2}; pairs of valence nucleons feel an additional repulsion from
long-range 3NF including a core nucleon~\cite{Otsuka:2009cs}. 
When this effect is included, with a chiral N$^2$LO 3NF and \emph{calculated} SPEs, the
dripline is at $^{24}$O~\cite{Holt:2011fj} (right panel and see Ref.~\cite{Otsuka:2009cs} for an earlier calculation with empirical SPEs).
The 3NF can also account for how the phenomenological shell model adjusts to reproduce
the same trends with $A$~\cite{Holt:2011fj}.

\begin{figure*}[tb]
\begin{center}
 \includegraphics[width=2.7in]{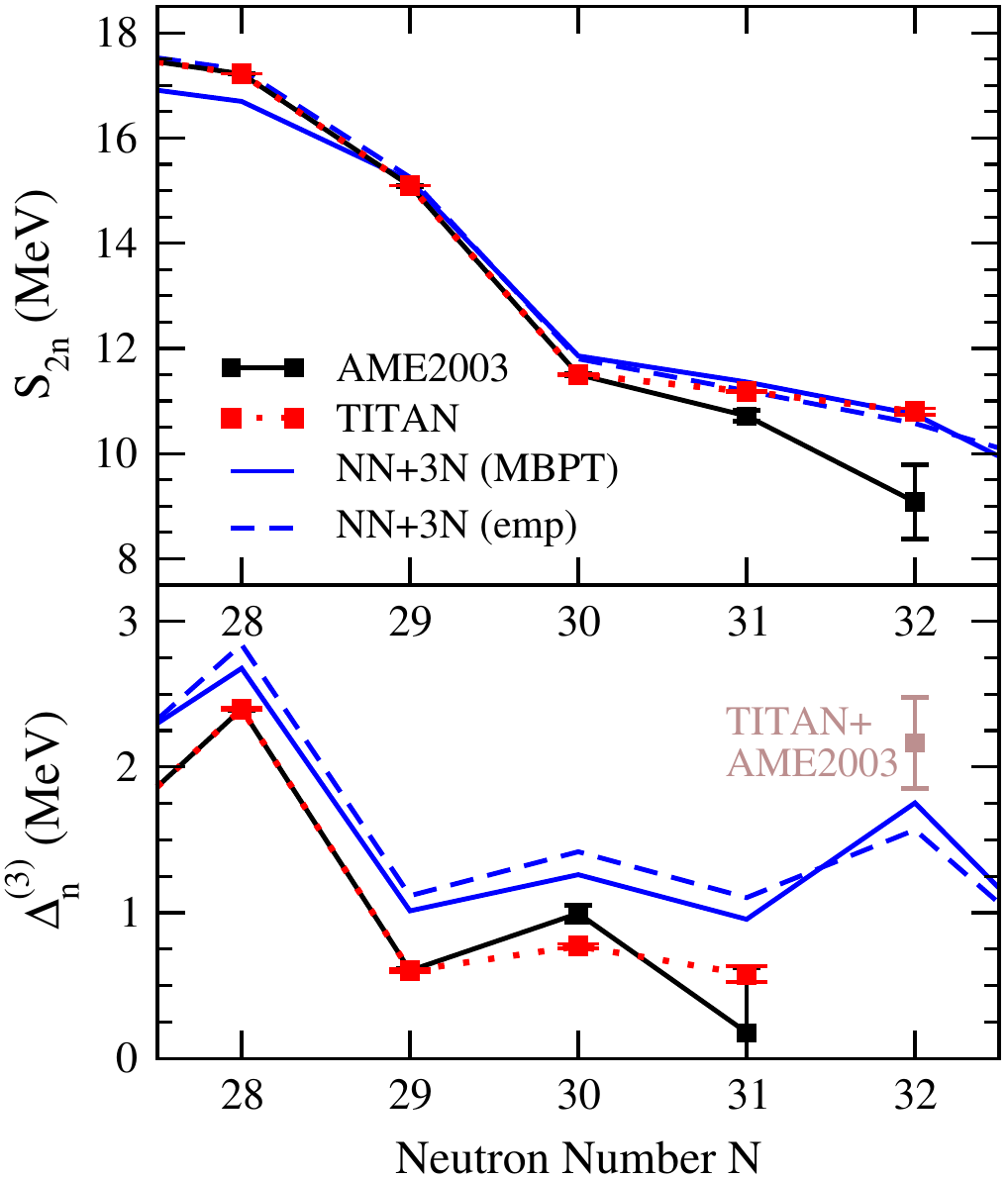}
   \caption{Predictions for two-neutron separation energy and pairing gaps in calcium
   isotopes including three-body forces compared to new experimental
   measurements~\cite{Gallant:2012as}.}
   \label{fig:calcium_sm}
\end{center}
\end{figure*}

When applied in the calcium isotopes, this microscopic MBPT method with 3NF
predicted that two-neutron
separation energies should be significantly larger than found
in previous experiments.
However, new high-precision measurements using a Penning trap show excellent agreement
with the MBPT predictions (see figure~\ref{fig:calcium_sm})~\cite{Gallant:2012as}.
Once again, the theoretical ingredients are SPEs and residual two-body interactions,
which are all calculated microscopically from NN and 3N forces~\cite{Gallant:2012as,Holt:2012fr}.
The NN forces are chiral EFT N$^3$LO interactions evolved with a smooth $\vlowk$ RG
to a low-momentum cutoff of $\Lambda = 2.0\fmi$ to improve the convergence of the MBPT.
Other calculations show the predictive power of the method for shell structure  and
pairing gaps~\cite{Holt:2013vqa}, excitation spectra~\cite{Holt:2011fj},
and properties of proton-rich nuclei~\cite{Holt:2012fr}.
On-going work seeks to extend the framework to include continuum effects for weakly bound
or unbound states, to develop nonperturbative methods for valence shell interactions~\cite{Tsukiyama:2012sm},
to relate to phenomenological
models, and to quantify theoretical uncertainties.

\subsection{Ab initio calculations with three-nucleon forces}

The frontier for RG-based ab initio calculations of finite nuclei using microscopic inter-nucleon forces is the inclusion of 3NF.  
The SRG has made possible the inclusion
of consistently evolved 3NF in a harmonic oscillator
basis~\cite{Jurgenson:2009qs,Jurgenson:2010wy}, which means 3NF are present in the initial
Hamiltonian but also induced as a result of RG evolution.  The NN+3N interaction
at lower resolution is found to have the same improved convergence properties for configuration interaction calculations as found earlier for NN-only calculations (e.g., see Refs.~\cite{Jurgenson:2013yya}).  Despite the softening, the factorial growth of basis spaces in the no-core shell model (NCSM) still limits calculations in a complete model space to light nuclei
(roughly up to $^{12}$C).  

\begin{figure}[t!]
\begin{center}
 \includegraphics[width=3.0in]{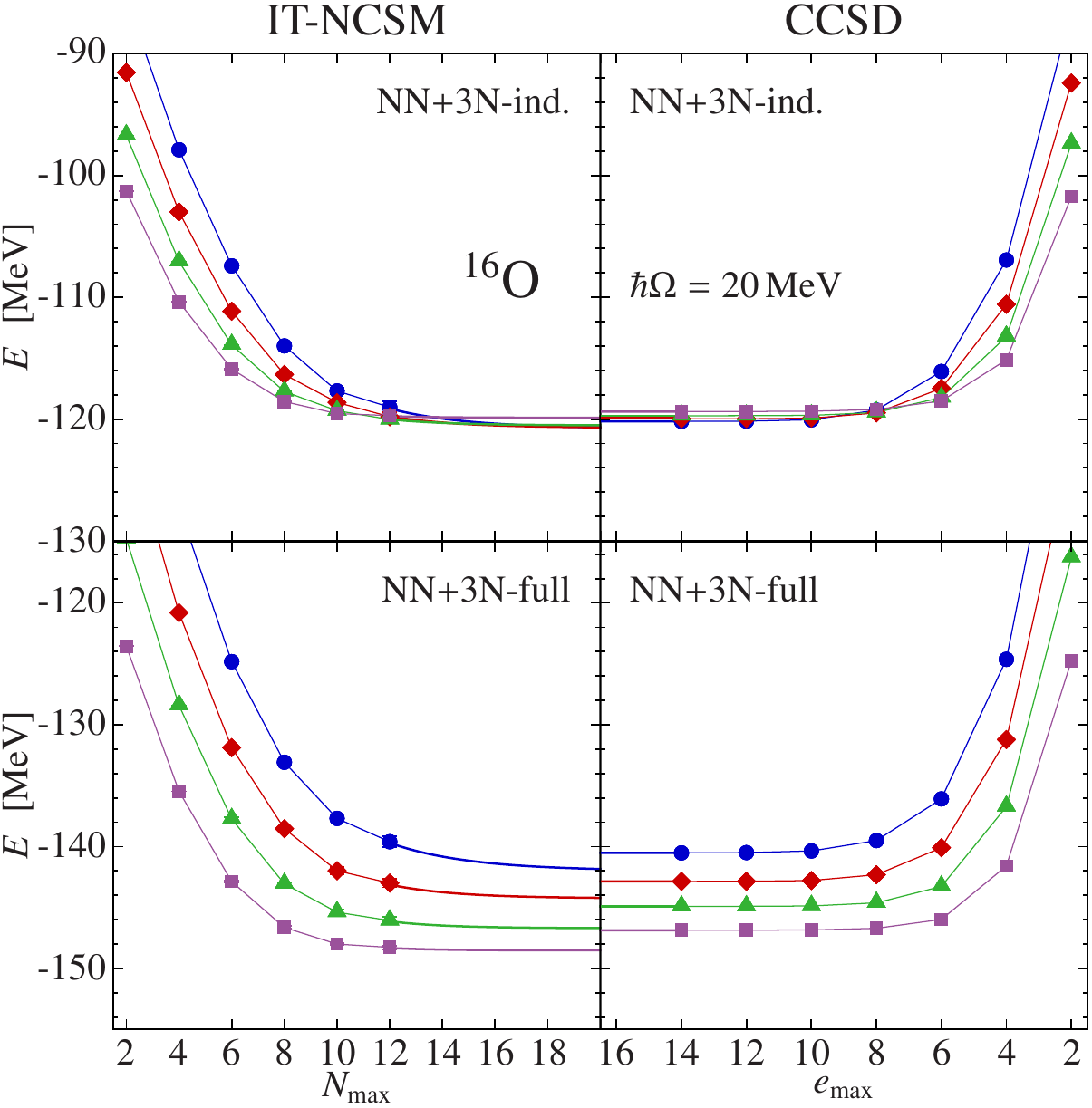}
   \caption{Comparison of IT-NCSM and CC calculations for $^{16}$O at four
   SRG resolutions without (above) and with (below) an initial three-body force (from Ref.~\cite{Roth:2011vt}).}
   \label{fig:roth1}
\end{center}
\end{figure}

To go to larger nuclei, Roth and collaborators have adapted importance truncation from quantum chemistry to nuclear calculations~\cite{Roth:2009cw}.  This method greatly reduces the size of the Hamiltonian matrix to be diagonalized by identifying the most important matrix elements.
While there are still some open questions
about uncertainties~\cite{Kruse:2013qaj}, results are very promising.  At the same time, RG-evolved 3NF has been added to nuclear coupled cluster (CC) calculations~\cite{Hagen:2007hi,Hagen:2007ew,Hagen:2010gd}, first in a normal-ordered
approximation~\cite{Roth:2011ar,Roth:2011vt} and then with the full 3NF interaction~\cite{Binder:2012mk}. 
In figure~\ref{fig:roth1}, importance truncated no-core shell model (IT-NCSM) 
results are compared to coupled cluster
calculations (at the CCSD level) for $^{16}$O.
In the top panels, an initial NN-only interaction is evolved to four different SRG resolutions,
including the induced 3NF~\cite{Roth:2011vt}.  The results are in good agreement and largely independent of the SRG flow parameter.

\begin{figure}[t!]
\begin{center}
 \includegraphics[clip=,width=3.0in]{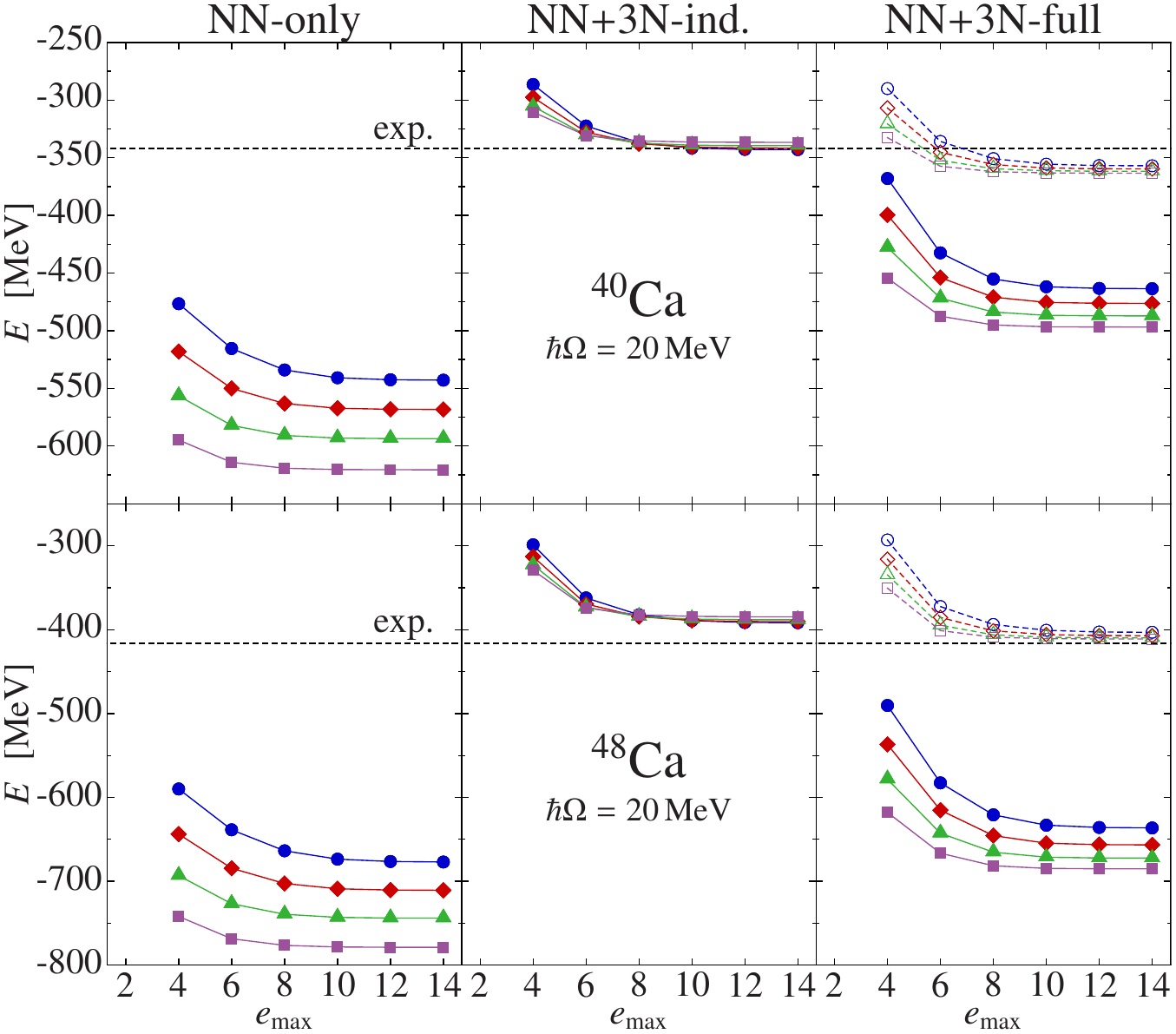}  
 \caption{Convergence of CC calculations for $^{40}$Ca and $^{48}$Ca with
 only NN (left), NN+3N-induced (middle), and NN+3N-full (right)~\cite{Roth:2011vt}.}
   \label{fig:roth2}
\end{center}
\end{figure}

However, the lower panels in figure~\ref{fig:roth1} show that while good convergence is still
found when initial 3NF are included, the flow is no longer unitary at the 10\,MeV level.
Detailed investigations~\cite{Roth:2011ar,Roth:2011vt} show that the long-range 3NF is the source of 
apparent large 4NF contributions for oxygen and heavier nuclei, causing a strong dependence on the flow parameter.  
However, by using a lower cutoff for the initial 3NF, cutoff independence is largely
restored and good agreement with experimental binding energies is achieved despite fitting only to
few-body properties~\cite{Roth:2011ar,Roth:2011vt}.
This is illustrated for two calcium isotopes in figure~\ref{fig:roth2}, with more examples
in Refs.~\cite{Roth:2011vt,Binder:2012mk}.
Work is in progress to check whether an alternative SRG 3NF evolution (e.g., in a momentum
basis) or the use of alternative SRG generators may be able to better control the
RG evolution of the initial 3NF.

\subsection{Ab-initio reactions with RG-evolved forces}

\begin{figure}[tb!]
\begin{center}
 \includegraphics[width=2.8in]{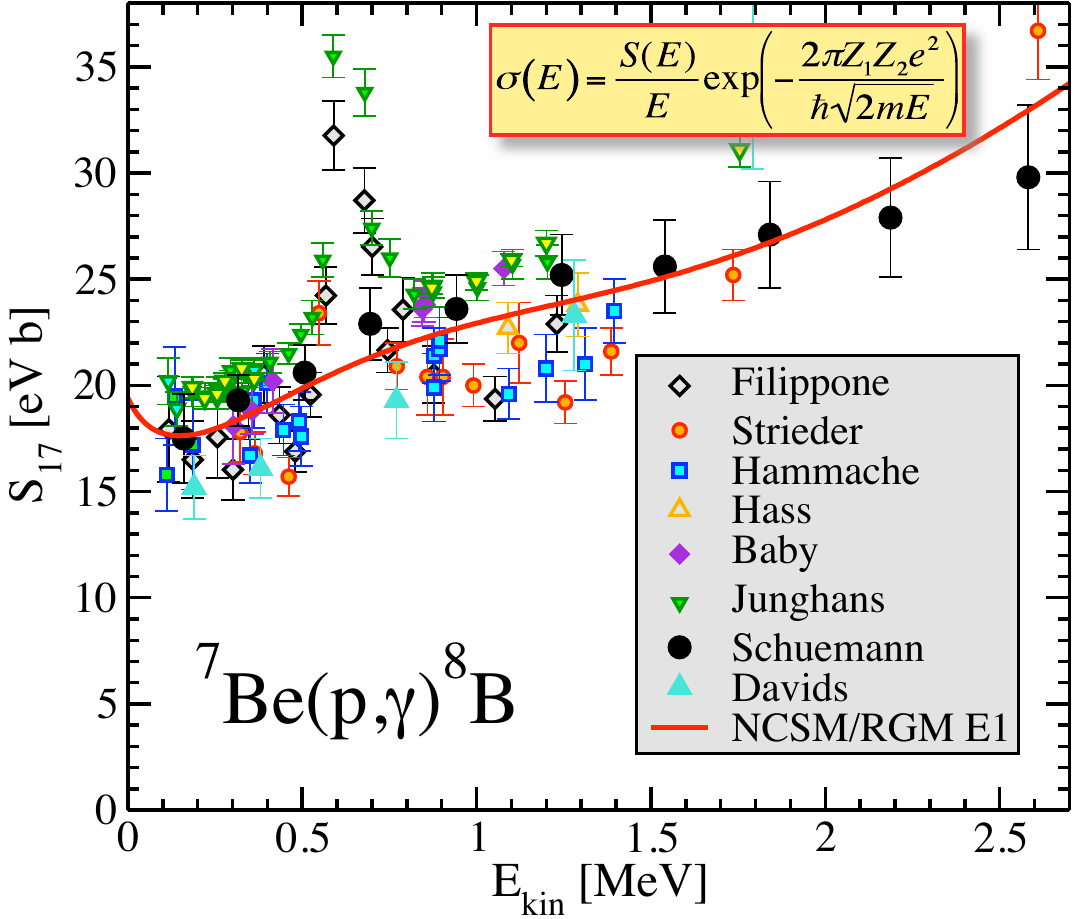}
   \caption{First ever ab initio calculations of $^7$Be$(p,\gamma){}^8$B
   astrophysical S-factor~\cite{Navratil:2011sa}.  Uses SRG-N3LO with $\lambda = 1.86\fmi$.}
   \label{fig:ncsmrgm1}
\end{center}
\end{figure}

\begin{figure}[tb!]
\begin{center}
 \includegraphics[width=2.5in]{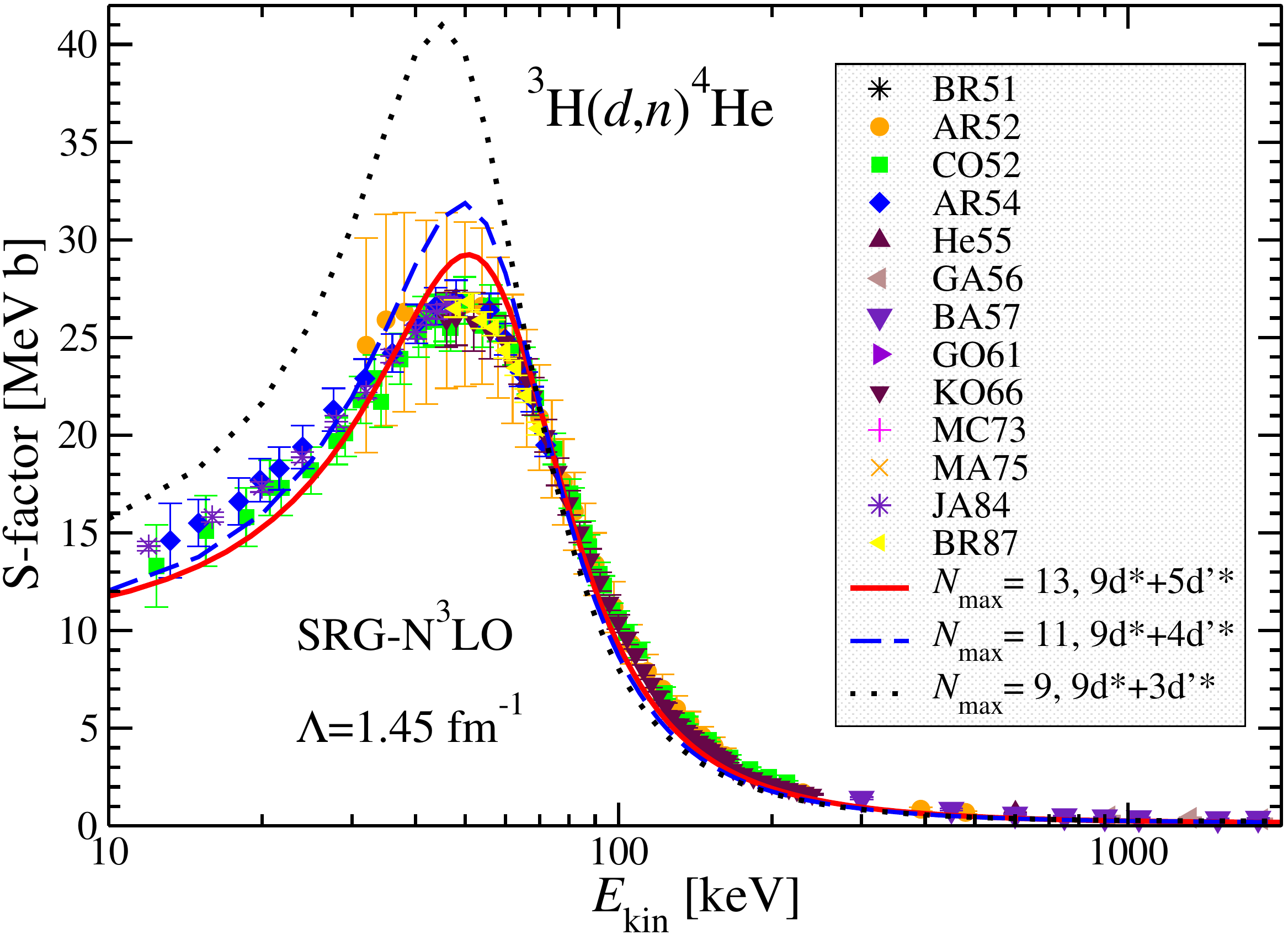}
   \caption{Calculated $^3$H$(d,n){}^4$He S-factor at three model space
   sizes compared to experimental
   data for an SRG-N$^3$LO NN potential with $\lambda = 1.45\fmi$ tuned
   to compensate for missing higher-order effects~\cite{Navratil:2011zs}.}
   \label{fig:ncsmrgm2}
\end{center}
\end{figure}

One of the principal aims of recent large-scale collaborations in low-energy 
nuclear physics (e.g., the UNEDF and NUCLEI projects~\cite{unedf:2011,Nam:2012gy,Bogner:2013pxa}) is to 
calculate reliable reaction cross sections for astrophysics, nuclear energy, and national security, for 
which extensions of standard phenomenology are insufficient. The interplay of structure and reactions is 
essential for a successful description of exotic nuclei as well. A powerful approach to implementing 
this interplay is the ab initio no-core shell model/resonating-group method (NCSM/RGM), which treats 
bound and scattering states within a unified framework using fundamental interactions between all 
nucleons~\cite{Navratil:2010ey,Navratil:2010jn}. Although only a few years under development,
a wide range of applications is already possible. Figure~\ref{fig:ncsmrgm1} shows the first-ever 
ab-initio calculation of the $^7$Be$(p,\gamma){}^8$B astrophysical S-factor~\cite{Navratil:2011sa}, a 
reaction important for solar neutrino physics. This calculation uses NCSM/RGM with an N$^3$LO NN interaction evolved
by the SRG to the special value of $\lambda = 1.86\fmi$, chosen to reproduce the observed separation
energy for $^8$B, which is important for an accurate reproduction of the low-energy behavior.
Both the normalization \emph{and} the shape of $S_{17}$ are predicted.
Other recent applications include the first ab initio many-body calculations of
$^3$H$(d,p){}^4$He and  $^3$H$(d,n){}^4$He fusion reactions, which reproduce the experimental
$Q$-value of both within 1\%~\cite{Navratil:2011zs}.
The convergence of the latter calculations is shown in figure~\ref{fig:ncsmrgm2}; 
the ability to tune the SRG interaction is again used to compensate for omitted
higher-order effects, such as the 3NF.
Calculations including consistent SRG-evolved 3NF will be available in the near future~\cite{Quaglioni:2012ue}.
%
Another related new development is the no-core shell model with continuum (NCSMC), which
is a unified approach to nuclear bound and continuum states~\cite{Baroni:2013fe}.
A recent proof-of-principle NCSMC calculation uses a realistic soft SRG-N$^3$LO 
nucleon-nucleon
potential to describe resonances in $^7$He~\cite{Baroni:2013fe}.

\begin{figure}[tb!]
\begin{center}
 \includegraphics[width=2.8in]{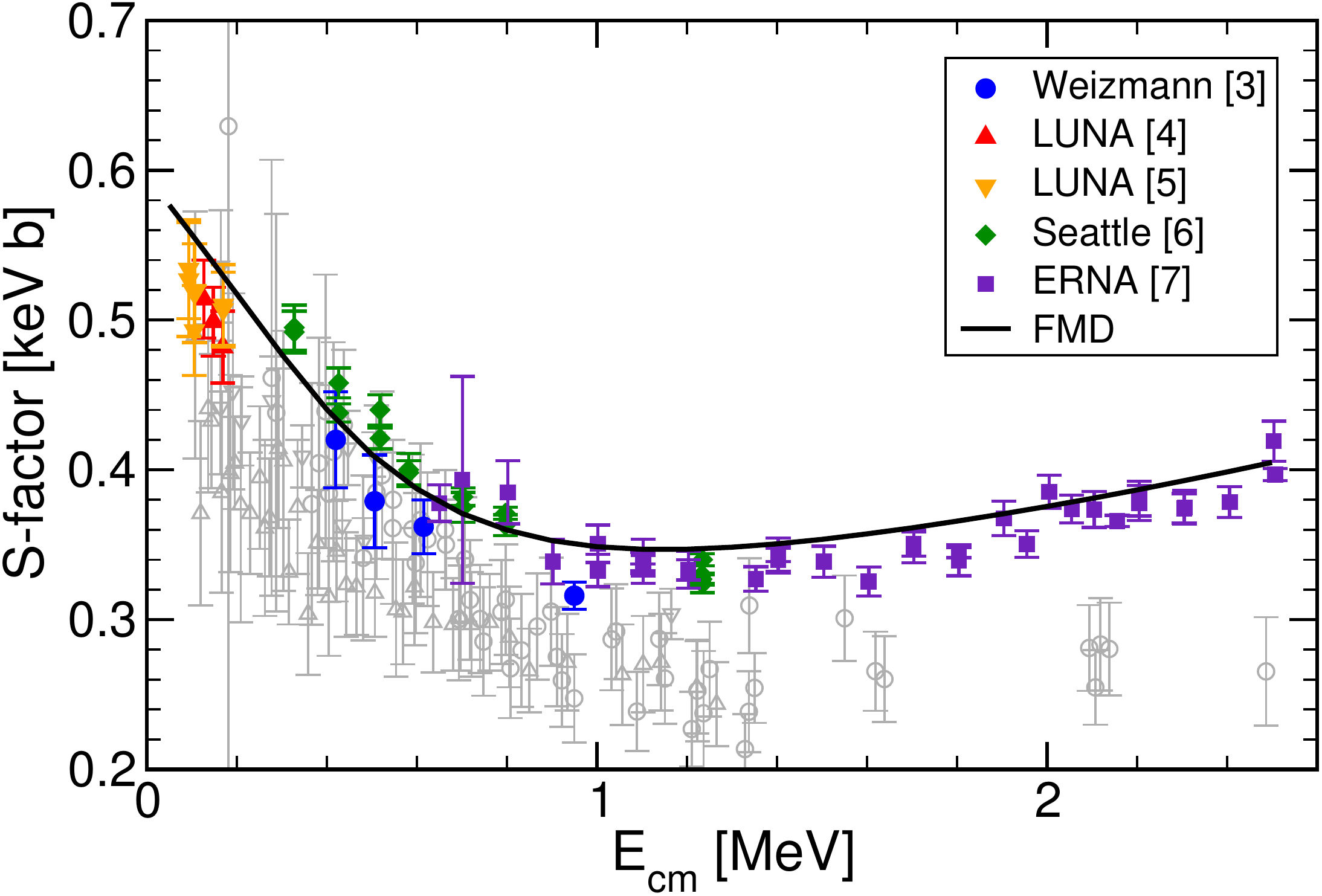}
   \caption{Astrophysical S-factor for $^3$He$(\alpha,\gamma){}^7$Be calculated using
   FMD~\cite{Neff:2012es,Neff:2010nm,Neff:2010gc}.}
   \label{fig:neff_Sfactor}
\end{center}
\end{figure}

An alternative approach to nuclear reactions of light nuclei
is the fermionic molecular dynamics (FMD) method,
which uses anti-symmetrized many-body states built from localized
(Gaussian) single-particle wave packets
to provide a fully microscopic calculation with both bound and scattering states
described consistently.  The effective interaction is derived by the
Unitary Correlation Operator Method (UCOM)~\cite{Roth:2010bm}, which eliminates short-range and tensor
correlations by unitary transformations guided by the SRG.
An example of the capabilities of FMD is shown in figure~\ref{fig:neff_Sfactor}, where the calculated
astrophysical S-factor for $^3$He$(\alpha,\gamma){}^7$Be is compared to new high-quality
experimental results~\cite{Neff:2012es,Neff:2010nm,Neff:2010gc}.  FMD is the only model that has been able to describe both the energy
dependence and normalization of the new data.

\subsection{In-medium Similarity Renormalization Group}

The in-medium SRG (IM-SRG) 
for nuclei, developed recently by Tsukiyama, Bogner, and
Schwenk~\cite{Tsukiyama:2010rj}, applies the RG flow equations 
in an $A$--body system using
a different reference state than the vacuum.
The key to
the IM-SRG is the use of \emph{normal-ordering} with respect to the
finite-density reference state. 
That is, starting from the
second-quantized Hamiltonian with two- and three-body interactions,
\bea
  H \ampseq \sum_{12} T_{12} \ad_1 a_2 + \frac{1}{(2!)^2} \sum_{1234}
  \, \langle12|V|34\rangle \ad_1\ad_2a_4a_3
  \nonumber \\
  \amps{} 
  \hspace*{-.2in}
  \null + \frac{1}{(3!)^2} \, \sum_{123456} \langle123|V^{(3)}|456\rangle
  \ad_1\ad_2\ad_3a_6a_5a_4 \,,
  \label{eq:Ham}
\eea
all operators are normal-ordered with respect to a finite-density 
Fermi vacuum $|\Phi\rangle$ (for example, the Hartree-Fock ground
state or the non-interacting Fermi sea in nuclear matter), as 
opposed to the zero-particle vacuum. Wick's theorem can then be
used to rewrite $H$ in normal-ordered form, which reshuffles the
contributions.  For example, the zero-, one-, and two-body normal-ordered
terms will now have contributions from the original three-body term
in Eq.~\eqref{eq:Ham},
which are in practice the dominant pieces.
Therefore, truncating the in-medium SRG
equations to two-body normal-ordered operators will
(to good approximation) evolve induced three- and higher-body interactions
through the density-dependent coefficients of the zero-, one-, and
two-body operators.
The appealing consequence is that,
unlike the free-space SRG evolution, the in-medium
SRG can approximately evolve
$3,...,A$-body operators using only two-body (or three-body) machinery.
However, also in contrast to the free-space SRG,
the in-medium evolution must be repeated for each nucleus or
density.

\begin{figure}[tbh!]
\begin{center}
 \includegraphics[width=3in]{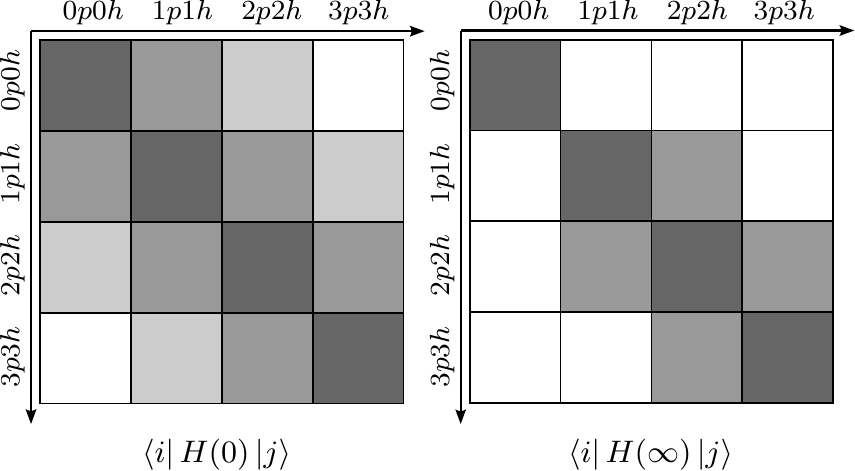}
   \caption{Schematic representation of the initial and final
   Hamiltonians in the many-body Hilbert space spanned by
   particle-hole excitations of the reference state~\cite{Hergert:2012nb}.}
   \label{fig:imsrg_schematic}
\end{center}
\end{figure}

The IM-SRG decouples the ground state of the many-body Hamiltonian from all excitations (see figure~\ref{fig:imsrg_schematic}) by 
means of a continuous unitary transformation, which is characterized by a suitable choice of dynamical generator. The method is implemented by solving a set of flow equations analogous to the free-space SRG approach. In principle, the IM-SRG is an exact method, but in practice a hierarchy of truncations is needed to close the set of equations; this hierarchy allows systematic improvements of the method. For practical applications, the IM-SRG has two appealing features.
First, most of the effects of higher-order induced many-body forces are 
automatically included (unlike free-space SRG evolution).
Second, it exhibits
polynomial scaling with the basis size, which allows us to perform calculations for nuclei that are not accessible by other ab initio many-body methods such as the NCSM or GFMC.

\begin{figure}[tbh!]
\begin{center}
 \includegraphics[width=2.5in]{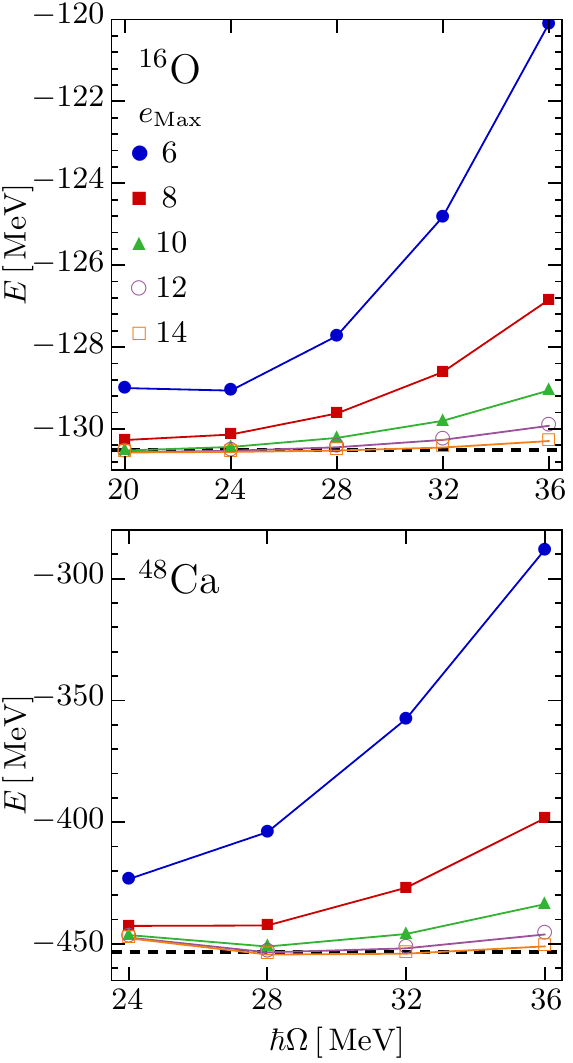}
   \caption{Examples of the convergence of the IM-SRG ground-state energy for different harmonic
   oscillator basis frequencies $\hbar \Omega$ and different basis sizes $e_{\rm{Max}}$
   for an initial NN+3N Hamiltonian, which is first
   evolved in free space to $\lambda=2.0\fmi$. It is evident that the results become 
   independent of the oscillator frequency with increasing basis size. See Ref.~\cite{Hergert:2012nb}
   for details.}
   \label{fig:imsrg_convergence}
\end{center}
\end{figure}

While the IM-SRG equations are of second
order in the interactions, the flow equations build up
non-perturbative physics through the evolution. In terms of diagrams, one can imagine
iterating the SRG equations in increments of the flow parameter $\delta s$. At each
additional increment $\delta s$, the interactions from the previous
step are inserted back into the right side of the SRG equations.
Iterating this procedure, one sees that the SRG accumulates
complicated particle-particle and particle-hole correlations to all orders
(see figure~2 in \cite{Hergert:2012nb}).
With an appropriate choice of generator, the Hamiltonian is
driven towards the diagonal, as indicated schematically
in figure~\ref{fig:imsrg_schematic}. This means that Hartree-Fock becomes
increasingly dominant with the off-diagonal matrix elements
being driven to zero~\cite{Kehrein:2006}.

\begin{figure}[tbh!]
 \begin{center}
  \includegraphics[width=0.5\textwidth]{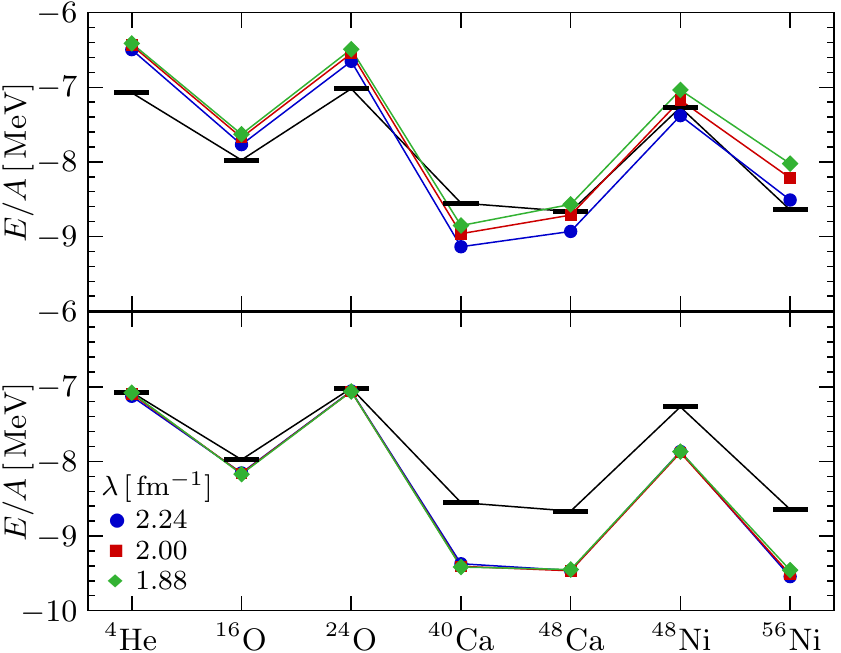}
    \caption{\label{fig:imsrg_closed_shell}
    Ground-state energies per nucleon of closed-shell nuclei using
    IM-SRG(2) at different SRG resolution scales $\lambda$~\cite{Hergert:2012nb}.
    Results with NN+3N-induced Hamiltonians are shown on top while the
    bottom includes initial 3NF.
    The black bars are experimental energies.
  } 
  \end{center}
\end{figure}

\begin{figure}[tbh!]
 \begin{center}
  \includegraphics[width=0.35\textwidth]{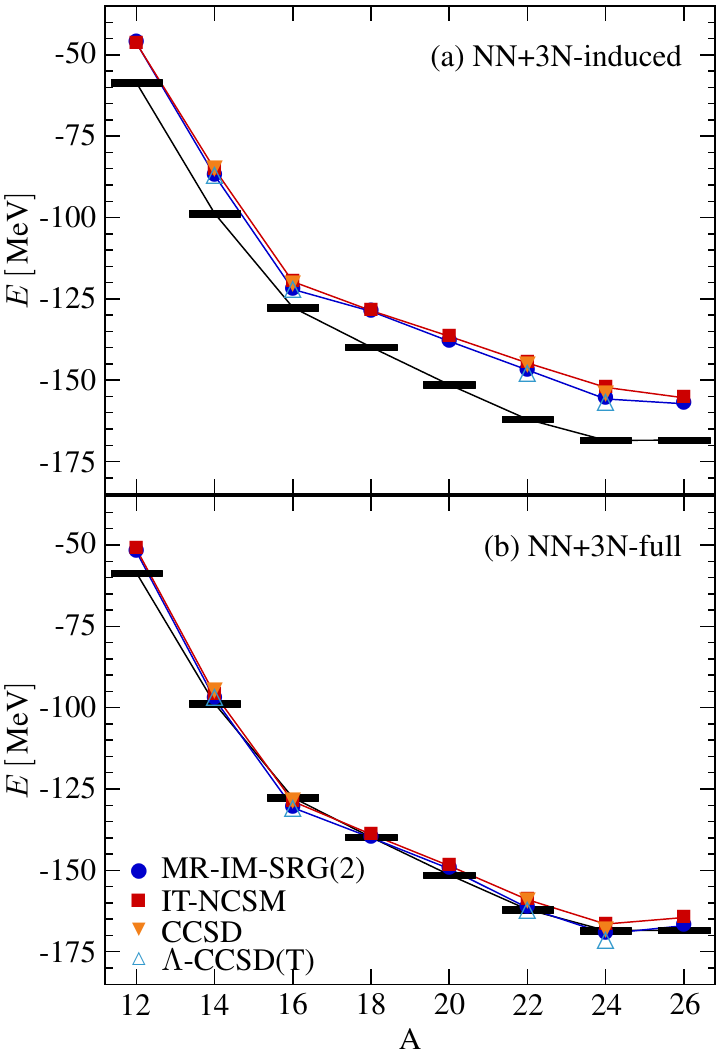}
    \caption{\label{fig:open}
    Ground-state energy of oxygen isotopes from the IM-SRG for different SRG parameters $\lambda$~\cite{Hergert:2013uja}. Top: Chiral NN Hamiltonian and induced 3N interaction (no initial 3N terms). Bottom: Consistently evolved chiral NN and 3N Hamiltonian.
    The pluses are experimental data from Ref.~\cite{Audi:2002rp}.
  } 
  \end{center}
\end{figure}

The in-medium SRG is well suited as an ab initio
method for finite nuclei. Figure~\ref{fig:imsrg_convergence} shows the 
rapid convergence of energy calculations.
Results for closed shell nuclei up to $^{56}$Ni are shown in 
figure~\ref{fig:imsrg_closed_shell}, which shows a striking improvement
in the isotopic trends with the inclusion of initial 3NF.
The IM-SRG truncated at the
normal-ordered two-body level gives results comparable to IT-NCSM and to
coupled-cluster calculations with some triples corrections ($\Lambda$-CCSD(T),
see Ref.~\cite{Hergert:2012nb}).

To allow systematic investigations of trends in ground-state energies (and other observables along complete isotopic chains), Hergert and collaborators have generalized the IM-SRG to the Multi-Reference IM-SRG (MR-IM-SRG)~\cite{Hergert:2013uja}. 
This approach allows the calculations to be extended to several hundred known spherical open-shell nuclei while retaining a polynomial numerical scaling. Figure~\ref{fig:open} shows a first application to the oxygen chain~\cite{Hergert:2013uja}, which is validated against CC and
IT-NCSM with the same Hamiltonians. 
The inclusion of an initial $3N$ Hamiltonian is seen to be needed
to obtain agreement with experimental data. 

Here are some in-progress extensions of the IM-SRG and MR-IM-SRG:
\begin{itemize}
  \item
   The calculation of excited states. There are several possibilities to achieve this, for instance, a modification of the currently used generator to decouple multiple states rather than just the ground state in the Hamiltonian's spectrum, or the adaptation of equations-of-motion methods like in coupled cluster (see e.g., \cite{shavitt2009many}). This would ultimately provide the capability to calculate transition densities, which can be used as input by the nuclear reaction community.
 \item   
As in the free-space SRG, all observables besides the Hamiltonian have to be evolved consistently, which can be implemented by evolving the creation and annihilation operators in which the flowing operators are represented rather than the matrix elements of these operators. The transformed basis operators, or alternatively transformed many-body density matrices, could then be used to calculate expectation values for any observable of interest in an economic fashion.
 \item
With a modified generator to allow for the decoupling of a pre-defined valence space, an evolved Hamiltonian obtained from a closed-shell IM-SRG calculation can be used as a microscopic input to traditional shell model approaches~\cite{Tsukiyama:2012sm}. 
Shell Model calculations with IM-SRG Hamiltonians will yield complete spectroscopic information, and are  in this sense complementary to the direct calculation of excited states in the IM-SRG framework. A combined IM-SRG/Shell Model approach is the most practical way to study deformed nuclei in the near future.
 \item
 IM-SRG is being adapted for nuclear matter by implementing it
 in a periodic box.  This will provide a non-perturbative
 assessment of infinite matter calculations.
\end{itemize} 

\subsection{Other applications and future directions}

We have only described a fraction of the recent and on-going work on finite nuclei
that exploits RG methods.  However, here we briefly describe some of the other important developments
that make use of low-momentum interactions (see also Refs.~\cite{Leidemann:2012hr,Barrett:2013nh}).  
 
\begin{itemize}

 \item
 In Ref.~\cite{Bacca:2012up}, the binding energy and radii of the two-neutron halo nucleus $^6$He have been studied in the hyperspherical harmonics
 approach based on low-momentum NN interactions. The RG evolution has been found to be essential to obtain converged results of the extended matter 
 radius and of the point-proton radius.

\item 
 In Ref.~\cite{Papadimitriou:2013ix}, the
 No-Core Gamow Shell Model (NCGSM), which treats bound, resonant, and scattering states
 equally, was first applied to study some well-bound and unbound states of the 
 helium isotopes~\cite{Papadimitriou:2013ix}.  
 The density matrix renormalization group (DMRG) method~\cite{Rotureau:2008rp}
 was used to solve the many-body Schr\"odinger equation.
 The $\vlowk$ RG was used to decouple high from low momentum to improve the convergence of the calculations.  RG-evolved 
 low-momentum interactions 
 are now a standard first choice for proof-of-principle or benchmark calculations using new techniques. 

\item 
 Time-dependent coupled-cluster theory has recently been studied in the framework of 
 nuclear physics~\cite{Pigg:2012sw}. 
Besides using a low-momentum SRG two-body interaction for their proof-of-principle computations,
the authors are able to
relate the real and imaginary time evolution of the Hamiltonian to SRG transformations.

\item
Microscopic calculations of pairing properties in mid- and heavy-mass nuclei are being
pursued using the ab-initio self-consistent Gorkov Green's function (SCGGF) framework based
on low-momentum interactions~\cite{Soma:2011aj,Duguet:2012te}. Because in practice a tractable truncation scheme must be implemented, RG-softened interactions are a key ingredient to make finite-order
schemes qualitatively and quantitatively usable.  Recent results using 3NF show
great promise in reproducing the physics of neutron driplines and truly open-shell systems,
increasing the number of medium-mass nuclei accessible by ab initio
methods from a few tens to a few hundreds~\cite{Barbieri:2012rd,Soma:2012zd,Cipollone:2013zma}.

 \item 
Recent work has critically examined the infrared and ultraviolet cutoffs imposed on few- and many-body systems by the use of harmonic oscillator basis expansions, which are common in nuclear physics~\cite{Coon:2012ab,Furnstahl:2012qg}.  While progress toward a theoretically founded understanding of universal infrared extrapolations has been made~\cite{More:2013rma}, the
ultraviolet situation is less clear.  The SRG offers a tool for studying ultraviolet extrapolations, including suggestive but as yet unexplained scaling with the SRG flow parameter~\cite{Furnstahl:2012qg}.

 \item 
Quantum Monte Carlo (QMC) methods such as Green's function or auxiliary field diffusion Monte Carlo
are powerful methods for nuclear structure calculations but have been restricted for technical reasons to local interactions.  This has precluded their application with low-momentum interactions.
Recent progress has been made toward relaxing this restriction, which has also demonstrated  universal behavior of propagators at large imaginary times~\cite{Lynn:2012fq}.  
  
 \item 
 Spectral distribution theory (SDT) has been applied to SRG-evolved Hamiltonians to study the 
nature of three-body SRG-induced interactions at an operator level~\cite{PhysRevC.85.044003}.
This approach reveals that the SRG-renormalized interaction is essentially two-body
driven, with the two-body part extractable in the SDT framework.
 
 \item 
 The role of long-distance symmetries within the
 context of SRG evolution has been explored recently~\cite{Timoteo:2011tt}.
 A particular SRG resolution scale is identified for which the Wigner SU(4) symmetry is almost
 perfectly realized at the two-body level.
 This motivates a search for similar symmetry patterns for many-body forces.
    
 \item 
The variable cutoff (or decoupling scale) implemented in RG methods is
a useful tool for analyzing scheme-dependent observables in nuclear 
structure\cite{Furnstahl:2010wd}. Recent applications have been made to ab initio spectroscopic
factors~\cite{Jensen:2010vj} and to effective single-particle energies (ESPEs)~\cite{Duguet:2011sq}.
In the latter case, varying the $\vlowk$ cutoff clearly identifies the scale dependence
of ESPEs and sets the stage for a future quantitative analysis, which will require the treatment of 3NF~\cite{Duguet:2011sq}.

\end{itemize}
We anticipate many new results from these and other RG-motivated investigations in the
near future.


\section{Correlations in nuclear systems}\label{Sect:Correlations}

\subsection{Evolution of operators as an RG frontier}

\begin{figure*}[tbh!]
\begin{center}
\includegraphics[width=0.45\textwidth]{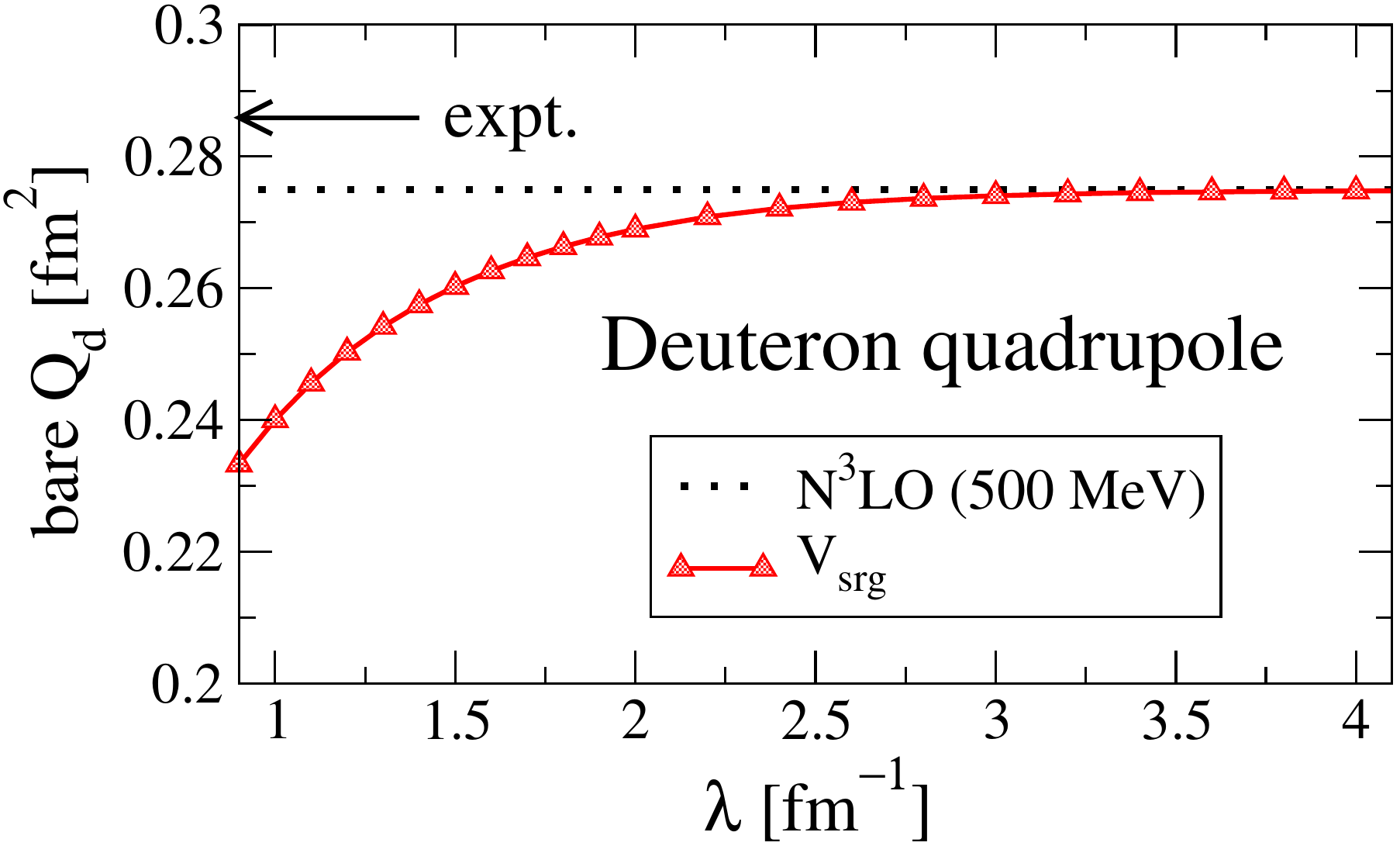}
   \caption{Expectation value in the deuteron of the unevolved
   quadrupole operator as a function of the
   SRG resolution $\lambda$~\cite{Jurgenson:2007td}.  The initial potential is the
   N$^3$LO (500 MeV) chiral NN interaction from Ref.~\cite{Entem:2003ft}.
   The experimental value of the electromagnetic quadrupole moment is
   marked with an arrow. }
   \label{fig:Qd_deuteron}
\end{center}
\end{figure*}

So far we have focused on the evolution of 
Hamiltonians, but an RG transformation will also modify the
operators associated with measurable quantities.
If we do not evolve the operator, then its matrix elements calculated
with the wave functions of the flowing Hamiltonian will change.
Consider, for example, the expectation of a quadrupole operator in the deuteron.
It is naturally defined in coordinate space~\cite{Bogner:2006vp}: 
\beq
\left\langle Q_{d}\right\rangle=\frac{1}{20}\int^{\infty}_{0}dr\,r^2 w(r)\left( \sqrt{8}\,u(r)-w(r) \right)
  \;,
  \label{eq:quadmom}
\eeq
where \textit{u} and \textit{w} are the \( ^{3}\mbox{S}_1\) and \(
^{3}\mbox{D}_1\) deuteron radial wave
functions. 
This expectation value as a function of SRG $\lambda$ is shown in figure~\ref{fig:Qd_deuteron}.
Because the quadrupole operator acts on long distance scales (that is, it
predominantly samples the large $r$ part of the relative wave function),
the variation with $\lambda$ is relatively small but increases more rapidly
at the lowest resolutions.  The result is never equal to the electromagnetic
quadrupole moment extracted from experiment because the operator in 
Eq.~\eqref{eq:quadmom} is only the leading-order piece of the full operator
that corresponds to this particular experimental quantity.  For each Hamiltonian
(either initial or at each evolved resolution), there will be a different
consistent operator.

To enable a general description of experiments, it is essential to be able to start with full operators consistent with the initial Hamiltonian
and then to evolve them maintaining this consistency.
By working within an EFT framework, we can ensure consistent initial
operators because the EFT provides a complete operator basis organized
hierarchically by power counting.
The second step can be technically difficult, especially
because we will inevitably induce many-body operators as we
evolve.
However, the technology recently developed for the SRG to evolve many-body
Hamiltonians (Section~\ref{sec:RG_tech}) can be adapted to evolve 
other operators at the same time.  

The SRG evolution with $\flow$ (recall $s = 1/\lambda^4$) 
of \emph{any} operator $O$ 
is given by:
\beq
  O_\flow = U_\flow O U^\dagger_\flow \;,
  \label{eq:Oflow}
\eeq
so $O_\flow$ evolves via
 \beq
   \frac{dO_\flow}{d\flow}
   = [ [G_\flow,H_\flow], O_\flow]  \;,
 \eeq
where we must use the same $G_\flow$ to evolve the Hamiltonian
and all other operators. 
While we can directly evolve any operator like this in parallel
to the evolution of the Hamiltonian, in practice it is more
efficient and numerically robust to either evolve the unitary
transformation $U_s$ itself:
\beq
  \frac{dU_\flow}{d\flow}
    = \eta_s U_s = [G_s,H_s] U_s
    \;,
\eeq
with initial value $U_{s=0} = 1$, or calculate it directly
from the eigenvectors of $H_{s=0}$ and $H_s$:
\beq
  U_s = \sum_i |\psi_i(s)\ra \la \psi_i(0)| \;.
\eeq
Then any operator is directly evolved to the desired $\flow$
by applying Eq.~\eqref{eq:Oflow} as a matrix multiplication.
The second method works well in practice.
As with the Hamiltonian, the two-body part of an operator is completely
determined by evolution in the $A=2$ space, the three-body part by
evolution in the $A=3$ space, and so on.

\begin{figure}[tbh!]
  \begin{center}
    \raisebox{0.85cm}{\includegraphics[scale=0.8]{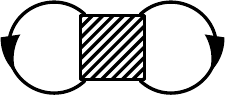}}
    \raisebox{1.2cm}{$+$}
    \includegraphics[scale=0.8]{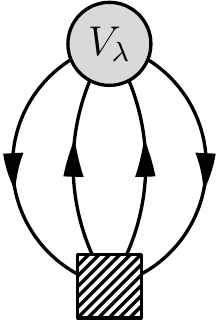}
    \raisebox{1.2cm}{$+$}
    \includegraphics[scale=0.8]{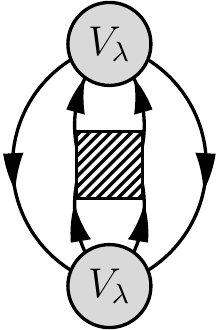}
    \raisebox{1.2cm}{$+$}
    \includegraphics[scale=0.8]{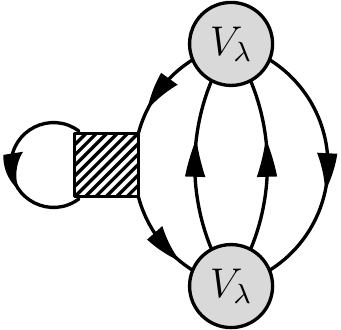}
    \caption{Diagrammatic contributions to the expectation value of
    a two-body operator (denoted by the shaded square)
    for the first three orders in MBPT with an NN-only potential.  }
    \label{fig:rel_mom_op}
  \end{center}
\end{figure}

One option to evaluate operator matrix elements is to apply many-body perturbation theory (MBPT). 
For example, the diagrams for the first three orders in MBPT of a two-body
operator (assumed to depend only on relative momenta)
are shown in figure~\ref{fig:rel_mom_op} assuming the Hamiltonian has only
two-body interactions. 
In effect, the operator is inserted into the MBPT expansion for the
energy in all possible ways (including proper symmetry factors).
Diagrammatic perturbation theory of this sort was recently used 
for an effective double-beta-decay
operator by Holt and Engel~\cite{Holt:2013tda}.
The challenge going forward is to include three-body interactions
(see figure~\ref{fig:EOS_diagrams}) and the induced three-body parts of SRG-evolved
operators.

For finite nuclei beyond the deuteron, the process of evolving and applying 
operators has several complications.  Imagine we start with a one-body operator
that we wish to evaluate in an $A$-particle nucleus. 
As we evolve the operator, first in a 2-particle basis, then a 3-particle basis,
and so on (until the desired level of truncation), 
$n$-body components will be induced and must
be kept if matrix elements are to be invariant.  These components must be
separated 
because they are be embedded in larger nuclei 
with different counting factors~\cite{Anderson:2010aq}.
In addition, we need in general to apply appropriate boosts to the operators
before embedding.
A flowchart summarizing the procedure is given as figure~49 in Ref.~\cite{Furnstahl:2012fn}.
Proof-of-principle calculations for this procedure are expected soon.

 The consistent evolution and application of operators is a frontier
for using RG in nuclear physics
and there are many opportunities for ground-breaking calculations.
Recent work with new operators include the calculation of the $^4$He total photo absorption cross section in Ref.~\cite{Schuster:2013sda}, which for the first time tests the consistency of the SRG approach with a continuum observable,
and the evaluation of neutrinoless double-beta decay with SRG-evolved interactions
in Ref.~\cite{Shukla:2011mb}.
Adapting the RG technology developed for evolving and evaluating Hamiltonians to 
extend these and related studies to use fully consistent evolved operators 
is an important goal.

\subsection{Scale dependence of short-range correlations}

\begin{figure}[tbh!]
  \begin{center}
    \includegraphics[scale=0.5]{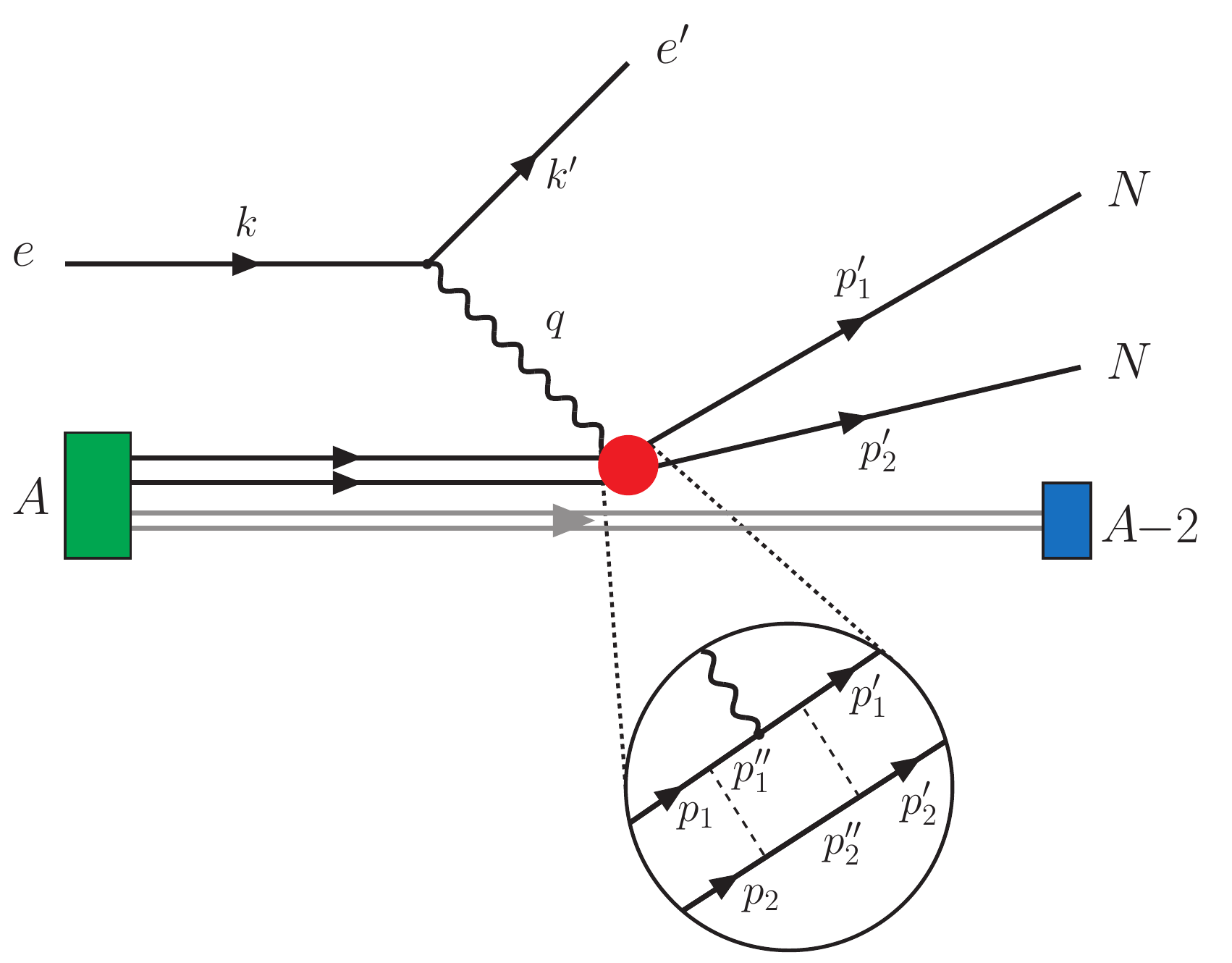}
  \end{center}
  \caption{  
  Illustration of different interpretations of deep-inelastic two-body knock-out 
  reactions: In the SRC picture NN interactions scatter two-body states in the initial nucleus with small initial 
  momenta $p_1$ and $p_2$ to states with large intermediate relative momenta $p''_1$ and $p''_2$, which are then knocked 
  out by the photon via a one-body interaction (see magnification of red vertex). The second interaction line represents a final state interaction.
  In general, the vertex function depends on the RG resolution scale. At low scales the wave functions are much simpler, but 
  the vertex is a more complex two-body operator.}
  \label{fig:src_diagram}
\end{figure}

\begin{figure*}[tbh!]
\begin{center}
\includegraphics[width=2.4in]{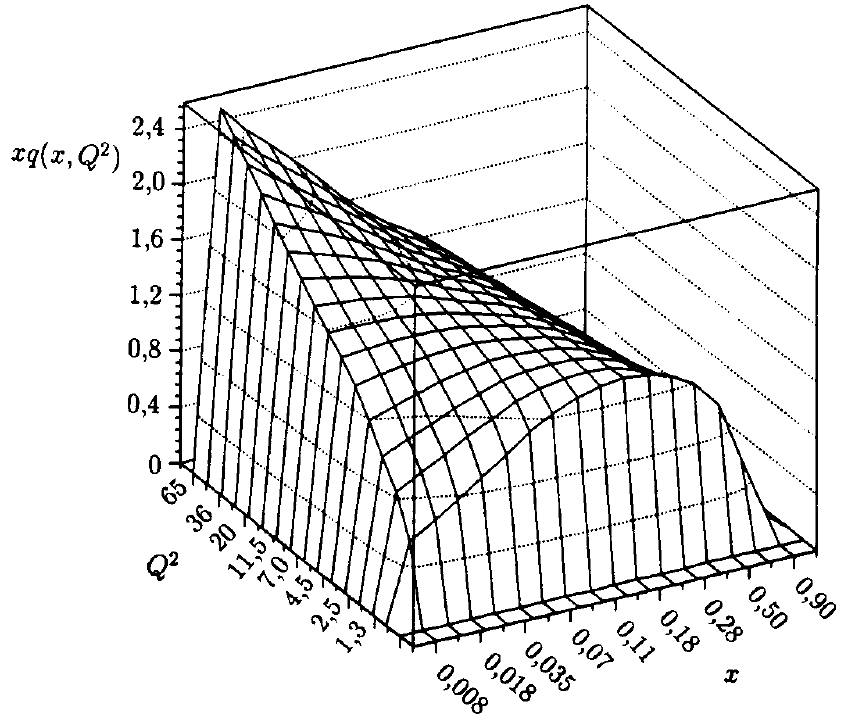}~~%
\includegraphics[width=2.5in]{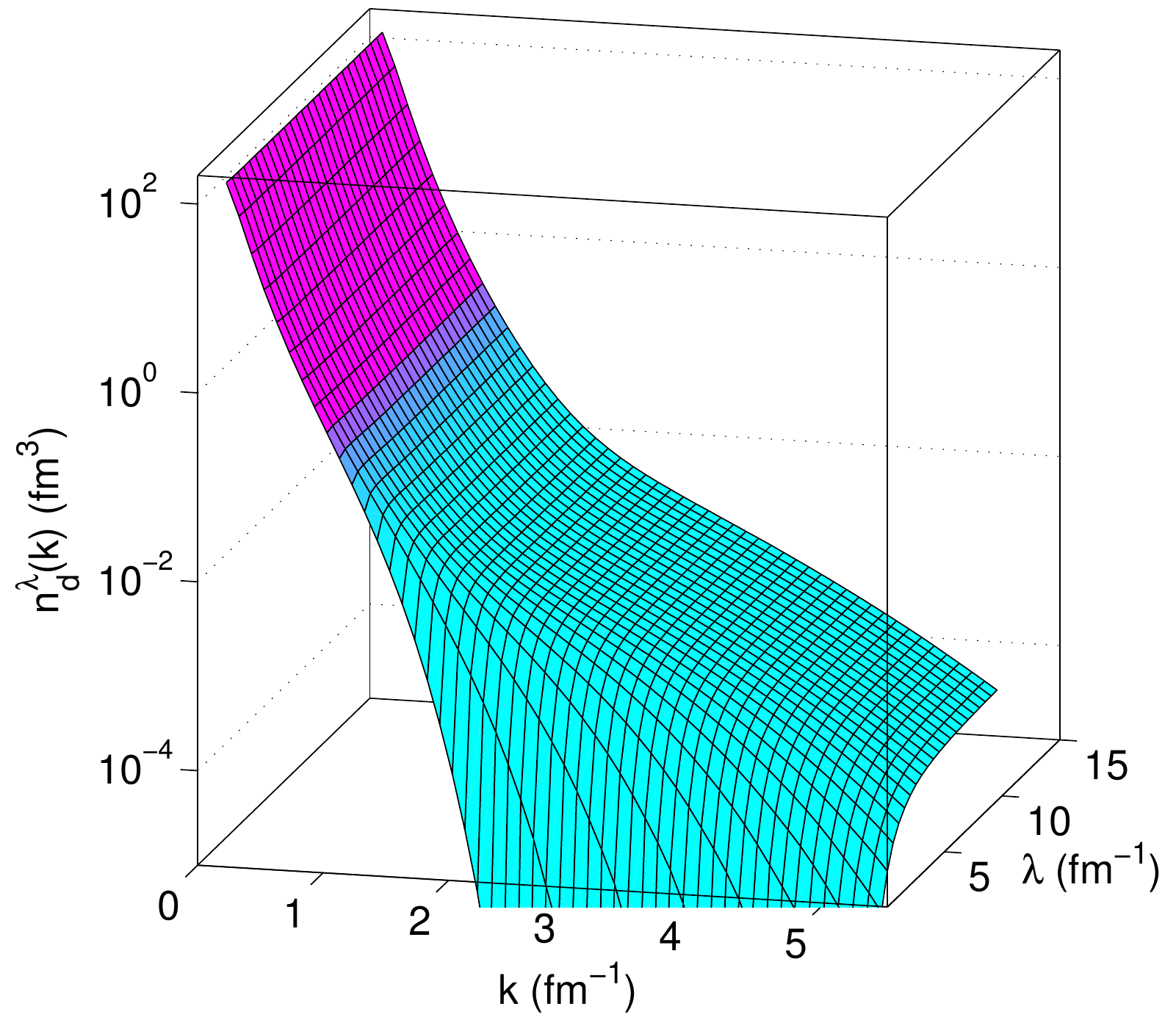}
   \caption{Quark parton distribution $x q(x,Q^2)$ as a function of $x$ and $Q^2$ 
   (left, from~\cite{Povh:706817}) and deuteron
   momentum distribution $n(k)$ at different SRG resolutions $\lambda$ (right).}
   \label{fig:mom_dists}
\end{center}
\end{figure*}

\begin{figure*}[tbh!]
\begin{center}
 \includegraphics[width=.45\textwidth]{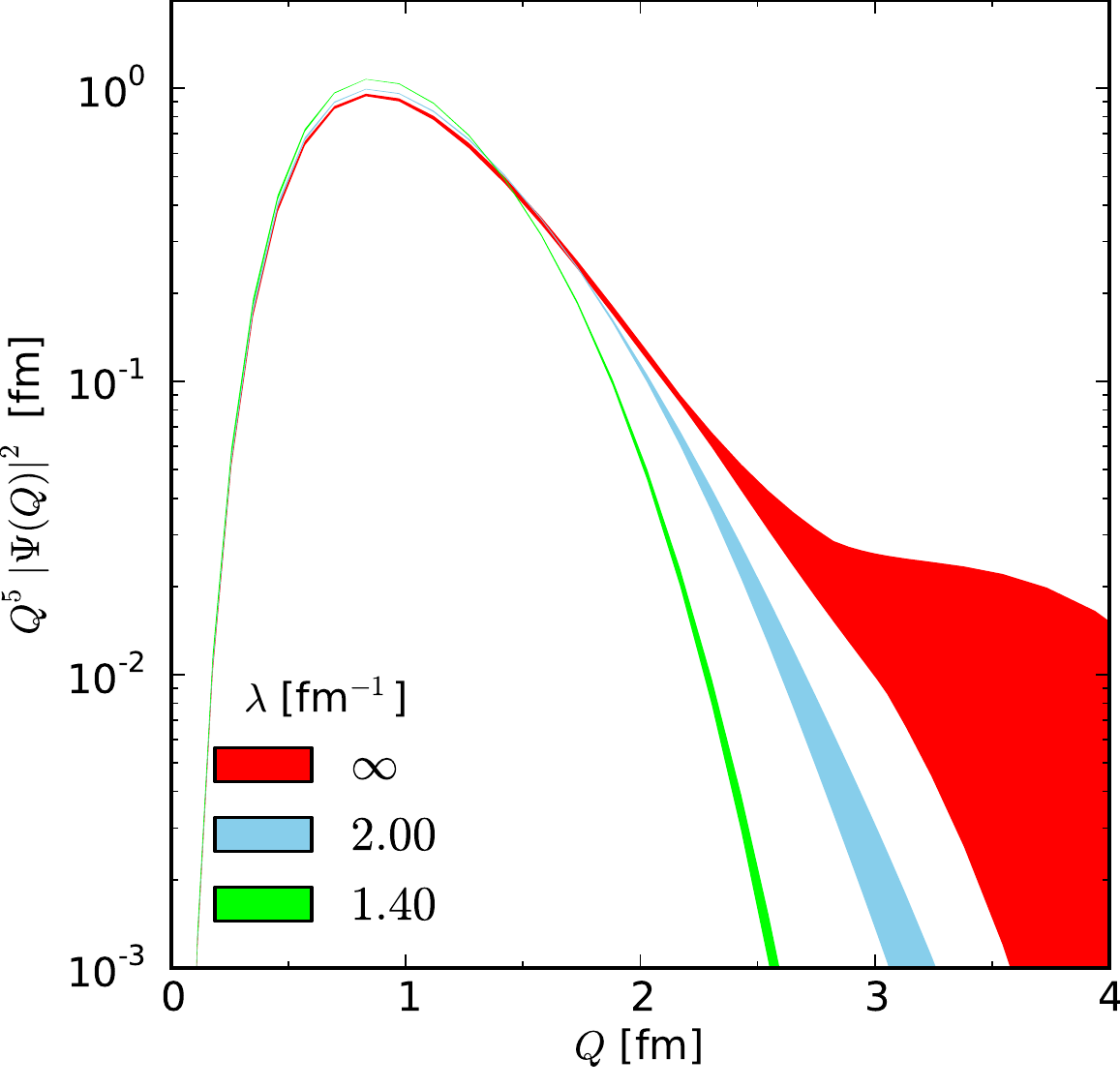}
   \caption{SRG evolution of the triton probability distribution as a function of hypermomentum $Q$ for several different SRG $\lambda$'s for a set of chiral EFT interactions. These are plotted as bands that span the range of the wave functions from different initial 2N+3N N2LO interactions. See Ref.~\cite{Wendt:2013bla} for details.}
   \label{fig:kyle_triton_wf}
\end{center}
\end{figure*}

\begin{figure*}[tbh!]
\begin{center}
 \includegraphics[width=5.0in]{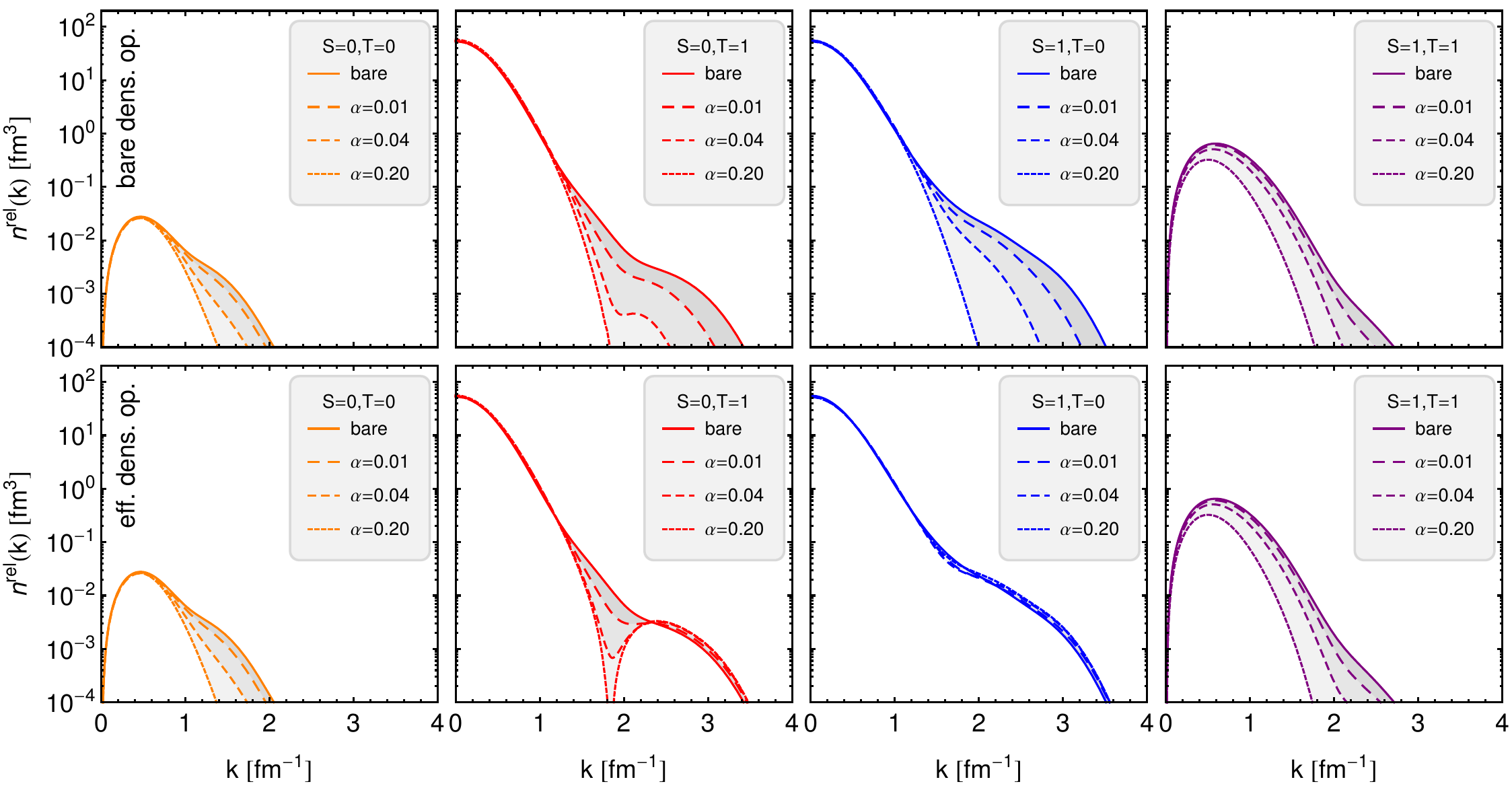}
   \caption{Two-body relative momentum distribution in $^4$He~\cite{Neff:2013private}.}
   \label{fig:neff_he4_n3lo_2body_density}
\end{center}
\end{figure*}

Recent experimental studies of proton knock-out reactions off nuclei at high-momentum transfer (see figure~\ref{fig:src_diagram}) have 
been explained by invoking short-range correlations (SRCs) in nuclear systems,
which are manifested as enhanced strength in relative momentum distributions
well above the nuclear Fermi momentum~\cite{Subedi:2008zz,Frankfurt:2008zv,Arrington:2011xs,Alvioli:2012qa}. 
Such explanations may seem at odds with RG evolution, which 
leads to many-body wave functions with highly suppressed SRCs. 
But the RG implies that
nuclear momentum distributions are scale (and scheme) dependent, just like QCD
parton distributions~\cite{Furnstahl:2010wd}.
This analogy is illustrated in figure~\ref{fig:mom_dists}.
In the left panel, the combination $x q(x,Q^2)$ measures the share of momentum carried by quarks in a proton within a particular $x$-interval~\cite{Povh:706817}.  This momentum distribution changes as a function
of the resolution scale $Q^2$ according to RG evolution equations.
In the right panel, the deuteron momentum distribution $n^{\lambda}(k)$
for an initial AV18 potential
(the choice of potential is a \emph{scheme} dependence) is SRG-evolved from $\lambda = \infty$
(corresponding to the initial potential) down to $\lambda = 1.5\fmi$.
It is evident that the high-momentum tail, which is identified with SRC physics,
is highly scale dependent and is essentially eliminated at lower resolution.

Recent calculations demonstrate the scale and scheme
dependence of the momentum distribution in few-body nuclei.
The probability distribution of the hypermomentum $Q$ of the triton is shown
in figure~\ref{fig:kyle_triton_wf}~\cite{Wendt:2013bla}.  
The high momentum components of the triton wave function are seen to disappear as the
SRG resolution scale $\lambda$ is lowered.  The spread of each band shows the scheme dependence
at each $\lambda$; that is, the dependence on the initial interaction.  As the interaction
becomes more universal with decreasing $\lambda$, the scheme dependence naturally decreases
as well.
A comparison of unevolved (``bare'') and evolved two-body momentum distribution in
$^4$He at four SRG resolutions
is shown in figure~\ref{fig:neff_he4_n3lo_2body_density}, where the densities are separated
into spin-isospin channels ($S$ and $T$)~\cite{Horiuchi:2013fza,Feldmeier:2011qy}.  The top row uses unevolved operators 
and illustrates the usual suppression of high-momentum components with SRG
evolution.  The bottom row uses operators consistently evolved but only at the two-body
level, i.e., induced three-body components are dropped~\cite{Neff:2013private}. 
If all components were kept,
the results would be identical at all resolutions, so the deviations from the initial
distributions indicate the 
expected size of three-body pieces at different momenta and in different channels.
For $S=1$, $T=0$, the result is almost unitary, while for $S=0$, $T=1$ deviations are
limited to a range of intermediate momenta, and there are significant deviations in
the other two channels.  Follow-up studies will soon be possible using evolved operators with
induced three-body components maintained.

\begin{figure*}[tbh!]
\begin{center}
\includegraphics[width=5.0in]{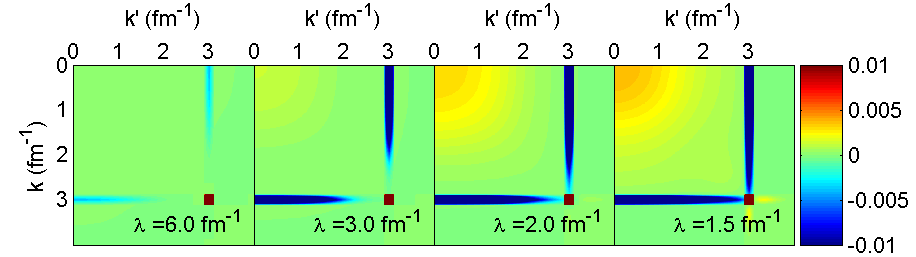}
 \includegraphics[width=5.0in]{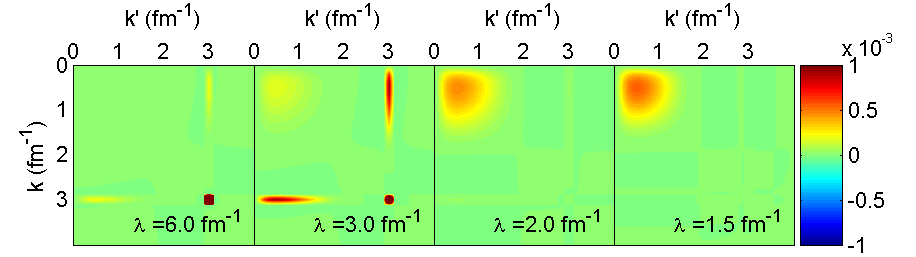} 
   \caption{The top plot is the integrand of the SRG-evolved operator $a^\dagger_q a_q$ in the deuteron channel evaluated at $q=3\fmi$. The bottom plot is the same
   operator but now sandwiched between the deuteron wave functions (see Ref.~\cite{Anderson:2010aq}).}
   \label{fig:mom_dist_integrands}
\end{center}
\end{figure*}

\begin{figure}[tbh!]
  \begin{center}
    \includegraphics[width=0.45\textwidth]{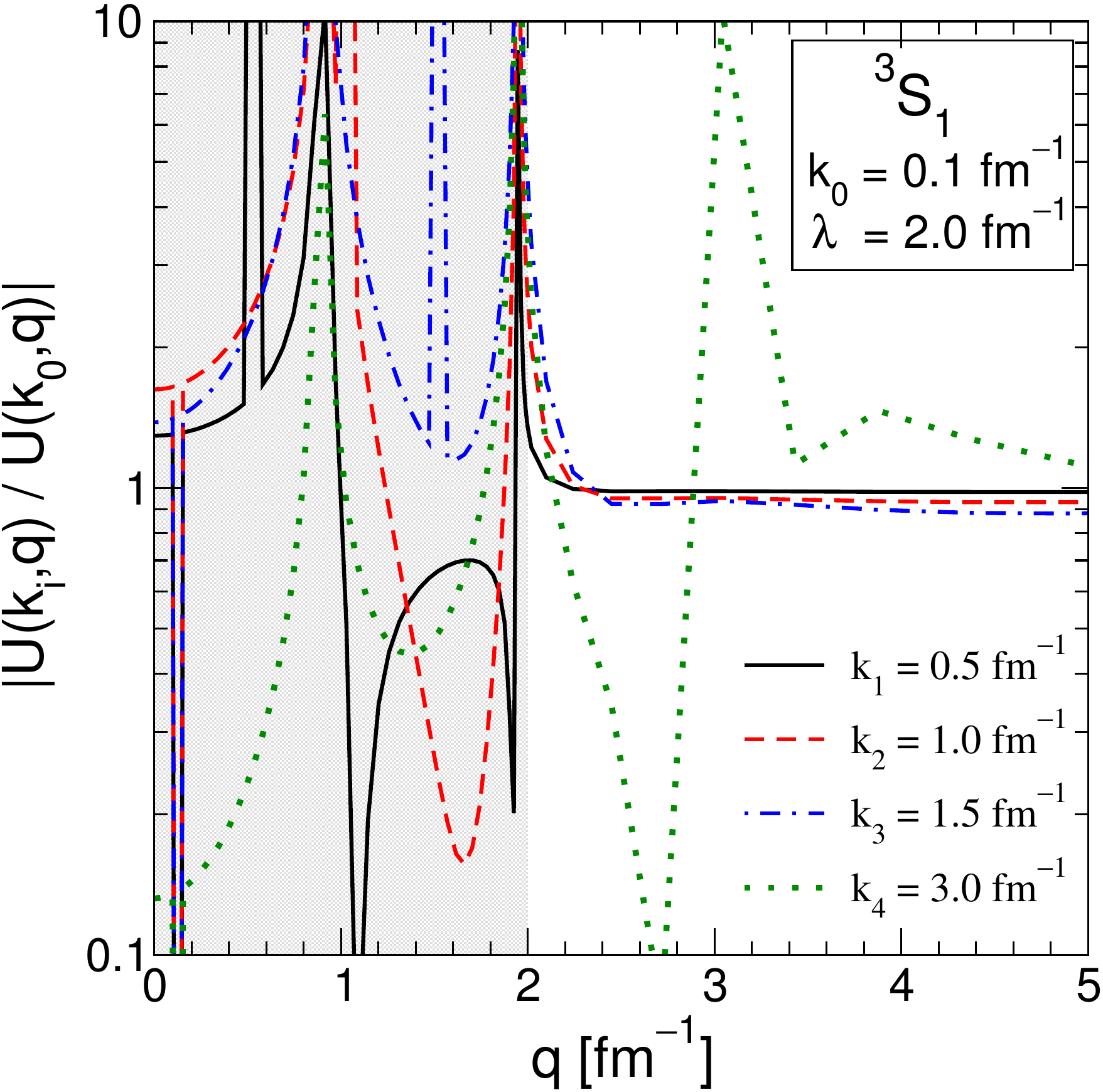}
  \end{center}
  \caption{Ratio of two-body SRG unitary transformations $U(k_i,q)/U(k_0,q)$ 
  at $\lambda = 2.0\fmi$
  for the $^3$S$_1$ channel, plotted as a
  function of $q$ at fixed $k_i$ and $k_0$. 
  Factorization is signaled by a plateau in this ratio; it is expected 
  when $q \gg \lambda$ (non-shaded region) \emph{and} $k_i \ll \lambda$
  (which is valid for $k_1$, $k_2$, and $k_3$ but not $k_4$). See 
  Ref.\cite{Anderson:2010aq} for details.}
  \label{fig:factorization}
\end{figure}

If there are no high momentum components at low RG resolution,
how do we interpret the physics in experiments such as in
figure~\ref{fig:src_diagram}?  The SRG unitary transformation approach means that cross sections should
be invariant under a change in resolution.  
Indeed, the evolution of the wave functions describing the structure is compensated
by the evolution of the operators describing the reaction.
This means that the relative
contributions of structure, currents, and initial/final state interactions
are not fixed. 
A key question is then:  What is the appropriate resolution scale for a given
process; e.g., can we minimize complications such as
final-state interactions? 

The changing physics of operators with decreasing resolution is illustrated
in figure~\ref{fig:mom_dist_integrands}.  The top row shows the integrand of the
simple operator $a^\dagger_q a_q$ at $q=3\fmi$ in the deuteron channel.
At high resolution ($\lambda=6\fmi$), it is essentially a one-body operator measuring
the momentum strength at $k=3\fmi$.  At lower resolutions, the one-body part
remains (one-body operators do not evolve for the usual SRG generator) but
smoothly distributed strength develops at low momentum.  When this operator is
sandwiched in the deuteron wave function (bottom row), the contribution at
$k=3\fmi$ fades away at lower resolutions because the wave function completely suppresses high momentum contributions, 
but it is precisely replaced by the smooth low momentum contribution, 
reproducing the original (unevolved) 
momentum distribution at all resolution scales~\cite{Anderson:2010aq}.

A similar trade-off of structure and reaction
happens within the magnified blob in figure~\ref{fig:src_diagram}. 
At high resolution, the major
contribution to two-nucleon knockout is when the two-body interaction couples
a low-momentum pair to high momentum (creating an SRC in the nuclear wave function), which is 
subsequently knocked out by the photon via the dominant \emph{one-body} interaction. At low 
resolution, the blob instead describes a \emph{two-body} operator vertex with two soft initial momenta and 
two hard final momenta, which represent the observed knocked out particles.

The low-resolution picture is also accompanied by a major simplification
from the scale separation of low and high-energy physics.
This is manifested by a corresponding factorization of the unitary
transformation, $U(k,q) \approx K(k)Q(q)$ for $k \ll \lambda$
and $q \gg \lambda$, which is demonstrated in figure~\ref{fig:factorization}.
The factorization of the unitary
transformation 
was shown  in Ref.~\cite{Anderson:2010aq} 
to follow from effective interaction methods
as well as the nonrelativistic
operator product expansion~\cite{Lepage:1997cs,Braaten:2008uh,Braaten:2008bi}.
 In Ref.~\cite{Bogner:2012zm}, Bogner and Roscher applied basic decoupling and scale-separation arguments
 to extend these results to arbitrary low-energy $A$-body states, showing 
 that
 the high-momentum tails of momentum distributions and static structure factors factorize into the product of a universal function of momentum fixed by two-body physics, and a state-dependent matrix element that is sensitive only to low-momentum structure of the many-body state. 
 This separation provides an alternative interpretation of phenomena like
nuclear scaling~\cite{Frankfurt:2008zv,Arrington:2011xs,Alvioli:2012qa}, because the universal part will cancel (to leading
order) in ratios of high-momentum tails or inclusive cross sections for different nuclei~\cite{Anderson:2010aq},
leading to characteristic plateaus.
The question under active investigation is whether 
this factorization can be exploited to \emph{quantitatively} calculate nuclear scaling
ratios as well as higher-order corrections.


\section{Summary and outlook}\label{Sect:Summary}

The general strategy of applying the renormalization group (RG) to low-energy physics is to
lower the resolution of inter-nucleon interactions while tracking dependence on it. 
High-resolution interactions contain strong coupling of low momenta to high momenta, 
which complicates solutions of the many-body problem for low-energy properties. 
An RG evolution leads to much fewer correlated wave functions at low resolution and to faster 
convergence of many-body methods. Current SRG flow equations decouple low and high momenta in 
the form of band or block diagonalization of the Hamiltonian matrix. They represent a
series of unitary transformations, in which observables are not altered but the physics interpretation 
can (and in general will) change. During the evolution to lower resolution non-local interactions 
and many-body operators are induced, which must be accommodated. In practice that means
the RG evolution is performed until few-body forces start to grow rapidly, so that no 
controlled many-body calculations are possible anymore.

The RG technology also provides new tools to estimate theoretical uncertainties.
The basic idea is that, in principle, observables should be unchanged
with RG evolution, i.e., be independent of the resolution scale. In practice however, there are 
approximations in the RG implementation and in calculating nuclear observables due to truncation 
of ``induced'' many-body forces/operators and from many-body approximations.  For nuclei
there can be dramatic changes even with apparently small changes in the resolution scale. These 
resolution-scale dependences can be used as diagnostics of approximations.
Recent applications of resolution-scale dependence include:
   \begin{itemize}
   \I using cutoff dependence at different orders in an EFT expansion to investigate the validity of 
    chiral power counting, which carries over to the corresponding RG-evolved interactions; 

   \I using the running of ground-state energies of nuclei with resolution scales
   to estimate errors, identify correlations and diagnosing missing many-body forces
   (e.g., Tjon lines)~\cite{Hammer:2010kp,Hammer:2011kg,Roth:2011vt,Binder:2012mk,Jurgenson:2013yya}; 
     
   \I validating MBPT convergence in calculations for infinite nuclear matter
   and setting lower bounds on the errors from uncertainties 
   in many-body interactions~\cite{Hebeler:2010xb,Hebeler:2013ri,Gezerlis:2013ipa}; 
      
   \I identifying and characterizing scheme-dependent observables,
   such as spectroscopic factors and effective single-particle 
   energies~\cite{Jensen:2010vj,Duguet:2011sq,Furnstahl:2010wd}. 
  \end{itemize}   

In this review, we have shown glimpses of the many promising applications of RG methods to nuclei.
Configuration interaction, coupled cluster, IM-SRG, and self-consistent Green's function
approaches using softened interactions converge faster, opening up new possibilities to extend the
limits of computational feasibility. Ground-breaking ab-initio reaction calculations are now possible.
Applications of low-momentum interactions to microscopic shell model calculations bring new understanding
to phenomenological results, highlighting the role of three-body forces. In-medium SRG offers a means 
to directly calculate effective shell-model interactions. Because many-body perturbation 
theory (possibly resummed) is feasible with the evolved interactions,
the door is opened to constructive nuclear density functional
theory~\cite{Bogner:2008kj,Drut:2009ce,Duguet:2010qw,Stoitsov:2010ha,Duguet:2012te,Holt:2013fwa}.

Despite many successes, there are also open questions and difficult problems in applying
RG to low-energy nuclear physics.  Here is a subset:
  \begin{itemize}
    \I Perhaps the most outstanding issue at present is the size and nature of
    four-body contributions in larger nuclei.  Calculations of induced
    four-body forces are now feasible and should provide direct tests
    in the near future. 

   \I
    More generally,
    we need quantitative power counting for evolved many-body operators.
    That is, how do we anticipate the size of contributions from
    induced many-body interactions and other operators? This is essential if 
    we are to have reliable estimates of theoretical errors, because truncations 
    are unavoidable. We need both analytic estimates to guide us as well as more extensive 
    numerical tests. Many of the same issues apply to chiral EFT; can the additional
    information available from SRG flow parameter dependence help with analyzing or even constructing EFT's?
    
    \I In Sections \ref{subsec:NN_evolv} and \ref{subsec:MBPT_finnucl} we presented calculations for infinite nuclear matter 
    and finite nuclei using different types of MBPT based on RG-evolved interactions. How can the different convergence patterns be 
    reconciled and what is the best measure to quantify the perturbativeness of nuclear interactions?

    \I Only a few possibilities for SRG generators have been
    considered so far for nuclear systems. Can other choices for the SRG $G_s$ operator 
    help to control the growth of many-body forces? Can a generator be found to drive non-local 
    potentials to a more local form, so they can be used with quantum Monte Carlo methods? Or can the SRG 
    equations be formulated to directly produce a local projection and a perturbative residual interaction?	
	
    \I An apparent close connection between the block-diagonal
    generator SRG and the ``standard'' $\vlowk$ RG has been established
    empirically, but a formal demonstration of the connection and
    its limits has not been made.	
	
    \I Are there other bases for SRG evolution that would be advantageous?
     Recent momentum-space implementations will provide
     necessary checks of evolution in the harmonic oscillator
     basis, and the evolved interactions in this form will be directly
     applied to test MBPT in infinite matter and to test nuclear
     scaling. The brand new use of hyperspherical coordinates~\cite{Wendt:2013bla} should
     be particularly useful for visualization of many-body forces. 
     
    \I There are many open questions and problems involving operators.
    These include formal issues such as the scaling of many-body
    operators and technical issues such as how to handle boosts of
    operators that are not galilean invariant. And there are simply
    many applications that are in there infancy (e.g., electroweak
    processes).
 
    \I The flow to universal form exhibited by two-body interactions
    has been clear from the beginning of RG applications to nuclei.
    The nature of this behavior for many-body interactions
    or for other operators is an still open question, although under active investigation.
	
    \I How does the SRG relate to functional RG equations and field theory methods?

    \I How can we use more of the power of the RG itself?

  \end{itemize}
Developments involving renormalization group methods in low-energy nuclear physics are occurring at a rapid pace, so we can expect steady progress on these challenges and opportunities.


\ack
We thank S.~Bogner, A.~Dyhdalo, A.~Gezerlis, H.~Hergert, J.~Men\'endez, S.~More, R.~Perry, R.~Roth, A.~Schwenk, V.~Som\`a, M.~Voskresenskaya, and K.~Wendt 
for helpful discussions and comments.
This work was supported in part by the National Science Foundation 
under Grant No.~PHY--1002478,
the U.S.\ Department of Energy under Grant No.~DE-SC0008533 (SciDAC-3/NUCLEI project), 
and an award of computational resources from the Ohio Supercomputer Center.


\section*{References}

\bibliographystyle{iopart-num}
\bibliography{srg_refs}


\end{document}